\newcommand{\cL}{{\cal L}}
\newcommand{\cO}{{\cal O}}
\newcommand{\nt}{\notag\\}
\newcommand{\q}{\theta}
\newcommand{\cA}{{\cal A}}
\newcommand{\cB}{{\cal B}}
\newcommand{\cC}{{\cal C}}

\newcommand{\cK}{{\cal K}}
\newcommand{\cI}{{\cal I}}
\newcommand{\ep}{\epsilon}

\renewcommand{\a}{{\alpha}}

\newcommand{\be}{\begin{equation}}
\newcommand{\ee}{\end{equation}}
\newcommand{\bea}{\begin{eqnarray}}
\newcommand{\eaa}{\end{eqnarray}}

\newcommand{\cG}{{\cal G}}

\renewcommand{\a}{\alpha}

\newcommand{\D}{\Delta}

\newcommand{\cN}{{\cal N}}
\newcommand{\cD}{{\cal D}}

\newcommand{\cT}{{\cal T}}

\newcommand{\tr}{{\rm tr}}

\newcommand{\p}[1]{(\ref{#1})}
\newcommand{\bt}[1]{{\bar t}}
\newcommand{\ts}{\textstyle}

\newcommand{\half}{{\ts \frac{1}{2}}}
\newcommand \vev [1] {\langle{#1}\rangle}

\newcommand\lr[1]{{\left({#1}\right)}}

\newcommand{\eea}{\end{eqnarray}}

\def\lb#1{\label{#1}}

\documentclass[12pt]{article}
\usepackage{epsfig,amssymb,amsmath,psfrag,subfigure}
\usepackage{cellspace} 
\usepackage{xcolor}
\usepackage{tikz}

\usepackage{hyperref}

\catcode`\@=11
\def \tr {\mathop{\rm tr}\nolimits}

\def\numberbysection{\@addtoreset{equation}{section}
                     \def\theequation{\thesection.\arabic{equation}}}

\numberbysection
 
\begin{document}
  
\thispagestyle{empty}

\null\vskip-53pt \hfill
\begin{minipage}[t]{50mm}
CERN-PH-TH-2015-294 \\
DCPT-15/67 \\
LAPTH-066/15 
\end{minipage}

\vskip1.1truecm
\begin{center}
\vskip 0.0truecm

 {\Large\bf
%\titleline
 All three-loop four-point correlators of half-BPS operators  in  planar $\cN=4$ SYM 
  }
\vskip 1.0truecm
%\vfill

{\bf  Dmitry Chicherin$^{a}$,   James Drummond$^{b}$, Paul Heslop$^{c}$, Emery Sokatchev$^{a,d}$ \\
}

\vskip 0.4truecm
%\addresses
 $^{a}$ {\it LAPTH\,\footnote{Laboratoire d'Annecy-le-Vieux de Physique Th\'{e}orique, UMR 5108},   Universit\'{e} de Savoie, CNRS,
B.P. 110,  F-74941 Annecy-le-Vieux, France\\
  \vskip .2truecm
$^{b}$ {\it School of Physics \& Astronomy, University of Southampton,\\
 Highfield, Southampton, SO17 1BJ, United Kingdom}\\
  \vskip .2truecm
$^{c}$ {\it  Mathematics Department, Durham University, 
Science Laboratories,
 \\
South Rd, Durham DH1 3LE,
United Kingdom} \\
 \vskip .2truecm
$^{d}$ Physics Department, Theory Unit, CERN,\\ CH -1211, Geneva 23, Switzerland 
} \\
\end{center}

\vskip .3truecm

\centerline{\bf Abstract} % \normalsize
\medskip
\noindent

We obtain  the planar correlation function of four half-BPS
operators of arbitrary weights, up to three loops. Our method exploits
only elementary properties of the integrand of the planar correlator, such as its symmetries and singularity structure. This allows us to write down a general ansatz for the integrand. The coefficients in the ansatz are fixed by means of a powerful light-cone OPE relation between correlators with different weights. Our result is formulated in terms of 
a limited number of functions built from known one-, two- and three-loop conformal integrals. These results are useful for checking recent integrability predictions for the OPE structure constants.  

\newpage

\thispagestyle{empty}

{\small \tableofcontents}

\newpage
\setcounter{page}{1}\setcounter{footnote}{0}

%%%%%%%%%

%%%%%%%%%

\section{Introduction}

The correlation functions of gauge invariant operators are the natural object  in a conformal field theory like $\cN=4$ super-Yang-Mills. Among them a privileged role is played by the correlators of half-BPS scalar operators. They form short superconformal multiplets whose lowest-weight states are annihilated by half of the Poincar\'e supercharges. The conformal dimensions and more generally, the two- and three-point correlation functions of the half-BPS operators are protected from quantum corrections, but the four-point functions are not. The OPE spectrum of two half-BPS operators is rich, coupling-dependent and generically contains unprotected (long) supermultiplets. Thus, the four-point correlators of half-BPS operators encode some genuinely dynamical information and hence are interesting objects to study.

Such correlators have attracted a lot of attention in the context of the AdS/CFT correspondence \cite{123}. In its simplest form it states that type IIB supergravity on an  $AdS_5 \times S^5$ background is dual to the limit of the gauge theory where the 't 
Hooft coupling $a=g^2 N_c/ (4 \pi^2)$ is infinite and the number of colours $N_c$ is large. The compactification of  type IIB supergravity on $S^5$ results in an infinite tower of (generically massive) Kaluza-Klein modes. According to the AdS/CFT conjecture, the half-BPS operators $\cO^{(k)}$ of dimension $k$ are dual to the KK modes transforming in the irrep $[0,k,0]$ of $SU(4)\sim SO(6)$.

Among all half-BPS operators the simplest and widely studied one is
that of minimal weight $k=2$. The corresponding supermultiplet
$\cT=\cO^{(2)}+ \ldots$ is very special, as it contains the conserved
R symmetry current, the stress-energy tensor and the Lagrangian of the ${\cal
  N}=4$ theory. It is dual to the graviton multiplet of the $AdS_5
\times S^5$ supergravity comprising the {\it massless} KK modes.  In
perturbation theory the loop corrections are generated by integrated
Lagrangian insertions. The {\it integrand} of the $\ell-$loop
correction to the $n-$point correlator of the stress-tensor multiplet
is most naturally obtained from the  correlator $\vev{\cT(1) \ldots
  \cT(n+\ell)}$ evaluated at the lowest perturbative (Born)
level. This approach was developed and successfully used for calculating the two-loop four-point function $\vev{\cO^{(2)}\cO^{(2)}\cO^{(2)}\cO^{(2)}}$ in \cite{Eden:2000mv}. More recently,  by exploiting a hidden permutation symmetry of the correlators $\vev{\cT(1) \ldots \cT(4+\ell)}$ the planar integrand of this four-point function  was found up to seven loops \cite{Eden:2011we,Eden:2012tu,Ambrosio:2013pba}.

Apart from the  simplest case of the stress-tensor supermultiplet,  the correlators of half-BPS operators of arbitrary weights deserve equal attention. From the AdS point of view, to start bringing out the flavor of the more involved 
ten-dimensional physics one has to go beyond the massless sector of the 
theory and consider new examples of supergravity-induced four-point 
correlators involving {BPS operators of higher dimension}. The first
steps in this direction were made in
\cite{Heslop:2002hp,Arutyunov:2002fh}. In~\cite{Arutyunov:2002fh} the four-point correlator $\vev{\cO^{(3)}\cO^{(3)}\cO^{(3)} \cO^{(3)}}$ was obtained at two loops and the matching AdS supergravity amplitude of {\it massive} KK states was constructed. Later on, the general case of  four half-BPS operators of equal  weights $\vev{\cO^{(k)}\cO^{(k)}\cO^{(k)} \cO^{(k)}}$ up to two loops was considered in \cite{Arutyunov:2003ae,Arutyunov:2003ad}.\footnote{Classes of correlators with different weights have also been studied in AdS supergravity in \cite{D'Hoker:2000dm,Berdichevsky:2007xd,Uruchurtu:2011wh}.} This study revealed a degeneracy phenomenon: in the large $N_c$ limit only one (at one loop) and two (at two loops) distinct functions of the conformally invariant cross-ratios described the whole variety of $SU(4)$ channels in these correlators. The degeneracy is lifted at strong coupling (AdS supergravity).

Besides the AdS/CFT duality, another good reason for studying four-point correlators  of {\it unequal} BPS weights comes from the recent advances in integrability. In the paper~\cite{Vieira:2013wya}, which generalizes the results of \cite{Escobedo:2010xs,Gromov:2012vu} to the non-compact case,  the three-point correlators of two half-BPS operators and one unprotected operator in the $SL(2)$ sector were studied in the one-loop approximation.
An integrability based conjecture was made for the values of the corresponding structure constants. Recently, this result was extended to two loops in \cite{Basso:2015zoa} and to three loops in \cite{Eden:2015ija,Basso:2015eqa}.
In the absence of direct calculations of the relevant three-point functions,  use can be made of the OPE of the four-point correlators of half-BPS operators, which produces sum rules for the structure constants. Such tests of the integrability conjecture are most sensitive if the  four-point correlators involve half-BPS operators of {\it different} weights. Two particular cases, the correlators $\vev{\cO^{(2)}\cO^{(2)}\cO^{(k)}\cO^{(k)}}$ and  $\vev{\cO^{(2)}\cO^{(3)}\cO^{(3)}\cO^{(4)}}$ have been computed to two loops in \cite{D'Alessandro:2005dq} and \cite{Chicherin:2014esa}, respectively. These results confirm  the prediction of \cite{Vieira:2013wya}. 
 Some preliminary three-loop results of the present paper have already been used by the authors of \cite{Basso:2015eqa} as a valuable check of their findings.
 
Finally, another motivation for the study of the whole class of correlators of half-BPS operators is the search for integrability directly at the level of the multipoint correlation functions. The recent advances in integrability give us strong evidence that the spectrum of anomalous dimensions (i.e., the two-point functions) and the OPE structure constants (i.e., the three-point functions) are integrable in planar $\cN=4$ SYM. It is well known that all correlation functions in a conformal theory can eventually be built from these two elementary ingredients. It is therefore reasonable to expect some kind of integrable structure in the higher-point functions as well. The results of the present work give indications in this direction.

In this paper we address the problem of finding the perturbative corrections to the four-point functions of half-BPS operators of arbitrary weights $\vev{\cO^{(k_1)}\cO^{(k_2)}\cO^{(k_3)}\cO^{(k_4)}}$ up to three loops. We apply and further develop the method proposed in \cite{Eden:2011we,Eden:2012tu}. The idea is not to compute such correlators using standard Feynman rules but rather to predict their {\it integrands}. As mentioned above, the integrand of the $\ell-$loop correction to the four-point function can be viewed as a $(4+\ell)-$point correlator with $\ell$ Lagrangian insertions, calculated at Born level. This is a rational function of the $(4+\ell)$ space-time points having certain simple properties. They follow from $\cN=4$ superconformal symmetry and also from the known short-distance physical singularities. This allows us to write down the most general ansatz in the form of a polynomial numerator with given conformal weights at each points, and a fixed universal denominator accounting for the expected singularities. We then classify all possible numerators. Their number is drastically reduced if we restrict ourselves only to {\it planar} configurations. 

The next step is to find a way to fix the arbitrary coefficients in the ansatz. Using the light-cone super-OPE of two half-BPS operators, we derive a very simple relation between two correlators with shifted weights at two points,  $k_1 k_2 k_3 k_4$ and $k_1+1, k_2+1, k_3 k_4$, in the limit where these two points become light-like separated.  Iterating this relation imposes many consistency conditions on the coefficients in our ansatz, for all possible values of the BPS weights. These conditions allow us to determine all the coefficients at two loops and all but one at three loops. The latter can be fixed by adapting the Euclidean logarithmic singularity criterion on the integrand elaborated in \cite{Eden:2012tu}. 

Our main result is that all possible correlators of four half-BPS operators, in the planar limit and up to three loops, are described by a limited number of conformally invariant functions (9 at two loops and 55 at three loops). This result, which we call {\it uniformity},  generalises the degeneracy of the one- and two-loop correlators with equal weights observed in \cite{Arutyunov:2003ae,Arutyunov:2003ad}. The various functions are made of a small number of one-, two- and three-loop planar conformal integrals, all of which have already appeared in the simplest correlator $\vev{\cO^{(2)}\cO^{(2)}\cO^{(2)}\cO^{(2)}}$ at three loops \cite{Eden:2011we}. When we convert the integrands that our method produces into conformal integrals, we use a number of identities for the latter \cite{Drummond:2006rz,Eden:2011we}. This reduces the basis of independent integrals in the final  result.  We would like to emphasise that, unlike the case $2222$ where planarity is automatic (absence of non-planar Feynman graphs), this is not so in the general case  $k_1 k_2 k_3 k_4$. So, planarity is a key ingredient in our construction. The fact that we are able to unambiguously predict the entire class of planar correlators of half-BPS operators to three loops, using only their elementary properties, can be interpreted as evidence for a new integrable structure. 

Having obtained an expression for all the three-loop correlators, %(under certain analytic assumptions?) 
we perform an OPE analysis of the results in perturbation theory. We focus on the leading twist contributions to each contributing $su(4)$ channel present in the joint OPE of $\mathcal{O}^{(k_1)} \mathcal{O}^{(k_2)}$ and $\mathcal{O}^{(k_3)} \mathcal{O}^{(k_4)}$ for many different values of $k_1,k_2,k_3,k_4$. We are able to verify predictions from \cite{Vieira:2013wya,Basso:2015zoa,Eden:2015ija,Basso:2015eqa} for three-point functions of two protected operators and one unprotected one. We also use this approach to formulate many consistency checks on the results obtained from the construction of the Born-level correlators. We are also able to relate the uniformity property of the Born-level correlators to the appearance of wrapping corrections to three-point functions in the approach of \cite{Basso:2015zoa}.

The paper is organised as follows. In Section 2 we give the basic definitions and recall some properties of the correlators we discuss. Then we summarise our two- and three-loop results in the form of two tables. The tables list the coefficients in front of the two- and three-loop conformal integrals that form a basis for all the correlators. The finite size of our tables reflects the fact the number of independent functions is limited (uniformity). We only display our results for the {\it integrals}, not the {\it integrands} due to size limitations. Section 3 contains a detailed description of the method we use to predict the integrand. We recall its basic properties and formulate the most general ansatz reflecting these properties. We then explain the role of planarity for drastically restricting the number of possible topologies of the integrands. The examination of the light-cone super-OPE of two half-BPS operators leads us to a powerful relation between pairs of correlators with shifted weights. In this section we also recall the Euclidean log criterion from \cite{Eden:2012tu} and the conformal integral identities from \cite{Drummond:2006rz,Eden:2011we}. Section 4 is devoted to an independent check of our results via the standard OPE analysis of the integrated correlation function. Section 5 contains our conclusions and possible further developments. Appendix A contains some details of the proof of the uniformity property. Appendix B summarises the necessary information on the superconformal OPE of two half-BPS operators.

\section{Generalities and summary of the results}

The lowest component of a half-BPS multiplet in ${\cal N}=4$ SYM is a real scalar field of dimension $k$ (with $k\geq2$) transforming in the irrep $[0,k,0]$ of the R symmetry group $SO(6)\sim SU(4)$.  In terms of the elementary fields it can be realised as a single-trace operator
\bea
\label{bps'}
\tr(\phi^{\{I_1}\ldots \phi^{I_k\}})\,.
\eea
Here $\phi^I$, $I=1,\ldots, 6$ are the ${\cal N}=4$ SYM scalars and
$\{,\}$ denotes  traceless and weighted symmetrization.
A convenient way of handling the SO(6) indices is to project the operator (\ref{bps'}) onto the highest weight state of the irrep $[0,k,0]$. This can be done with the help of a complex null vector $Y^I$ ($Y^I Y^I=0$):
\bea
\label{bps}
{\cal O}^{(k)}(x,y)= Y^{I_1}\ldots Y^{I_k}\tr(\phi^{I_1}\ldots \phi^{I_k})\, .
\eea

We start by summarizing the general properties of the four-point correlator of
half-BPS operators in the ${\cal N}=4$ SYM theory, 
\begin{align}\label{gampl}
 \cG_{k_1 k_2 k_3 k_4} & = \langle \cO^{(k_1)}(x_1,Y_1) \,\cO^{(k_2)}(x_2,Y_2)\, \cO^{(k_3)}(x_3,Y_3)\, \cO^{(k_4)}(x_4,Y_4)\rangle \, , 
\end{align}
where $k_1\geq k_2\geq  k_3\geq  k_4\geq 2$ are the weights of the four half-BPS operators. The allowed combinations of four label are those for which $\sum_{i=1}^4 k_i = 2n$ and $k_1 \leq k_2+k_3+k_4$, so that it is possible to connect the four points with free propagators without leaving any scalars unpaired. For our purposes, a  further restriction comes from the fact that the so-called `extremal' (with $k_1 = k_2+k_3+k_4$) and `next-to-extremal'  (with $k_1 = k_2+k_3+k_4-2$) correlators are protected \cite{D'Hoker:1999ea,Eden:1999kw,Eden:2000gg}, i.e. for them  $\cG^{\rm loop}$ does not exist. This amounts to requiring $ k_i < \sum_{j\neq i} k_j-2$.

The correlator \p{gampl} splits into two parts, 
\begin{align}\label{freeandloop}
\cG_{k_1 k_2 k_3 k_4} = \cG^0_{k_1 k_2 k_3 k_4} +\cG^{\rm loop}_{k_1 k_2 k_3 k_4}\,.
\end{align}
The first part is a rational function of the space-time coordinates and corresponds to the Born (free) approximation. The second part includes all the loop corrections which involve non-trivial functions originating from Feynman integrals. 

The expression for $ \cG^0$ is a polynomial in the elementary propagators (Wick contractions) of two free scalars 
\begin{align}\label{25}
d_{ij} = d_{ji} \equiv 4\pi^2\vev{\phi(x_i,y_i) \phi(x_j,y_j)} =  \frac{y^2_{ij}}{x^2_{ij}}\,,
\end{align}
where $y^2_{ij} = Y_i\cdot Y_j$ and $x^2_{ij}=(x_i-x_j)^2$.  Then we can write the general expression 
\begin{align}\label{26}
\cG^0_{k_1 k_2 k_3 k_4} = \sum_{\{a_{ij}\}}\left(\prod_{1 \leq i < j \leq 4} (d_{ij})^{a_{ij}} \right)C_{\{a_{ij}\}}\,,
\end{align}
where $a_{ij}=a_{ji} \geq 0$ (with $i\neq j$) are  integers such that $\sum_{j\neq i} a_{ij} = k_i$ for each $i=1,\ldots,4$. The sum in \p{26} goes over all possible partitions $\{a_{ij}\}$ satisfying the above condition. Each term in the sum has the required conformal and R-symmetry weights at each of the four points. The coefficients $C_{\{a_{ij}\}}$ are numbers obtained by calculating the colour and combinatorial factors of the different free Feynman diagrams. Here is a simple example:
\begin{align}\label{}
\cG^0_{2222}= \frac{N_c^2}{(4 \pi^2)^4} \,\
\left( d_{12} d_{23}d_{34}d_{14}  \, + \, 
 d_{12}d_{24}d_{34}d_{13}  \, + \, 
 d_{13}d_{23}d_{24}d_{14}  \right)\,,
\end{align}
where we have displayed only the connected part and the colour factor
is given for $N_c>>1$.\footnote{Our colour convention is $\tr(t_a t_b)
  = {\delta_{ab}}/{2}$.}

In principle, the interacting (loop) part of the correlator $\cG^{\rm loop}$ has a structure similar to \p{26}. The main difference is that the constant coefficients $C_{\{a_{ij}\}}$ are replaced by functions of the two independent conformally invariant cross-ratios
\begin{align}\label{cr}
u=\frac{x^2_{12}\, x^2_{34}}{x^2_{13}\, x^2_{24}}\,, \qquad
v=\frac{x^2_{14}\, x^2_{23}}{x^2_{13}\, x^2_{24}}\,.
\end{align}
Thus, in general we can write
\begin{align}\label{29}
\cG^{\rm loop}_{k_1 k_2 k_3 k_4} = \sum_{\{a_{ij}\}}\prod_{1 \leq i < j \leq 4} (d_{ij})^{a_{ij}} F_{\{a_{ij}\}}(u,v)\,,
\end{align}
where each function admits a perturbative expansion in the `t Hooft coupling $a=g^2 N_c/ (4 \pi^2)$,
\begin{align}\label{}
F_{\{a_{ij}\}}(u,v) = \sum_{\ell \geq 1} a^\ell F^{(\ell)}_{\{a_{ij}\}}(u,v)\,.
\end{align}

$\cN=4$ superconformal symmetry puts additional restrictions on the
coefficient functions in \p{29}. According to the `partial
non-renormalisation' theorem of Ref. \cite{Eden:2000bk,Heslop:2002hp} (for 
alternative derivations see also \cite{Nirschl:2004pa,Eden:2011we}), the interacting
part of the correlator takes the {\it factorised form} 
\begin{align}\label{211}
\cG^{\rm loop}_{k_1 k_2 k_3 k_4} = C_{k_1 k_2 k_3 k_4}\, R(1,2,3,4)\times \sum_{\{b_{ij}\}}\left(\prod_{1 \leq i < j \leq 4} (d_{ij})^{b_{ij}} \right) \frac{F_{\{b_{ij}\}}(u,v)}{x^2_{13} x^2_{24}}\,,
\end{align} 
where
\begin{align}\label{219}
C_{k_1 k_2 k_3 k_4} =   \frac{1}{2}\, \left(\frac{N_c}{2}\right)^{\half \sum k_i -2}\,  \frac{k_1 k_2 k_3 k_4}{(4\pi^2)^{\frac{1}{2}\sum k_i}}
\end{align} 
is a normalisation factor and $R$ is a {\it universal rational prefactor} carrying $SU(4)$ weight 2 and conformal weight 1 at each point. 
Explicitly,
\begin{align}\label{R4}
R(1,2,3,4) & =  d_{12}^2 d_{34}^2 x_{12}^2 x_{34}^2 +  d_{13}^2 d_{24}^2 x_{13}^2 x_{24}^2 +  d_{14}^2 d_{23}^2 x_{14}^2 x_{23}^2
\notag\\ 
& +d_{12} d_{23} d_{34} d_{14} ( x_{13}^2 x_{24}^2 -x_{12}^2 x_{34}^2 - x_{14}^2 x_{23}^2) \notag \\
& +d_{12} d_{13} d_{24} d_{34} ( x_{14}^2 x_{23}^2 -x_{12}^2 x_{34}^2 - x_{13}^2 x_{24}^2) \notag \\
& +d_{13} d_{14} d_{23} d_{24} ( x_{12}^2 x_{34}^2 -x_{14}^2 x_{23}^2 - x_{13}^2 x_{24}^2)\,,
\end{align}
which is fully symmetric in the points $1,2,3,4$. The denominator $x^2_{13} x^2_{24}$ supplies the missing conformal weights, so that the functions $F_{\{b_{ij}\}}(u,v)$ are conformally invariant.  The new partitions $\{ b_{ij} \}$ in \p{211} satisfy the modified conditions $\sum_{j\neq i} b_{ij} = k_i-2$ for each $i=1,\ldots,4$. For the purpose of presentation we organise $\{ b_{ij} \}$ into sextuples of integers,
\begin{align}\label{}
\{ b_{ij} \}=\{ b_{12},  b_{13},  b_{14},  b_{23},  b_{24},  b_{34} \}\, .
\end{align}

The simplest example again is 
\begin{align}\label{213}
\cG^{\rm loop}_{2222} = \frac{2 N_c^2}{(4\pi^2)^4}\, R(1,2,3,4)\times \frac{F(u,v)}{x^2_{13} x^2_{24}}\,.
\end{align}
Here $\{b_{ij}\} = \{0,0,0,0,0,0\}$ and the dynamical information is encoded in the single 
 function $F(u,v)$. This is not the only case where the sum on the right-hand side of \p{211} contains only one term.
There are several infinite families of such correlators.
A straightforward generalization of $\cG^{\rm loop}_{2222}$ 
is the correlator $\cG^{\rm loop}_{kk22}$ with $k \geq 2$. In this case there is a unique $y$-structure 
encoded by the sextuple $\{b_{ij}\} = \{k-2,0,0,0,0,0\}$.

More generally, the correlators with weights 
$k_1=a+b+c+2$, $k_2=a+2$, $k_3=b+2$, $k_4=c+2$ (or equivalently, $k_1 = k_2+k_3+k_4-4$) 
are characterised by the unique sextuple $\{b_{ij}\}=\{a,b,c,0,0,0\}$. 
Such correlators are known as `near extremal' \cite{D'Hoker:2000dm} or  `next-next-to-extremal' \cite{Uruchurtu:2011wh}. Another three-parameter family of correlators with a unique
$y$-structure are those containing one (or more) weight-two operator. For example, if $k_4=2$  the unique sextuple is $\{b_{ij}\}=\{a,b,0,c,0,0\}$
corresponding to weights $k_1=a+b+2$, $k_2=a+c+2$, $k_3=b+c+2$, $k_4=2$.

In addition to the conformal and R-symmetry properties, the correlator may be further restricted by the permutation symmetry of the external points. If two or more of the labels $k_i$ are equal, the operators $\cO^{(k_i)}$ are identical and the correlator must be invariant under the permutations of the  corresponding points. This symmetry organises the propagator structures $\prod (d_{ij})^{b_{ij}}$ and the coefficient functions $F_{\{b_{ij}\}}$ into equivalence classes. 

A further and less obvious symmetry takes place if some $k_i=2$. In this case $\cO^{(2)}$ is the superconformal primary of the energy-momentum supermultiplet, which also contains the Lagrangian of the theory. This results in a rather powerful permutation symmetry between (some of) the external points and the Lagrangian insertion points (see \cite{Eden:2011we} for details).

\subsection{Summary of the results}\label{s2.1}

In this subsection we summarise our results for all possible choices of
the four labels $k_i$, up to three loops. We restrict ourselves to the planar limit $N_c \to \infty$
{and planarity of the resulting correlator graphs will be a key input.}

The generic expression for the conformally invariant functions $F_{\{b_{ij}\}}(u,v)$ is given in terms of a set of one-, two- and three-loop integrals (with the cross-ratios defined in \p{cr}): 
\begin{eqnarray}
{F^{(1)}}/{x^2_{13} x^2_{24}} & =&  g_{1234}\,  \notag
\\[2mm] \notag
{F^{(2)}}/{x^2_{13} x^2_{24}}  & =& c_h^1 h_{1 2;3 4} + c_h^2 h_{1 3;2 4}    +c_h^3 h_{1 4;2 3}  
  + 
 \frac12 \lr{c_{gg}^1 x_{12}^2x_{34}^2+ c_{gg}^2 x_{13}^2 x_{24}^2+c_{gg}^3  x_{14}^2x_{23}^2}  [g_{1 2 3 4}]^2\,  \notag
\\[2mm] \notag
{F^{(3)}}/{x^2_{13} x^2_{24}}  & = & c_{gh}^1 x^2_{12} x^2_{34} \, (g\times h)_{1 2;3 4} \, + \,
c_{gh}^2 x^2_{13} x^2_{24} \, (g\times h)_{1 3;2 4} \, + \, c_{gh}^3  x^2_{14} x^2_{23} \, (g\times h)_{1 4;2 3} 
\nonumber \\[2mm]
& + & c_L^1 L_{1 2;3 4} \, + \, c_L^2 L_{1 3;2 4} \, + \, c_L^3 L_{1 4;2 3}   \, + \,  c_E^1 E_{1 2;3 4} \, + \, c_E^2 E_{1 3;2 4} \, + \, c_E^3 E_{1 4;2 3}  
 \nonumber \\[2mm] \label{loop3F} 
& + &  \frac12 (c_H^1 + c_H^2 1/v) H_{1 2;3 4} \, + \,  \frac12(c_H^3 +c_H^4 u/v) H_{1 3;2 4} \, + \,\frac12(c_H^5 +c_H^6 u) H_{1 4;2 3}  \, , 
\end{eqnarray}
where the conformal integrals are defined as follows:
\begin{eqnarray} \label{eq:14}
 g_{1234}  &=& -\frac{1}{4 \pi^2}
\int \frac{d^4x_5}{x_{15}^2 x_{25}^2 x_{35}^2 x_{45}^2}   \nonumber \\
 h_{1 2;3 4}  &=&  \frac{x^2_{34}}{(4 \pi^2)^2}
\int \frac{d^4x_5 \, d^4x_6}{(x_{15}^2 x_{35}^2 x_{45}^2) x_{56}^2
(x_{26}^2 x_{36}^2 x_{46}^2)}   \nonumber \\
E_{1 2;3 4} & = & \frac{x^2_{23} x^2_{24}}{(-4 \pi^2)^3}
\int \frac{d^4x_5 \, d^4x_6 \, d^4x_7 \ x^2_{16}}{(x_{15}^2 x_{25}^2 x_{35}^2)
x_{56}^2 (x_{26}^2 x_{36}^2 x^2_{46}) x^2_{67} (x_{17}^2 x_{27}^2 x_{47}^2)}
 \nonumber \\
L_{1 2;3 4} & = & \frac{x^4_{34}}{(-4 \pi^2)^3}
\int \frac{d^4x_5 \, d^4x_6 \, d^4x_7}{(x_{15}^2 x_{35}^2 x_{45}^2) x_{56}^2
(x_{36}^2 x_{46}^2) x^2_{67} (x_{27}^2 x_{37}^2 x_{47}^2)}  \nonumber \\
  {(g\times h)}_{1 2;3 4} &=&{x_{12}^2 x_{34}^4\over (-4\pi^2)^3}\int 
   \frac{d^4x_5d^4x_6 d^4x_7}{(x_{15}^2  x_{25}^2  x_{35}^2 x_{45}^2) 
 (x_{16}^2  x_{36}^2x_{46}^2) 
    (x_{27}^2  x_{37}^2   x_{47}^2) x_{67}^2} \nonumber\\
H_{1 2;3 4} & = &   \frac{x_{41}^2 x_{23}^2
    x_{34}^2 
}{(-4 \pi^2)^3}
\int \frac{d^4x_5 \, d^4x_6 \, d^4x_7 \ x^2_{57}}{(x_{15}^2 x_{25}^2 x_{35}^2
x^2_{45}) x_{56}^2 (x_{36}^2 x^2_{46}) x^2_{67} (x_{17}^2 x_{27}^2 x^2_{37}
x_{47}^2)} \ .
\end{eqnarray}

The one-loop correlators are always the same independently of the partition $\{b_{ij}\}$, only the normalisation factor \p{219}  changes (see \cite{Arutyunov:2003ae} for the case of equal weights). Our two- and three-loop results are presented in the form of two tables where the values of the numerical coefficients in front of the various integrals in \p{loop3F} are listed.   Some of the two-loop results  in table  \ref{tab1} were obtained in the past through direct Feynman graph calculations \cite{Eden:2000mv,Arutyunov:2002fh,Arutyunov:2003ad,D'Alessandro:2005dq,Uruchurtu:2011wh,Chicherin:2014esa}. The three-loop result for the case $\cG_{2222}$ was first obtained in Ref.~\cite{Eden:2011we} by a method similar to the one used in the present paper. The other results shown in Table  \ref{tab2} are new.

We show that the number of functions that encode the quantum corrections of all the correlators 
at two and three loops is finite. There are 9 independent functions at two loops and 55 functions at three loops.
These functions $F_{\{b_{ij}\}}$ have the form \p{loop3F} with the numerical coefficients listed in tables \ref{tab1}, \ref{tab2}. Some of the lines in the tables are double, which means that the two sextuples come with the same function. 

We say that a pair of sextuples are two-loop-equivalent, $\{b_{ij}\} \sim \{b'_{ij}\}$, if 
some of the entries $b_{ij}, b'_{ij} \geq 1$ are different but all the entries $b_{ij}, b'_{ij} =0$ are the same. The corresponding two-loop functions are equal, $F^{(2)}_{\{b_{ij}\}} = F^{(2)}_{\{b'_{ij}\}}$.
Similarly,  a pair of sextuples are three-loop-equivalent, $\{b_{ij}\} \sim \{b'_{ij}\}$, 
if only their entries $b_{ij}, b'_{ij} \geq 2$ can possibly differ but all the entries $b_{ij}, b'_{ij} =0,1$ are the same. The corresponding three-loop functions are equal, $F^{(3)}_{\{b_{ij}\}} = F^{(3)}_{\{b'_{ij}\}}$.

\newcommand{\mysextuple}[6]{$\{ \makebox[12pt][c]{$#1$},\makebox[12pt][c]{$#2$},\makebox[12pt][c]{$#3$},
\makebox[12pt][c]{$#4$},\makebox[12pt][c]{$#5$},\makebox[12pt][c]{$#6$} \}$}

\newcommand*{\diam}{0.5pt}
\newcommand*{\len}{0.4}

\setlength\cellspacetoplimit{2pt}
\setlength\cellspacebottomlimit{2pt}

To extract a particular correlator $\mathcal{G}^{\rm loop}_{k_1 k_2 k_3 k_4}$ from the tables, we first need to enumerate 
all the relevant $y$-structures encoded by the sextuples $\{b_{ij}\}$ in  \p{211}, satsifying the conditions $\sum_{j\neq i} b_{ij} = k_i-2$ for each $i=1,\ldots,4$. . The coefficients of the various integrals making up the functions $F_{\{b_{ij}\}}$ (one representative of each crossing equivalence class) 
are then listed in the tables.
\begin{table}
\begin{center}
\begin{tabular}{|Sc|Sc|Sc|Sc|Sc|Sc|Sc|Sc|}
\hline
 $\{b_{ij}\}$ & 
 \begin{tikzpicture}
\filldraw[black] (0,0) circle (\diam) node[anchor=east]{$\scriptscriptstyle 4$}; 
\filldraw[black] (0,\len) circle (\diam) node[anchor=east]{$\scriptscriptstyle 1$};
\filldraw[black] (\len,0) circle (\diam) node[anchor=west]{$\scriptscriptstyle 3$};
\filldraw[black] (\len,\len) circle (\diam) node[anchor=west]{$\scriptscriptstyle 2$}; 
 \draw[dashed] (0,\len) -- (\len,\len) -- (\len,0) -- (0,0) -- (0,\len);
 \draw[dashed] (0,\len) -- (\len,0);
 \draw[dashed] (0,0) -- (\len,\len);
\end{tikzpicture} & $c_{gg}^1$ & $c_{gg}^2$ & $c_{gg}^3$ & $c_h^1$ & $c_h^2$ & $c_h^3$ \\  \hline
 \mysextuple{0}{0}{0}{0}{0}{0} & 
\begin{tikzpicture}
\filldraw[black] (0,0) circle (\diam); 
\filldraw[black] (0,\len) circle (\diam);
\filldraw[black] (\len,0) circle (\diam);
\filldraw[black] (\len,\len) circle (\diam); 
\end{tikzpicture} & 1 & 1 & 1 & 2 & 2 & 2 \\  \hline
 \mysextuple{\beta_1}{0}{0}{0}{0}{0} & 
\begin{tikzpicture}
\filldraw[black] (0,0) circle (\diam); 
\filldraw[black] (0,\len) circle (\diam);
\filldraw[black] (\len,0) circle (\diam);
\filldraw[black] (\len,\len) circle (\diam);
 \draw (0,\len) -- (\len,\len);
\end{tikzpicture}& 0 & 1 & 1 & 1 & 2 & 2 \\  \hline
 \mysextuple{\beta_1}{\beta_2}{0}{0}{0}{0} & 
\begin{tikzpicture}
\filldraw[black] (0,0) circle (\diam); 
\filldraw[black] (0,\len) circle (\diam);
\filldraw[black] (\len,0) circle (\diam);
\filldraw[black] (\len,\len) circle (\diam);
 \draw (0,\len) -- (\len,\len);
 \draw (0,\len) -- (\len,0);
\end{tikzpicture} & 0 & 0 & 1 & 1 & 1 & 2 \\  \hline
\begin{tabular}{c}
 \mysextuple{ \beta_1 }{ \beta_2 }{ 0 }{ \beta_3 }{ 0 }{ 0 }\\
 \mysextuple{ \beta_1 }{ \beta_2 }{ \beta_3}{  0 }{ 0 }{ 0 }
\end{tabular} &
\begin{tabular}{c}
\begin{tikzpicture}
\filldraw[black] (0,0) circle (\diam); 
\filldraw[black] (0,\len) circle (\diam);
\filldraw[black] (\len,0) circle (\diam);
\filldraw[black] (\len,\len) circle (\diam);
 \draw (0,\len) -- (\len,\len) -- (\len,0) -- (0,\len);
\end{tikzpicture}  
\\ 
\begin{tikzpicture}
\filldraw[black] (0,0) circle (\diam); 
\filldraw[black] (0,\len) circle (\diam);
\filldraw[black] (\len,0) circle (\diam);
\filldraw[black] (\len,\len) circle (\diam);
 \draw (0,0) -- (0,\len) -- (\len,\len);
 \draw (0,\len) -- (\len,0);
\end{tikzpicture}
\end{tabular} & 0 & 0 & 0 & 1 & 1 & 1 \\  \hline
 \mysextuple{0}{0}{\beta_1}{\beta_2}{0}{0} &
 \begin{tikzpicture}
\filldraw[black] (0,0) circle (\diam); 
\filldraw[black] (0,\len) circle (\diam);
\filldraw[black] (\len,0) circle (\diam);
\filldraw[black] (\len,\len) circle (\diam);
 \draw (0,0) -- (0,\len);
 \draw (\len,\len) -- (\len,0); 
\end{tikzpicture} & 1 & 1 & 0 & 2 & 2 & 0 \\  \hline
 \mysextuple{\beta_1}{0}{\beta_2}{\beta_3}{0}{0} &
\begin{tikzpicture}
\filldraw[black] (0,0) circle (\diam); 
\filldraw[black] (0,\len) circle (\diam);
\filldraw[black] (\len,0) circle (\diam);
\filldraw[black] (\len,\len) circle (\diam);
 \draw (0,0) -- (0,\len) -- (\len,\len) -- (\len,0); 
\end{tikzpicture} & 0 & 1 & 0 & 1 & 2 & 0 \\  \hline
 \mysextuple{\beta_1}{\beta_2}{\beta_3}{\beta_4}{0}{0} &
\begin{tikzpicture}
\filldraw[black] (0,0) circle (\diam); 
\filldraw[black] (0,\len) circle (\diam);
\filldraw[black] (\len,0) circle (\diam);
\filldraw[black] (\len,\len) circle (\diam);
 \draw (0,0) -- (0,\len) -- (\len,\len) -- (\len,0); 
 \draw (0,\len) -- (\len,0);
\end{tikzpicture} & 0 & 0 & 0 & 1 & 1 & 0 \\  \hline
 \mysextuple{0}{\beta_1}{\beta_2}{\beta_3}{\beta_4}{0} & 
\begin{tikzpicture}
\filldraw[black] (0,0) circle (\diam); 
\filldraw[black] (0,\len) circle (\diam);
\filldraw[black] (\len,0) circle (\diam);
\filldraw[black] (\len,\len) circle (\diam);
 \draw (0,0) -- (0,\len);
 \draw (\len,\len) -- (\len,0); 
 \draw (0,\len) -- (\len,0);
 \draw (0,0) -- (\len,\len);
\end{tikzpicture} & 1 & 0 & 0 & 2 & 0 & 0 \\  \hline
 \mysextuple{\beta_1}{\beta_2}{\beta_3}{\beta_4}{\beta_5}{0} &
\begin{tikzpicture}
\filldraw[black] (0,0) circle (\diam); 
\filldraw[black] (0,\len) circle (\diam);
\filldraw[black] (\len,0) circle (\diam);
\filldraw[black] (\len,\len) circle (\diam);
 \draw (0,0) -- (0,\len) -- (\len,\len) -- (\len,0); 
 \draw (0,\len) -- (\len,0);
 \draw (0,0) -- (\len,\len);
\end{tikzpicture} & 0 & 0 & 0 & 1 & 0 & 0 \\  \hline
 \mysextuple{\beta_1}{\beta_2}{\beta_3}{\beta_4}{\beta_5}{\beta_6} & 
\begin{tikzpicture}
\filldraw[black] (0,0) circle (\diam); 
\filldraw[black] (0,\len) circle (\diam);
\filldraw[black] (\len,0) circle (\diam);
\filldraw[black] (\len,\len) circle (\diam);
 \draw (0,\len) -- (\len,\len) -- (\len,0) -- (0,0) -- (0,\len);
 \draw (0,\len) -- (\len,0);
 \draw (0,0) -- (\len,\len);
\end{tikzpicture} & 0 & 0 & 0 & 0 & 0 & 0 \\  \hline
\end{tabular}
\end{center}
\caption{  \label{tab1} Numerical coefficients specifying the two-loop functions $F^{(2)}_{\{b_{ij}\}}$, Eq.~\p{loop3F}.
All possible sextuples $\{b_{ij}\}$ (up to crossing permutations) are listed.
The parameters $\beta_i\geq 1$  in the different lines are independent. The graphs depict the $y$-structures encoded by the sextuples $\{b_{ij}\}$. A line between  points $i$ and $j$
corresponds to $(y_{ij}^2)^{b_{ij}}$ with $b_{ij} \geq 1$.}
\end{table}

Let us consider a couple of examples. 
In the correlator $\mathcal{G}_{3 3 2 2}$ there is a unique $y$-structure $y_{12}^2$ corresponding to $\{1,0,0,0,0,0\}$.
At two loops we find $F^{(2)}_{\{1,0,0,0,0,0\}}$ in the 2nd line of table \ref{tab1}
and at three loops $F^{(3)}_{\{1,0,0,0,0,0\}}$ in the 2nd line of table \ref{tab2}.
In the correlator $\mathcal{G}_{4 4 4 4}$ there are six $y$-structures which break down into two equivalence classes under crossing symmetry
%% \begin{align}
%% &y_{14}^4 y_{23}^4 \to \{0,0,2,2,0,0\}  \;\;,\;\; y_{13}^4 y_{24}^4 \to  \{0,2,0,0,2,0\}\;\;,\;\; y_{12}^4 y_{34}^4 \to \{2,0,0,0,0,2\} 
%% \notag\\
%% &y_{13}^2 y_{23}^2 y_{24}^2 y_{14}^2 \to \{0,1,1,1,1,0\} \;\;,\;\;
%% y_{12}^2 y_{23}^2 y_{34}^2 y_{14}^2 \to \{1,0,1,1,0,1\} \;\;,\;\; 
%% y_{12}^2 y_{24}^2 y_{34}^2 y_{13}^2 \to \{1,1,0,0,1,1\}  
%% \notag
%% \end{align}
\begin{align*}
\begin{array}{c|c||c|c||c|c}
y_{14}^4 y_{23}^4 & \{0,0,2,2,0,0\}  & y_{13}^4 y_{24}^4 & \{0,2,0,0,2,0\} & y_{12}^4 y_{34}^4 & \{2,0,0,0,0,2\} 
\rule[-1.2ex]{0pt}{0pt} \\ \hline \hline
y_{13}^2 y_{23}^2 y_{24}^2 y_{14}^2 & \{0,1,1,1,1,0\} &
y_{12}^2 y_{23}^2 y_{34}^2 y_{14}^2 & \{1,0,1,1,0,1\} & 
y_{12}^2 y_{24}^2 y_{34}^2 y_{13}^2 & \{1,1,0,0,1,1\} \rule{0pt}{2.6ex} 
\end{array}
\end{align*}
%% $y_{12}^4 y_{34}^4$, $y_{13}^4 y_{24}^4$, $y_{14}^4 y_{23}^4$ 
%% and $y_{12}^2 y_{23}^2 y_{34}^2 y_{14}^2$, $y_{12}^2 y_{24}^2 y_{34}^2 y_{13}^2$, $y_{13}^2 y_{23}^2 y_{24}^2 y_{14}^2$.
%% The corresponding indices are $\{2,0,0,0,0,2\}$, $\{0,2,0,0,2,0\}$, $\{0,0,2,2,0,0\}$ and
%% $\{1,0,1,1,0,1\}$, $\{1,1,0,0,1,1\}$, $\{0,1,1,1,1,0\}$.
At two loops we find $F^{(2)}_{\{0,0,2,2,0,0\}}$ in the 5th line of table \ref{tab1} and
$F^{(2)}_{\{0,1,1,1,1,0\}}$ in the 8th line. The remaining four functions are obtained by crossing 
from the previous two. At three loops we find $F^{(3)}_{\{0,0,2,2,0,0\}}$ in the 20th line of table \ref{tab2} and
$F^{(3)}_{\{0,1,1,1,1,0\}}$ in the 28th line.

\setlength\cellspacetoplimit{2pt}
\setlength\cellspacebottomlimit{2pt}

\renewcommand*{\len}{0.3}

\begin{table}
%\begin{center} 
\hspace{-1cm}
\begin{tabular}{|Sc|Sc|Sc|Sc|Sc|Sc|Sc|Sc|Sc|Sc|Sc|Sc|Sc|Sc|Sc|Sc|Sc|}
\hline
$\{b_{ij}\}$ & \begin{tikzpicture}
\filldraw[black] (0,0) circle (\diam) node[anchor=east]{$\scriptscriptstyle 4$}; 
\filldraw[black] (0,\len) circle (\diam) node[anchor=east]{$\scriptscriptstyle 1$};
\filldraw[black] (\len,0) circle (\diam) node[anchor=west]{$\scriptscriptstyle 3$};
\filldraw[black] (\len,\len) circle (\diam) node[anchor=west]{$\scriptscriptstyle 2$}; 
 \draw[dashed] (0,\len) -- (\len,\len) -- (\len,0) -- (0,0) -- (0,\len);
 \draw[dashed] (0,\len) -- (\len,0);
 \draw[dashed] (0,0) -- (\len,\len);
\end{tikzpicture} & $c^1_{gh}$ & $c_{gh}^2$ & $c_{gh}^3$ & $c^1_L$ & $c_L^2$ & $c_L^3$ & $c_E^1$ & $c_E^2$
   & $c_E^3$ & $c_H^1$ & $c_H^2$ & $c_H^3$ & $c_H^4$ & $c_H^5$ & $c_H^6$ \\   \hline
 \mysextuple{0}{0}{0}{0}{0}{0} & 
\begin{tikzpicture}
\filldraw[black] (0,0) circle (\diam); 
\filldraw[black] (0,\len) circle (\diam);
\filldraw[black] (\len,0) circle (\diam);
\filldraw[black] (\len,\len) circle (\diam); 
\end{tikzpicture} & 2 & 2 & 2 & 6 & 6 & 6 & 4 & 4 & 4 & 2 & 2 & 2 & 2 & 2 & 2 \\  \hline
 \mysextuple{1}{0}{0}{0}{0}{0} &
 \begin{tikzpicture}
\filldraw[black] (0,0) circle (\diam); 
\filldraw[black] (0,\len) circle (\diam);
\filldraw[black] (\len,0) circle (\diam);
\filldraw[black] (\len,\len) circle (\diam);
\draw (0,\len) -- (\len,\len); 
\end{tikzpicture} & -1 & 2 & 2 & 2 & 6 & 6 & 4 & 2 & 2 & 1 & 1 & 2 & 0 & 2 & 0 \\  \hline
 \mysextuple{\beta_1}{0}{0}{0}{0}{0} &
 \begin{tikzpicture}
\filldraw[black] (0,0) circle (\diam); 
\filldraw[black] (0,\len) circle (\diam);
\filldraw[black] (\len,0) circle (\diam);
\filldraw[black] (\len,\len) circle (\diam);
\draw[very thick] (0,\len) -- (\len,\len); 
\end{tikzpicture} & 0 & 2 & 2 & 3 & 6 & 6 & 4 & 2 & 2 & 1 & 1 & 2 & 0 & 2 & 0 \\  \hline
 \mysextuple{1}{1}{0}{0}{0}{0} &
 \begin{tikzpicture}
\filldraw[black] (0,0) circle (\diam); 
\filldraw[black] (0,\len) circle (\diam);
\filldraw[black] (\len,0) circle (\diam);
\filldraw[black] (\len,\len) circle (\diam);
\draw (0,\len) -- (\len,\len); 
\draw (0,\len) -- (\len,0);
\end{tikzpicture}  & -1 & -1 & 2 & 2 & 2 & 6 & 2 & 2 & 1 & 1 & 0 & 1 & 0 & 0 & 0 \\  \hline
 \mysextuple{\beta_1}{1}{0}{0}{0}{0} & 
 \begin{tikzpicture}
\filldraw[black] (0,0) circle (\diam); 
\filldraw[black] (0,\len) circle (\diam);
\filldraw[black] (\len,0) circle (\diam);
\filldraw[black] (\len,\len) circle (\diam);
\draw[very thick] (0,\len) -- (\len,\len); 
\draw (0,\len) -- (\len,0);
\end{tikzpicture} & 0 & -1 & 2 & 3 & 2 & 6 & 2 & 2 & 1 & 1 & 0 & 1 & 0 & 0 & 0 \\  \hline
 \mysextuple{\beta_1}{\beta_2}{0}{0}{0}{0} &
  \begin{tikzpicture}
\filldraw[black] (0,0) circle (\diam); 
\filldraw[black] (0,\len) circle (\diam);
\filldraw[black] (\len,0) circle (\diam);
\filldraw[black] (\len,\len) circle (\diam);
\draw[very thick] (0,\len) -- (\len,\len); 
\draw[very thick] (0,\len) -- (\len,0);
\end{tikzpicture} & 0 & 0 & 2 & 3 & 3 & 6 & 2 & 2 & 1 & 1 & 0 & 1 & 0 & 0 & 0 \\  \hline
\begin{tabular}{c}
 \mysextuple{ 1 }{ 1 }{ 0 }{ 1 }{ 0 }{ 0} \\
 \mysextuple{ 1 }{ 1 }{ 1 }{ 0 }{ 0 }{ 0} 
\end{tabular} & 
\begin{tabular}{c}
\begin{tikzpicture}
\filldraw[black] (0,0) circle (\diam); 
\filldraw[black] (0,\len) circle (\diam);
\filldraw[black] (\len,0) circle (\diam);
\filldraw[black] (\len,\len) circle (\diam);
\draw (0,\len) -- (\len,\len) -- (\len,0) -- (0,\len);
\end{tikzpicture} \\
\begin{tikzpicture}
\filldraw[black] (0,0) circle (\diam); 
\filldraw[black] (0,\len) circle (\diam);
\filldraw[black] (\len,0) circle (\diam);
\filldraw[black] (\len,\len) circle (\diam);
\draw (0,\len) -- (\len,\len);
\draw (0,\len) -- (\len,0);
\draw (0,\len) -- (0,0);
\end{tikzpicture}
\end{tabular} 
& -1 & -1 & -1 & 2 & 2 & 2 & 1 & 1 & 1 & 0 & 0 & 0 & 0 & 0 & 0 \\  \hline
\begin{tabular}{c}
\mysextuple{ \beta_1 }{ 1 }{ 0 }{ 1 }{ 0 }{ 0 }\\
\mysextuple{ \beta_1 }{ 1 }{ 1 }{ 0 }{ 0 }{ 0 }
\end{tabular} & 
\begin{tabular}{c}
\begin{tikzpicture}
\filldraw[black] (0,0) circle (\diam); 
\filldraw[black] (0,\len) circle (\diam);
\filldraw[black] (\len,0) circle (\diam);
\filldraw[black] (\len,\len) circle (\diam);
\draw[very thick] (0,\len) -- (\len,\len);
\draw (\len,\len) -- (\len,0) -- (0,\len);
\end{tikzpicture} \\
\begin{tikzpicture}
\filldraw[black] (0,0) circle (\diam); 
\filldraw[black] (0,\len) circle (\diam);
\filldraw[black] (\len,0) circle (\diam);
\filldraw[black] (\len,\len) circle (\diam);
\draw[very thick] (0,\len) -- (\len,\len);
\draw (0,\len) -- (\len,0);
\draw (0,\len) -- (0,0);
\end{tikzpicture}
\end{tabular} & 0 & -1 & -1 & 3 & 2 & 2 & 1 & 1 & 1 & 0 & 0 & 0 & 0 & 0 & 0 \\  \hline
\begin{tabular}{c}
\mysextuple{ \beta_1 }{ \beta_2 }{ 0 }{ 1 }{ 0 }{ 0 }\\
\mysextuple{ \beta_1 }{ \beta_2 }{ 1 }{ 0 }{ 0 }{ 0 }
\end{tabular} &
\begin{tabular}{c}
\begin{tikzpicture}
\filldraw[black] (0,0) circle (\diam); 
\filldraw[black] (0,\len) circle (\diam);
\filldraw[black] (\len,0) circle (\diam);
\filldraw[black] (\len,\len) circle (\diam);
\draw[very thick] (0,\len) -- (\len,\len);
\draw (\len,\len) -- (\len,0);
\draw[very thick] (\len,0) -- (0,\len);
\end{tikzpicture} \\
\begin{tikzpicture}
\filldraw[black] (0,0) circle (\diam); 
\filldraw[black] (0,\len) circle (\diam);
\filldraw[black] (\len,0) circle (\diam);
\filldraw[black] (\len,\len) circle (\diam);
\draw[very thick] (0,\len) -- (\len,\len);
\draw[very thick] (0,\len) -- (\len,0);
\draw (0,\len) -- (0,0);
\end{tikzpicture}
\end{tabular} & 0 & 0 & -1 & 3 & 3 & 2 & 1 & 1 & 1 & 0 & 0 & 0 & 0 & 0 & 0 \\  \hline
\begin{tabular}{c}
\mysextuple{ \beta_1 }{ \beta_2 }{ 0 }{ \beta_3 }{ 0 }{ 0 }\\
\mysextuple{ \beta_1 }{ \beta_2 }{ \beta_3 }{ 0 }{ 0 }{ 0 }
\end{tabular} &
 \begin{tabular}{c}
\begin{tikzpicture}
\filldraw[black] (0,0) circle (\diam); 
\filldraw[black] (0,\len) circle (\diam);
\filldraw[black] (\len,0) circle (\diam);
\filldraw[black] (\len,\len) circle (\diam);
\draw[very thick] (0,\len) -- (\len,\len);
\draw[very thick] (\len,\len) -- (\len,0);
\draw[very thick] (\len,0) -- (0,\len);
\end{tikzpicture} \\
\begin{tikzpicture}
\filldraw[black] (0,0) circle (\diam); 
\filldraw[black] (0,\len) circle (\diam);
\filldraw[black] (\len,0) circle (\diam);
\filldraw[black] (\len,\len) circle (\diam);
\draw[very thick] (0,\len) -- (\len,\len);
\draw[very thick] (0,\len) -- (\len,0);
\draw[very thick] (0,\len) -- (0,0);
\end{tikzpicture}
\end{tabular}  & 0 & 0 & 0 & 3 & 3 & 3 & 1 & 1 & 1 & 0 & 0 & 0 & 0 & 0 & 0 \\  \hline
 \mysextuple{0}{0}{1}{1}{0}{0} &
 \begin{tikzpicture}
\filldraw[black] (0,0) circle (\diam); 
\filldraw[black] (0,\len) circle (\diam);
\filldraw[black] (\len,0) circle (\diam);
\filldraw[black] (\len,\len) circle (\diam);
\draw (0,\len) -- (0,0);
\draw (\len,\len) -- (\len,0);
\end{tikzpicture} & 2 & 2 & 0 & 6 & 6 & -2 & 0 & 0 & 4 & 0 & 2 & 0 & 2 & 0 & 0 \\  \hline
 \mysextuple{1}{0}{1}{1}{0}{0} &
 \begin{tikzpicture}
\filldraw[black] (0,0) circle (\diam); 
\filldraw[black] (0,\len) circle (\diam);
\filldraw[black] (\len,0) circle (\diam);
\filldraw[black] (\len,\len) circle (\diam);
\draw (0,0) -- (0,\len) -- (\len,\len) -- (\len,0);
\end{tikzpicture} & -1 & 2 & 0 & 2 & 6 & -2 & 0 & 0 & 2 & 0 & 1 & 0 & 0 & 0 & 0 \\  \hline
 \mysextuple{\beta_1}{0}{1}{1}{0}{0} &
 \begin{tikzpicture}
\filldraw[black] (0,0) circle (\diam); 
\filldraw[black] (0,\len) circle (\diam);
\filldraw[black] (\len,0) circle (\diam);
\filldraw[black] (\len,\len) circle (\diam);
\draw (0,\len) -- (0,0);
\draw (\len,\len) -- (\len,0);
\draw[very thick] (0,\len) -- (\len,\len);
\end{tikzpicture} & 0 & 2 & 0 & 3 & 6 & -2 & 0 & 0 & 2 & 0 & 1 & 0 & 0 & 0 & 0 \\  \hline
 \mysextuple{1}{1}{1}{1}{0}{0} &
 \begin{tikzpicture}
\filldraw[black] (0,0) circle (\diam); 
\filldraw[black] (0,\len) circle (\diam);
\filldraw[black] (\len,0) circle (\diam);
\filldraw[black] (\len,\len) circle (\diam);
\draw (0,0) -- (0,\len) -- (\len,\len) -- (\len,0);
\draw (0,\len) -- (\len,0);
\end{tikzpicture} & -1 & -1 & 0 & 2 & 2 & -2 & 0 & 0 & 1 & 0 & 0 & 0 & 0 & 0 & 0 \\  \hline
 \mysextuple{\beta_1}{1}{1}{1}{0}{0} &
 \begin{tikzpicture}
\filldraw[black] (0,0) circle (\diam); 
\filldraw[black] (0,\len) circle (\diam);
\filldraw[black] (\len,0) circle (\diam);
\filldraw[black] (\len,\len) circle (\diam);
\draw (0,0) -- (0,\len);
\draw[very thick]  (0,\len) -- (\len,\len);
\draw (\len,\len) -- (\len,0);
\draw (0,\len) -- (\len,0);
\end{tikzpicture} & 0 & -1 & 0 & 3 & 2 & -2 & 0 & 0 & 1 & 0 & 0 & 0 & 0 & 0 & 0 \\  \hline
 \mysextuple{\beta_1}{\beta_2}{1}{1}{0}{0} &
 \begin{tikzpicture}
\filldraw[black] (0,0) circle (\diam); 
\filldraw[black] (0,\len) circle (\diam);
\filldraw[black] (\len,0) circle (\diam);
\filldraw[black] (\len,\len) circle (\diam);
\draw (0,0) -- (0,\len);
\draw[very thick]  (0,\len) -- (\len,\len);
\draw (\len,\len) -- (\len,0);
\draw[very thick] (0,\len) -- (\len,0);
\end{tikzpicture} & 0 & 0 & 0 & 3 & 3 & -2 & 0 & 0 & 1 & 0 & 0 & 0 & 0 & 0 & 0 \\  \hline
\begin{tabular}{c}
\mysextuple{ 1 }{ 1 }{ 1 }{ \beta_1 }{ 0 }{ 0 }\\
\mysextuple{ 1 }{ 1 }{ \beta_1 }{ 1 }{ 0 }{ 0 }
\end{tabular} &
\begin{tabular}{c}
\begin{tikzpicture}
\filldraw[black] (0,0) circle (\diam); 
\filldraw[black] (0,\len) circle (\diam);
\filldraw[black] (\len,0) circle (\diam);
\filldraw[black] (\len,\len) circle (\diam);
\draw (0,0) -- (0,\len);
\draw  (0,\len) -- (\len,\len);
\draw[very thick] (\len,\len) -- (\len,0);
\draw (0,\len) -- (\len,0);
\end{tikzpicture}
\\
\begin{tikzpicture}
\filldraw[black] (0,0) circle (\diam); 
\filldraw[black] (0,\len) circle (\diam);
\filldraw[black] (\len,0) circle (\diam);
\filldraw[black] (\len,\len) circle (\diam);
\draw[very thick] (0,0) -- (0,\len);
\draw (0,\len) -- (\len,\len);
\draw (\len,\len) -- (\len,0);
\draw (0,\len) -- (\len,0);
\end{tikzpicture}
\end{tabular} & -1 & -1 & 0 & 2 & 2 & -1 & 0 & 0 & 1 & 0 & 0 & 0 & 0 & 0 & 0 \\  \hline
\begin{tabular}{c}
\mysextuple{ \beta_1 }{ 1 }{ 1 }{ \beta_2 }{ 0 }{ 0 }\\
\mysextuple{ \beta_1 }{ 1 }{ \beta_2 }{ 1 }{ 0 }{ 0 }
\end{tabular} &
\begin{tabular}{c}
\begin{tikzpicture}
\filldraw[black] (0,0) circle (\diam); 
\filldraw[black] (0,\len) circle (\diam);
\filldraw[black] (\len,0) circle (\diam);
\filldraw[black] (\len,\len) circle (\diam);
\draw (0,0) -- (0,\len);
\draw[very thick]  (0,\len) -- (\len,\len);
\draw[very thick] (\len,\len) -- (\len,0);
\draw (0,\len) -- (\len,0);
\end{tikzpicture}
\\
\begin{tikzpicture}
\filldraw[black] (0,0) circle (\diam); 
\filldraw[black] (0,\len) circle (\diam);
\filldraw[black] (\len,0) circle (\diam);
\filldraw[black] (\len,\len) circle (\diam);
\draw[very thick] (0,0) -- (0,\len);
\draw[very thick] (0,\len) -- (\len,\len);
\draw (\len,\len) -- (\len,0);
\draw (0,\len) -- (\len,0);
\end{tikzpicture}
\end{tabular} & 0 & -1 & 0 & 3 & 2 & -1 & 0 & 0 & 1 & 0 & 0 & 0 & 0 & 0 & 0 \\  \hline
\begin{tabular}{c}
\mysextuple{ \beta_1 }{ \beta_2 }{ 1 }{ \beta_3 }{ 0 }{ 0 }\\
\mysextuple{ \beta_1 }{ \beta_2 }{ \beta_3 }{ 1 }{ 0 }{ 0 }
\end{tabular} &
\begin{tabular}{c}
\begin{tikzpicture}
\filldraw[black] (0,0) circle (\diam); 
\filldraw[black] (0,\len) circle (\diam);
\filldraw[black] (\len,0) circle (\diam);
\filldraw[black] (\len,\len) circle (\diam);
\draw (0,0) -- (0,\len);
\draw[very thick]  (0,\len) -- (\len,\len);
\draw[very thick] (\len,\len) -- (\len,0);
\draw[very thick] (0,\len) -- (\len,0);
\end{tikzpicture}
\\
\begin{tikzpicture}
\filldraw[black] (0,0) circle (\diam); 
\filldraw[black] (0,\len) circle (\diam);
\filldraw[black] (\len,0) circle (\diam);
\filldraw[black] (\len,\len) circle (\diam);
\draw[very thick] (0,0) -- (0,\len);
\draw[very thick] (0,\len) -- (\len,\len);
\draw (\len,\len) -- (\len,0);
\draw[very thick] (0,\len) -- (\len,0);
\end{tikzpicture}
\end{tabular} & 0 & 0 & 0 & 3 & 3 & -1 & 0 & 0 & 1 & 0 & 0 & 0 & 0 & 0 & 0 \\  \hline
 \mysextuple{0}{0}{\beta_1}{\beta_2}{0}{0} &
 \begin{tikzpicture}
\filldraw[black] (0,0) circle (\diam); 
\filldraw[black] (0,\len) circle (\diam);
\filldraw[black] (\len,0) circle (\diam);
\filldraw[black] (\len,\len) circle (\diam);
\draw[very thick] (0,\len) -- (0,0);
\draw[very thick] (\len,\len) -- (\len,0);
\end{tikzpicture} & 2 & 2 & 0 & 6 & 6 & 0 & 0 & 0 & 4 & 0 & 2 & 0 & 2 & 0 & 0 \\  \hline
 \mysextuple{1}{0}{\beta_1}{\beta_2}{0}{0} &
 \begin{tikzpicture}
\filldraw[black] (0,0) circle (\diam); 
\filldraw[black] (0,\len) circle (\diam);
\filldraw[black] (\len,0) circle (\diam);
\filldraw[black] (\len,\len) circle (\diam);
\draw[very thick] (0,\len) -- (0,0);
\draw[very thick] (\len,\len) -- (\len,0);
\draw (0,\len) -- (\len,\len);
\end{tikzpicture} & -1 & 2 & 0 & 2 & 6 & 0 & 0 & 0 & 2 & 0 & 1 & 0 & 0 & 0 & 0 \\  \hline
 \mysextuple{\beta_1}{0}{\beta_2}{\beta_3}{0}{0} &
 \begin{tikzpicture}
\filldraw[black] (0,0) circle (\diam); 
\filldraw[black] (0,\len) circle (\diam);
\filldraw[black] (\len,0) circle (\diam);
\filldraw[black] (\len,\len) circle (\diam);
\draw[very thick] (0,\len) -- (0,0);
\draw[very thick] (\len,\len) -- (\len,0);
\draw[very thick] (0,\len) -- (\len,\len);
\end{tikzpicture} & 0 & 2 & 0 & 3 & 6 & 0 & 0 & 0 & 2 & 0 & 1 & 0 & 0 & 0 & 0 \\  \hline
 \mysextuple{1}{1}{\beta_1}{\beta_2}{0}{0} &
  \begin{tikzpicture}
\filldraw[black] (0,0) circle (\diam); 
\filldraw[black] (0,\len) circle (\diam);
\filldraw[black] (\len,0) circle (\diam);
\filldraw[black] (\len,\len) circle (\diam);
\draw[very thick] (0,\len) -- (0,0);
\draw[very thick] (\len,\len) -- (\len,0);
\draw (0,\len) -- (\len,\len);
\draw (0,\len) -- (\len,0);
\end{tikzpicture} & -1 & -1 & 0 & 2 & 2 & 0 & 0 & 0 & 1 & 0 & 0 & 0 & 0 & 0 & 0 \\  \hline
 \mysextuple{\beta_1}{1}{\beta_2}{\beta_3}{0}{0} &
 \begin{tikzpicture}
\filldraw[black] (0,0) circle (\diam); 
\filldraw[black] (0,\len) circle (\diam);
\filldraw[black] (\len,0) circle (\diam);
\filldraw[black] (\len,\len) circle (\diam);
\draw[very thick] (0,\len) -- (0,0);
\draw[very thick] (\len,\len) -- (\len,0);
\draw[very thick] (0,\len) -- (\len,\len);
\draw (0,\len) -- (\len,0);
\end{tikzpicture} & 0 & -1 & 0 & 3 & 2 & 0 & 0 & 0 & 1 & 0 & 0 & 0 & 0 & 0 & 0 \\  \hline
 \mysextuple{\beta_1}{\beta_2}{\beta_3}{\beta_4}{0}{0} &
 \begin{tikzpicture}
\filldraw[black] (0,0) circle (\diam); 
\filldraw[black] (0,\len) circle (\diam);
\filldraw[black] (\len,0) circle (\diam);
\filldraw[black] (\len,\len) circle (\diam);
\draw[very thick] (0,\len) -- (0,0);
\draw[very thick] (\len,\len) -- (\len,0);
\draw[very thick] (0,\len) -- (\len,\len);
\draw[very thick] (0,\len) -- (\len,0);
\end{tikzpicture} & 0 & 0 & 0 & 3 & 3 & 0 & 0 & 0 & 1 & 0 & 0 & 0 & 0 & 0 & 0 \\  \hline
 \mysextuple{1}{\beta_1}{0}{0}{1}{0} &
 \begin{tikzpicture}
\filldraw[black] (0,0) circle (\diam); 
\filldraw[black] (0,\len) circle (\diam);
\filldraw[black] (\len,0) circle (\diam);
\filldraw[black] (\len,\len) circle (\diam);
\draw (\len,\len) -- (0,0);
\draw (0,\len) -- (\len,\len);
\draw[very thick] (0,\len) -- (\len,0);
\end{tikzpicture} & -1 & 0 & 2 & 2 & -1 & 6 & 0 & 2 & 0 & 1 & 0 & 0 & 0 & 0 & 0 \\  \hline
 \mysextuple{\beta_1}{\beta_2}{0}{0}{1}{0} &
 \begin{tikzpicture}
\filldraw[black] (0,0) circle (\diam); 
\filldraw[black] (0,\len) circle (\diam);
\filldraw[black] (\len,0) circle (\diam);
\filldraw[black] (\len,\len) circle (\diam);
\draw (\len,\len) -- (0,0);
\draw[very thick] (0,\len) -- (\len,\len);
\draw[very thick] (0,\len) -- (\len,0);
\end{tikzpicture} & 0 & 0 & 2 & 3 & -1 & 6 & 0 & 2 & 0 & 1 & 0 & 0 & 0 & 0 & 0 \\  \hline
 \mysextuple{0}{1}{1}{1}{1}{0} &
 \begin{tikzpicture}
\filldraw[black] (0,0) circle (\diam); 
\filldraw[black] (0,\len) circle (\diam);
\filldraw[black] (\len,0) circle (\diam);
\filldraw[black] (\len,\len) circle (\diam);
\draw (0,\len) -- (0,0) -- (\len,\len) -- (\len,0) -- (0,\len);
\end{tikzpicture} & 2 & 0 & 0 & 6 & -2 & -2 & 0 & 0 & 0 & 0 & 0 & 0 & 0 & 0 & 0 \\  \hline
 \mysextuple{1}{1}{1}{1}{1}{0} &
 \begin{tikzpicture}
\filldraw[black] (0,0) circle (\diam); 
\filldraw[black] (0,\len) circle (\diam);
\filldraw[black] (\len,0) circle (\diam);
\filldraw[black] (\len,\len) circle (\diam);
\draw (0,\len) -- (0,0) -- (\len,\len) -- (\len,0) -- (0,\len);
\draw (0,\len) -- (\len,\len);
\end{tikzpicture} & -1 & 0 & 0 & 2 & -2 & -2 & 0 & 0 & 0 & 0 & 0 & 0 & 0 & 0 & 0 \\  \hline
\end{tabular}
%\end{center}
\end{table}

\begin{table}
\hspace{-1cm}
\begin{tabular}{|Sc|Sc|Sc|Sc|Sc|Sc|Sc|Sc|Sc|Sc|Sc|Sc|Sc|Sc|Sc|Sc|Sc|}
\hline
$\{b_{ij}\}$ & 
\begin{tikzpicture}
\filldraw[black] (0,0) circle (\diam) node[anchor=east]{$\scriptscriptstyle 4$}; 
\filldraw[black] (0,\len) circle (\diam) node[anchor=east]{$\scriptscriptstyle 1$};
\filldraw[black] (\len,0) circle (\diam) node[anchor=west]{$\scriptscriptstyle 3$};
\filldraw[black] (\len,\len) circle (\diam) node[anchor=west]{$\scriptscriptstyle 2$}; 
 \draw[dashed] (0,\len) -- (\len,\len) -- (\len,0) -- (0,0) -- (0,\len);
 \draw[dashed] (0,\len) -- (\len,0);
 \draw[dashed] (0,0) -- (\len,\len);
\end{tikzpicture} & $c^1_{gh}$ & $c_{gh}^2$ & $c_{gh}^3$ & $c^1_L$ & $c_L^2$ & $c_L^3$ & $c_E^1$ & $c_E^2$
   & $c_E^3$ & $c_H^1$ & $c_H^2$ & $c_H^3$ & $c_H^4$ & $c_H^5$ & $c_H^6$ \\   \hline
 \mysextuple{\beta_1}{1}{1}{1}{1}{0} & 
  \begin{tikzpicture}
\filldraw[black] (0,0) circle (\diam); 
\filldraw[black] (0,\len) circle (\diam);
\filldraw[black] (\len,0) circle (\diam);
\filldraw[black] (\len,\len) circle (\diam);
\draw (0,\len) -- (0,0) -- (\len,\len) -- (\len,0) -- (0,\len);
\draw[very thick] (0,\len) -- (\len,\len);
\end{tikzpicture} & 0 & 0 & 0 & 3 & -2 & -2 & 0 & 0 & 0 & 0 & 0 & 0 & 0 & 0 & 0 \\  \hline
 \mysextuple{1}{\beta_1}{1}{1}{1}{0} & 
 \begin{tikzpicture}
\filldraw[black] (0,0) circle (\diam); 
\filldraw[black] (0,\len) circle (\diam);
\filldraw[black] (\len,0) circle (\diam);
\filldraw[black] (\len,\len) circle (\diam);
\draw (0,\len) -- (0,0) -- (\len,\len) -- (\len,0);
\draw[very thick] (\len,0)  -- (0,\len);
\draw (0,\len) -- (\len,\len);
\end{tikzpicture} & -1 & 0 & 0 & 2 & -1 & -2 & 0 & 0 & 0 & 0 & 0 & 0 & 0 & 0 & 0 \\  \hline
 \mysextuple{\beta_1}{\beta_2}{1}{1}{1}{0} &
 \begin{tikzpicture}
\filldraw[black] (0,0) circle (\diam); 
\filldraw[black] (0,\len) circle (\diam);
\filldraw[black] (\len,0) circle (\diam);
\filldraw[black] (\len,\len) circle (\diam);
\draw (0,\len) -- (0,0) -- (\len,\len) -- (\len,0);
\draw[very thick] (\len,0)  -- (0,\len);
\draw[very thick] (0,\len) -- (\len,\len);
\end{tikzpicture} & 0 & 0 & 0 & 3 & -1 & -2 & 0 & 0 & 0 & 0 & 0 & 0 & 0 & 0 & 0 \\  \hline	
\begin{tabular}{c}
 \mysextuple{ 1 }{ \beta_1 }{ 1 }{ \beta_2 }{ 1 }{ 0 }\\
 \mysextuple{ 1 }{ \beta_1 }{ \beta_2 }{ 1 }{ 1 }{ 0 }
\end{tabular} &
\begin{tabular}{c}
\begin{tikzpicture}
\filldraw[black] (0,0) circle (\diam); 
\filldraw[black] (0,\len) circle (\diam);
\filldraw[black] (\len,0) circle (\diam);
\filldraw[black] (\len,\len) circle (\diam);
\draw (0,\len) -- (0,0) -- (\len,\len);
\draw[very thick] (\len,\len) -- (\len,0);
\draw[very thick] (\len,0)  -- (0,\len);
\draw (0,\len) -- (\len,\len);
\end{tikzpicture}
\\
\begin{tikzpicture}
\filldraw[black] (0,0) circle (\diam); 
\filldraw[black] (0,\len) circle (\diam);
\filldraw[black] (\len,0) circle (\diam);
\filldraw[black] (\len,\len) circle (\diam);
\draw[very thick]  (0,\len) -- (0,0);
\draw (0,0) -- (\len,\len);
\draw (\len,\len) -- (\len,0);
\draw[very thick] (\len,0)  -- (0,\len);
\draw (0,\len) -- (\len,\len);
\end{tikzpicture}
\end{tabular} & -1 & 0 & 0 & 2 & -1 & -1 & 0 & 0 & 0 & 0 & 0 & 0 & 0 & 0 & 0 \\  \hline
\begin{tabular}{c}
 \mysextuple{ \beta_1 }{ \beta_2 }{ 1 }{ \beta_3 }{ 1 }{ 0 }\\
 \mysextuple{ \beta_1 }{ \beta_2 }{ \beta_3 }{ 1 }{ 1 }{ 0 }
\end{tabular} & 
\begin{tabular}{c}
\begin{tikzpicture}
\filldraw[black] (0,0) circle (\diam); 
\filldraw[black] (0,\len) circle (\diam);
\filldraw[black] (\len,0) circle (\diam);
\filldraw[black] (\len,\len) circle (\diam);
\draw (0,\len) -- (0,0) -- (\len,\len);
\draw[very thick] (\len,\len) -- (\len,0);
\draw[very thick] (\len,0)  -- (0,\len);
\draw[very thick] (0,\len) -- (\len,\len);
\end{tikzpicture}
\\
\begin{tikzpicture}
\filldraw[black] (0,0) circle (\diam); 
\filldraw[black] (0,\len) circle (\diam);
\filldraw[black] (\len,0) circle (\diam);
\filldraw[black] (\len,\len) circle (\diam);
\draw[very thick]  (0,\len) -- (0,0);
\draw (0,0) -- (\len,\len);
\draw (\len,\len) -- (\len,0);
\draw[very thick] (\len,0)  -- (0,\len);
\draw[very thick] (0,\len) -- (\len,\len);
\end{tikzpicture}
\end{tabular} & 0 & 0 & 0 & 3 & -1 & -1 & 0 & 0 & 0 & 0 & 0 & 0 & 0 & 0 & 0 \\  \hline	
 \mysextuple{0}{1}{\beta_1}{\beta_2}{1}{0} &
 \begin{tikzpicture}
\filldraw[black] (0,0) circle (\diam); 
\filldraw[black] (0,\len) circle (\diam);
\filldraw[black] (\len,0) circle (\diam);
\filldraw[black] (\len,\len) circle (\diam);
\draw[very thick]  (0,\len) -- (0,0);
\draw (0,0) -- (\len,\len);
\draw[very thick] (\len,\len) -- (\len,0);
\draw (\len,0)  -- (0,\len);
\end{tikzpicture} & 2 & 0 & 0 & 6 & -2 & 0 & 0 & 0 & 0 & 0 & 0 & 0 & 0 & 0 & 0 \\  \hline
 \mysextuple{1}{1}{\beta_1}{\beta_2}{1}{0} &
 \begin{tikzpicture}
\filldraw[black] (0,0) circle (\diam); 
\filldraw[black] (0,\len) circle (\diam);
\filldraw[black] (\len,0) circle (\diam);
\filldraw[black] (\len,\len) circle (\diam);
\draw[very thick]  (0,\len) -- (0,0);
\draw (0,0) -- (\len,\len);
\draw[very thick] (\len,\len) -- (\len,0);
\draw (\len,0)  -- (0,\len);
\draw (0,\len)  -- (\len,\len);
\end{tikzpicture} & -1 & 0 & 0 & 2 & -2 & 0 & 0 & 0 & 0 & 0 & 0 & 0 & 0 & 0 & 0 \\  \hline
 \mysextuple{\beta_1}{1}{\beta_2}{\beta_3}{1}{0} &
 \begin{tikzpicture}
\filldraw[black] (0,0) circle (\diam); 
\filldraw[black] (0,\len) circle (\diam);
\filldraw[black] (\len,0) circle (\diam);
\filldraw[black] (\len,\len) circle (\diam);
\draw[very thick]  (0,\len) -- (0,0);
\draw (0,0) -- (\len,\len);
\draw[very thick] (\len,\len) -- (\len,0);
\draw (\len,0)  -- (0,\len);
\draw[very thick] (0,\len)  -- (\len,\len);
\end{tikzpicture} & 0 & 0 & 0 & 3 & -2 & 0 & 0 & 0 & 0 & 0 & 0 & 0 & 0 & 0 & 0 \\  \hline
 \mysextuple{1}{\beta_1}{\beta_2}{\beta_3}{1}{0} &
 \begin{tikzpicture}
\filldraw[black] (0,0) circle (\diam); 
\filldraw[black] (0,\len) circle (\diam);
\filldraw[black] (\len,0) circle (\diam);
\filldraw[black] (\len,\len) circle (\diam);
\draw[very thick]  (0,\len) -- (0,0);
\draw (0,0) -- (\len,\len);
\draw[very thick] (\len,\len) -- (\len,0);
\draw[very thick] (\len,0)  -- (0,\len);
\draw (0,\len)  -- (\len,\len);
\end{tikzpicture} & -1 & 0 & 0 & 2 & -1 & 0 & 0 & 0 & 0 & 0 & 0 & 0 & 0 & 0 & 0 \\  \hline
 \mysextuple{\beta_1}{\beta_2}{\beta_3}{\beta_4}{1}{0} &
 \begin{tikzpicture}
\filldraw[black] (0,0) circle (\diam); 
\filldraw[black] (0,\len) circle (\diam);
\filldraw[black] (\len,0) circle (\diam);
\filldraw[black] (\len,\len) circle (\diam);
\draw[very thick]  (0,\len) -- (0,0);
\draw (0,0) -- (\len,\len);
\draw[very thick] (\len,\len) -- (\len,0);
\draw[very thick] (\len,0)  -- (0,\len);
\draw[very thick] (0,\len)  -- (\len,\len);
\end{tikzpicture} & 0 & 0 & 0 & 3 & -1 & 0 & 0 & 0 & 0 & 0 & 0 & 0 & 0 & 0 & 0 \\  \hline
 \mysextuple{0}{\beta_1}{\beta_2}{\beta_3}{\beta_4}{0} &
 \begin{tikzpicture}
\filldraw[black] (0,0) circle (\diam); 
\filldraw[black] (0,\len) circle (\diam);
\filldraw[black] (\len,0) circle (\diam);
\filldraw[black] (\len,\len) circle (\diam);
\draw[very thick]  (0,\len) -- (0,0);
\draw[very thick] (0,0) -- (\len,\len);
\draw[very thick] (\len,\len) -- (\len,0);
\draw[very thick] (\len,0)  -- (0,\len);
\end{tikzpicture} & 2 & 0 & 0 & 6 & 0 & 0 & 0 & 0 & 0 & 0 & 0 & 0 & 0 & 0 & 0 \\  \hline
 \mysextuple{1}{\beta_1}{\beta_2}{\beta_3}{\beta_4}{0} &
 \begin{tikzpicture}
\filldraw[black] (0,0) circle (\diam); 
\filldraw[black] (0,\len) circle (\diam);
\filldraw[black] (\len,0) circle (\diam);
\filldraw[black] (\len,\len) circle (\diam);
\draw[very thick]  (0,\len) -- (0,0);
\draw[very thick] (0,0) -- (\len,\len);
\draw[very thick] (\len,\len) -- (\len,0);
\draw[very thick] (\len,0)  -- (0,\len);
\draw (0,\len)  -- (\len,\len);
\end{tikzpicture} & -1 & 0 & 0 & 2 & 0 & 0 & 0 & 0 & 0 & 0 & 0 & 0 & 0 & 0 & 0 \\  \hline
 \mysextuple{\beta_1}{\beta_2}{\beta_3}{\beta_4}{\beta_5}{0} &
 \begin{tikzpicture}
\filldraw[black] (0,0) circle (\diam); 
\filldraw[black] (0,\len) circle (\diam);
\filldraw[black] (\len,0) circle (\diam);
\filldraw[black] (\len,\len) circle (\diam);
\draw[very thick]  (0,\len) -- (0,0);
\draw[very thick] (0,0) -- (\len,\len);
\draw[very thick] (\len,\len) -- (\len,0);
\draw[very thick] (\len,0)  -- (0,\len);
\draw[very thick] (0,\len)  -- (\len,\len);
\end{tikzpicture} & 0 & 0 & 0 & 3 & 0 & 0 & 0 & 0 & 0 & 0 & 0 & 0 & 0 & 0 & 0 \\  \hline
 \mysextuple{\beta_1}{0}{0}{0}{0}{1} &
 \begin{tikzpicture}
\filldraw[black] (0,0) circle (\diam); 
\filldraw[black] (0,\len) circle (\diam);
\filldraw[black] (\len,0) circle (\diam);
\filldraw[black] (\len,\len) circle (\diam);
\draw  (0,0) -- (\len,0);
\draw[very thick] (0,\len)  -- (\len,\len);
\end{tikzpicture} & 0 & 2 & 2 & -1 & 6 & 6 & 4 & 0 & 0 & 0 & 0 & 2 & 0 & 2 & 0 \\  \hline
 \mysextuple{\beta_1}{0}{1}{1}{0}{1} &
 \begin{tikzpicture}
\filldraw[black] (0,0) circle (\diam); 
\filldraw[black] (0,\len) circle (\diam);
\filldraw[black] (\len,0) circle (\diam);
\filldraw[black] (\len,\len) circle (\diam);
\draw  (0,0) -- (\len,0);
\draw[very thick] (0,\len)  -- (\len,\len);
\draw (0,\len) -- (0,0);
\draw (\len,\len) -- (\len,0);
\end{tikzpicture} & 0 & 2 & 0 & -1 & 6 & -2 & 0 & 0 & 0 & 0 & 0 & 0 & 0 & 0 & 0 \\  \hline
 \mysextuple{\beta_1}{0}{\beta_2}{\beta_3}{0}{1} &
 \begin{tikzpicture}
\filldraw[black] (0,0) circle (\diam); 
\filldraw[black] (0,\len) circle (\diam);
\filldraw[black] (\len,0) circle (\diam);
\filldraw[black] (\len,\len) circle (\diam);
\draw (0,0) -- (\len,0);
\draw[very thick] (0,\len)  -- (\len,\len);
\draw[very thick] (0,\len) -- (0,0);
\draw[very thick] (\len,\len) -- (\len,0);
\end{tikzpicture} & 0 & 2 & 0 & -1 & 6 & 0 & 0 & 0 & 0 & 0 & 0 & 0 & 0 & 0 & 0 \\  \hline
 \mysextuple{\beta_1}{\beta_2}{0}{0}{1}{1} &
 \begin{tikzpicture}
\filldraw[black] (0,0) circle (\diam); 
\filldraw[black] (0,\len) circle (\diam);
\filldraw[black] (\len,0) circle (\diam);
\filldraw[black] (\len,\len) circle (\diam);
\draw (\len,0) -- (0,0) -- (\len,\len);
\draw[very thick] (0,\len)  -- (\len,\len);
\draw[very thick] (0,\len) -- (\len,0);
\end{tikzpicture} & 0 & 0 & 2 & -1 & -1 & 6 & 0 & 0 & 0 & 0 & 0 & 0 & 0 & 0 & 0 \\  \hline
 \mysextuple{1}{1}{1}{1}{1}{1} &
 \begin{tikzpicture}
\filldraw[black] (0,0) circle (\diam); 
\filldraw[black] (0,\len) circle (\diam);
\filldraw[black] (\len,0) circle (\diam);
\filldraw[black] (\len,\len) circle (\diam);
\draw (0,0) -- (0,\len) -- (\len,\len) -- (\len,0) -- (0,0);
\draw (0,\len)  -- (\len,0);
\draw (0,0) -- (\len,\len);
\end{tikzpicture} & 0 & 0 & 0 & -2 & -2 & -2 & 0 & 0 & 0 & 0 & 0 & 0 & 0 & 0 & 0 \\  \hline
 \mysextuple{\beta_1}{1}{1}{1}{1}{1} &
  \begin{tikzpicture}
\filldraw[black] (0,0) circle (\diam); 
\filldraw[black] (0,\len) circle (\diam);
\filldraw[black] (\len,0) circle (\diam);
\filldraw[black] (\len,\len) circle (\diam);
\draw (0,0) -- (0,\len);
\draw[very thick] (0,\len) -- (\len,\len);
\draw (\len,\len) -- (\len,0) -- (0,0);
\draw (0,\len)  -- (\len,0);
\draw (0,0) -- (\len,\len);
\end{tikzpicture} & 0 & 0 & 0 & -1 & -2 & -2 & 0 & 0 & 0 & 0 & 0 & 0 & 0 & 0 & 0 \\  \hline
 \mysextuple{\beta_1}{\beta_2}{1}{1}{1}{1} &
 \begin{tikzpicture}
\filldraw[black] (0,0) circle (\diam); 
\filldraw[black] (0,\len) circle (\diam);
\filldraw[black] (\len,0) circle (\diam);
\filldraw[black] (\len,\len) circle (\diam);
\draw (0,0) -- (0,\len);
\draw[very thick] (0,\len) -- (\len,\len);
\draw (\len,\len) -- (\len,0) -- (0,0);
\draw[very thick] (0,\len)  -- (\len,0);
\draw (0,0) -- (\len,\len);
\end{tikzpicture} & 0 & 0 & 0 & -1 & -1 & -2 & 0 & 0 & 0 & 0 & 0 & 0 & 0 & 0 & 0 \\  \hline
\begin{tabular}{c}
 \mysextuple{ \beta_1 }{ \beta_2 }{ 1 }{ \beta_3 }{ 1 }{ 1 }\\
 \mysextuple{ \beta_1 }{ \beta_2 }{ \beta_3 }{ 1 }{ 1 }{ 1 }
\end{tabular} &
\begin{tabular}{c}
\begin{tikzpicture}
\filldraw[black] (0,0) circle (\diam); 
\filldraw[black] (0,\len) circle (\diam);
\filldraw[black] (\len,0) circle (\diam);
\filldraw[black] (\len,\len) circle (\diam);
\draw (0,0) -- (0,\len);
\draw[very thick] (0,\len) -- (\len,\len);
\draw[very thick] (\len,\len) -- (\len,0);
\draw (\len,0) -- (0,0);
\draw[very thick] (0,\len)  -- (\len,0);
\draw (0,0) -- (\len,\len);
\end{tikzpicture}
\\
\begin{tikzpicture}
\filldraw[black] (0,0) circle (\diam); 
\filldraw[black] (0,\len) circle (\diam);
\filldraw[black] (\len,0) circle (\diam);
\filldraw[black] (\len,\len) circle (\diam);
\draw[very thick] (0,0) -- (0,\len);
\draw[very thick] (0,\len) -- (\len,\len);
\draw (\len,\len) -- (\len,0) -- (0,0);
\draw[very thick] (0,\len)  -- (\len,0);
\draw (0,0) -- (\len,\len);
\end{tikzpicture}
\end{tabular} & 0 & 0 & 0 & -1 & -1 & -1 & 0 & 0 & 0 & 0 & 0 & 0 & 0 & 0 & 0 \\  \hline
 \mysextuple{1}{1}{\beta_1}{\beta_2}{1}{1} &
  \begin{tikzpicture}
\filldraw[black] (0,0) circle (\diam); 
\filldraw[black] (0,\len) circle (\diam);
\filldraw[black] (\len,0) circle (\diam);
\filldraw[black] (\len,\len) circle (\diam);
\draw[very thick] (0,0) -- (0,\len);
\draw (0,\len) -- (\len,\len);
\draw[very thick] (\len,\len) -- (\len,0);
\draw (\len,0) -- (0,0);
\draw (0,\len)  -- (\len,0);
\draw (0,0) -- (\len,\len);
\end{tikzpicture} & 0 & 0 & 0 & -2 & -2 & 0 & 0 & 0 & 0 & 0 & 0 & 0 & 0 & 0 & 0 \\  \hline
 \mysextuple{\beta_1}{1}{\beta_2}{\beta_3}{1}{1} &
 \begin{tikzpicture}
\filldraw[black] (0,0) circle (\diam); 
\filldraw[black] (0,\len) circle (\diam);
\filldraw[black] (\len,0) circle (\diam);
\filldraw[black] (\len,\len) circle (\diam);
\draw[very thick] (0,0) -- (0,\len);
\draw[very thick] (0,\len) -- (\len,\len);
\draw[very thick] (\len,\len) -- (\len,0);
\draw (\len,0) -- (0,0);
\draw (0,\len)  -- (\len,0);
\draw (0,0) -- (\len,\len);
\end{tikzpicture} & 0 & 0 & 0 & -1 & -2 & 0 & 0 & 0 & 0 & 0 & 0 & 0 & 0 & 0 & 0 \\  \hline
 \mysextuple{\beta_1}{\beta_2}{\beta_3}{\beta_4}{1}{1} &
 \begin{tikzpicture}
\filldraw[black] (0,0) circle (\diam); 
\filldraw[black] (0,\len) circle (\diam);
\filldraw[black] (\len,0) circle (\diam);
\filldraw[black] (\len,\len) circle (\diam);
\draw[very thick] (0,0) -- (0,\len);
\draw[very thick] (0,\len) -- (\len,\len);
\draw[very thick] (\len,\len) -- (\len,0);
\draw (\len,0) -- (0,0);
\draw[very thick] (0,\len)  -- (\len,0);
\draw (0,0) -- (\len,\len);
\end{tikzpicture} & 0 & 0 & 0 & -1 & -1 & 0 & 0 & 0 & 0 & 0 & 0 & 0 & 0 & 0 & 0 \\  \hline
 \mysextuple{1}{\beta_1}{\beta_2}{\beta_3}{\beta_4}{1} &
 \begin{tikzpicture}
\filldraw[black] (0,0) circle (\diam); 
\filldraw[black] (0,\len) circle (\diam);
\filldraw[black] (\len,0) circle (\diam);
\filldraw[black] (\len,\len) circle (\diam);
\draw[very thick] (0,0) -- (0,\len);
\draw (0,\len) -- (\len,\len);
\draw[very thick] (\len,\len) -- (\len,0);
\draw (\len,0) -- (0,0);
\draw[very thick] (0,\len)  -- (\len,0);
\draw[very thick] (0,0) -- (\len,\len);
\end{tikzpicture} & 0 & 0 & 0 & -2 & 0 & 0 & 0 & 0 & 0 & 0 & 0 & 0 & 0 & 0 & 0 \\  \hline
 \mysextuple{\beta_1}{\beta_2}{\beta_3}{\beta_4}{\beta_5}{1} &
 \begin{tikzpicture}
\filldraw[black] (0,0) circle (\diam); 
\filldraw[black] (0,\len) circle (\diam);
\filldraw[black] (\len,0) circle (\diam);
\filldraw[black] (\len,\len) circle (\diam);
\draw[very thick] (0,0) -- (0,\len);
\draw[very thick] (0,\len) -- (\len,\len);
\draw[very thick] (\len,\len) -- (\len,0);
\draw (\len,0) -- (0,0);
\draw[very thick] (0,\len)  -- (\len,0);
\draw[very thick] (0,0) -- (\len,\len);
\end{tikzpicture} & 0 & 0 & 0 & -1 & 0 & 0 & 0 & 0 & 0 & 0 & 0 & 0 & 0 & 0 & 0 \\  \hline
 \mysextuple{\beta_1}{\beta_2}{\beta_3}{\beta_4}{\beta_5}{\beta_6} &
 \begin{tikzpicture}
\filldraw[black] (0,0) circle (\diam); 
\filldraw[black] (0,\len) circle (\diam);
\filldraw[black] (\len,0) circle (\diam);
\filldraw[black] (\len,\len) circle (\diam); 
 \draw[very thick] (0,\len) -- (\len,\len) -- (\len,0) -- (0,0) -- (0,\len);
 \draw[very thick] (0,\len) -- (\len,0);
 \draw[very thick] (0,0) -- (\len,\len);
\end{tikzpicture} & 0 & 0 & 0 & 0 & 0 & 0 & 0 & 0 & 0 & 0 & 0 & 0 & 0 & 0 & 0 \\  \hline
\end{tabular}
\caption{  \label{tab2}  Numerical coefficients specifying the three-loop functions $F^{(3)}_{\{b_{ij}\}}$, Eq.~\p{loop3F}.
All possible sextuples $\{b_{ij}\}$ (up to crossing permutations) are listed. 
The parameters $\beta_i\geq 2$  in the different lines are independent.
A thin line between points $i$ and $j$ corresponds to $y_{ij}^2$, i.e. $b_{ij} = 1$, and 
a thick line   to $(y_{ij}^{2})^{b_{ij}}$ with $b_{ij} \geq 2$.}
\end{table}

\newpage

\section{Description of the method}\label{s3}

The results listed in the tables have been obtained by {using similar ideas to those employed in} Refs.~\cite{Eden:2011we,Eden:2012tu}
for constructing (the integrand of) the correlator $\cG_{2222}$. In the case of different BPS weights there appear some important new ingredients. Here we give a brief summary of the method and explain the new key points. 

\subsection{General properties of the integrand}\label{s31}

The loop corrections \p{211} are obtained by the Lagrangian insertion procedure. It amounts to computing  
the Born-level $(4+\ell)$-point correlator with $\ell$ Lagrangian insertions and then 
integrating over the coordinates of the insertion points,
\begin{align} \lb{loops}
&\cG^{\ell}_{k_1 k_2 k_3 k_4} =   \int \frac{d^4 x_5 \ldots d^4 x_{4+\ell}}{\ell !(-4\pi^2)^{\ell}}\, G^{\ell}_{k_1 k_2 k_3 k_4}  \\
&G^{\ell}_{k_1 k_2 k_3 k_4} = \langle \cO^{(k_1)}(1)\,\cO^{(k_2)}(2)\,\cO^{(k_3)}(3)\,\cO^{(k_4)}(4)\,
\cL(5) \ldots \cL(4+\ell)\rangle_{\text{Born}}\ .   \label{integrand}
\end{align} 
Thus the problem is reduced to determining the correlator \p{integrand}.

The cases where one or more $k_i=2$ are special. The half-BPS scalar operator $\cO^{(2)}$ and the Lagrangian $\cL$ are members of the same $\cN=4$ supermultiplet, the chiral truncation $\cT$ of the stress-tensor supermultiplet,
\begin{align}\label{33}
\cT(x,y,\rho) = \cO^{(2)}(x,y) + \ldots +\rho^4 \cL(x)\,,
\end{align}
where $\rho^a_\a = \q^a_\a + \q^{a'}_\a y^a_{a'}$ is the $SU(4)$ harmonic projection of the chiral odd variable $\q^A_\a$.\footnote{The complex four-vector $y^a_{a'}$ is part of the $SO(6)$ null vector $Y^I=(1,y, \sqrt{-1-y^2})$.} This projection carries $U(1)$ charge $(+1)$. The operator $ \cO^{(2)}$ (as well as the whole supermultiplet $\cT$) has charge $(+4)$ in the same units. The Lagrangian $\cL$ is chargeless and hence independent of the harmonic variable $y$ ($SU(4)$ singlet).  The half-BPS operators $\cO^{(k)}(x,y)$ of conformal weight $k>2$ are the bottom components of other analytic superfields depending  on $\rho$, carrying $U(1)$ charge $2k$. Thus, the integrand of the loop corrections is given by  the component 
$(\rho_1)^0(\rho_2)^0(\rho_3)^0 (\rho_4)^0 (\rho_5)^4 \ldots (\rho_{4+\ell})^4$ of the super-correlator 
\be \lb{supercor}
 \langle \cO^{(k_1)}(1)\,\cO^{(k_2)}(2)\,\cO^{(k_3)}(3)\,\cO^{(k_4)}(4)\,
\cT(5) \ldots \cT(4+\ell)\rangle_{\text{Born}}\ ,
\ee
evaluated in the Born approximation.
It is invariant under the permutations of the points  $(5, \ldots ,4+\ell)$, 
i.e. it has  $S_{\ell}$ symmetry.  In the special case where $p$ of the $k_i=2$ this symmetry is enhanced to $S_{p+\ell}$. If some operators have equal weights  $k_i=k_j\neq2$, there is an additional permutation symmetry of those points. The most symmetric case $\cG^{\ell}_{2222}$ was studied in \cite{Eden:2011we,Eden:2012tu}, where the maximal $S_{4+\ell}$ symmetry proved to be extremely helpful in constructing the integrand. In the general case $\cG^{\ell}_{k_1 k_2 k_3 k_4}$ we have less symmetry but are nevertheless able to determine the integrand up to three loops, as explained below. 

Superconformal symmetry imposes restrictions on the form of the correlators $\cG^{\ell}$.
According to the partial non-renormalisation theorem, the loop corrections 
to any four-point correlator of scalar half-BPS operators are proportional to the rational function $R(1,2,3,4)$ defined in \p{R4}. 
It is convenient to turn $R$ into a polynomial 
multiplying it by the permutation invariant factor $x_{12}^2 x_{13}^2 x_{14}^2 x_{23}^2 x_{24}^2 x_{34}^2$. 
The prefactor $R$ has $U(1)$ charge (+4) at each point whereas the correlator $\cG_{k_1 k_2 k_3 k_4}$ bears charges  $2k_i \geq 4$ at each point. 
The difference of $U(1)$ charges between $\cG^{\ell}$ and $R$ can be compensated by a product of propagator factors, 
\be \lb{Gell}
G^{\ell}_{k_1 k_2 k_3 k_4} = C_{k_1 k_2 k_3 k_4}\, \times \cI \times \,  \sum_{\{b_{ij}\}} \left( \prod_{1 \leq i < j \leq 4} (d_{ij})^{b_{ij}}\right) \,  f^\ell_{\{b_{ij}\}}(x_1,\ldots,x_{4+\ell})\,.
\ee
where $\cI = R  \, x_{12}^2 x_{13}^2 x_{14}^2 x_{23}^2 x_{24}^2 x_{34}^2$ is a polynomial in both $y$ and $x$. Explicitly this polynomial is given by
\begin{align}\label{intril}
  \cI&:=
  x_{14}^4 x_{23}^4 y_{12}^2 y_{13}^2 y_{24}^2 y_{34}^2 +x_{12}^2 x_{14}^2 x_{34}^2 x_{23}^2 y_{13}^4 y_{24}^4+x_{13}^2 x_{14}^2 x_{24}^2
  x_{23}^2 y_{12}^4 y_{34}^4\notag\\
&-x_{12}^2 x_{14}^2 x_{34}^2 x_{23}^2 y_{13}^2 y_{14}^2 y_{23}^2 y_{24}^2  -x_{13}^2 x_{14}^2 x_{24}^2 x_{23}^2
   y_{12}^2 y_{14}^2 y_{23}^2 y_{34}^2 -x_{13}^2 x_{14}^2 x_{24}^2 x_{23}^2 y_{12}^2 y_{13}^2 y_{24}^2 y_{34}^2 \notag \\ &-x_{12}^2 x_{14}^2 x_{34}^2
   x_{23}^2 y_{12}^2 y_{13}^2 y_{24}^2 y_{34}^2 +x_{12}^2 x_{13}^2 x_{24}^2 x_{34}^2 y_{14}^4 y_{23}^4+x_{12}^4 x_{34}^4 y_{13}^2 y_{14}^2
   y_{23}^2 y_{24}^2\notag\\ 
 &-x_{12}^2 x_{13}^2 x_{24}^2 x_{34}^2 y_{13}^2 y_{14}^2 y_{23}^2 y_{24}^2+x_{13}^4 x_{24}^4 y_{12}^2 y_{14}^2 y_{23}^2
   y_{34}^2-x_{12}^2 x_{13}^2 x_{24}^2 x_{34}^2 y_{12}^2 y_{14}^2 y_{23}^2 y_{34}^2\ .
\end{align}

This expression accounts for the $y-$dependence of the integrand of the correlator $\cG_{k_1 k_2 k_3 k_4}$. The $x-$coordinate 
part is not completely fixed by the superconformal symmetry.
It is encoded in the $(4+\ell)-$point rational functions $f^\ell_{\{b_{ij}\}}(x)$  having the crossing symmetry $S_{\ell}$ (or higher, depending on the weights $k_1 k_2 k_3 k_4$) of \p{supercor}. 
They can be written in the form
\begin{align}\label{3.10}
 f^\ell_{\{b_{ij}\}} &= \frac{P^{\ell}_{\{b_{ij}\}}(x_1,\ldots,x_{4+\ell})}{\prod_{1\leq p<q \leq 4+\ell} x_{pq}^2}\,,
\end{align} 
where $P^{\ell}_{\{b_{ij}\}}$ are polynomials of conformal weight $(1-\ell)$ at each point. To justify the singularity structure of this correlator we need to consider the OPE of the various operators (see Sect.~\ref{s332}). 

All possible numerator terms up to three loops were analysed
in~\cite{Eden:2011we}. There we had an additional permutation symmetry
-- not present in the current more general situation -- which meant
that all terms came with the same coefficient. Here the terms which
can appear are the same as there, but the coefficients are different.

At two and three  loops then we can write the general ansatz as
\begin{align}
  P^2_{\{b_{ij}\}}(x_1, \dots x_6) &\ =\ \sum_{\sigma \in S_6/\text{auto}}
  a^{(2)}_{\{b_{ij},\sigma\}}  x^2_{\sigma_1 \sigma_2}x^2_{\sigma_3
                                     \sigma_4}x^2_{\sigma_5 \sigma_6}
                                     \notag \\
  P^3_{\{b_{ij}\}}(x_1, \dots x_7) &\ =\ \sum_{\sigma \in S_7/\text{auto}}
  a^{(3)}_{\{b_{ij},\sigma\}}  x^4_{\sigma_1 \sigma_2}x^2_{\sigma_3
  \sigma_4}x^2_{\sigma_4 \sigma_5} x^2_{\sigma_5
  \sigma_6}x^2_{\sigma_6 \sigma_7}x^2_{\sigma_7 \sigma_1}\ ,\label{eq:2}
\end{align}
and all that remains is to determine the coefficients
$a^{(\ell)}_{\{b_{ij},\sigma\}}$.
 Here the sum is over all permutations
of $S_{4+\ell}$ which are inequivalent when acting on the  monomial.
So for example, clearly 
  $x_{12}^2x_{34}^2x_{56}^2=x_{21}^2x_{34}^2x_{56}^2$, so in the two-loop case  the
  identity permutation and the  permutation $(12)$ give the
  same monomial and we only sum over one of the two. This is the same as modding out by the automorphism group of the corresponding graph which explains our notation $S_{4+\ell}/$auto.
  Furthermore we also
  explicitly symmetrise over permutations of the integration variables,
  which further trivially identifies coefficients.

  To each term in $f_{\{b_{ij}\}}$ we can draw the corresponding graph
  \begin{align}\label{eq:9}
    \begin{tabular}{cc}
      \includegraphics[width=5cm]{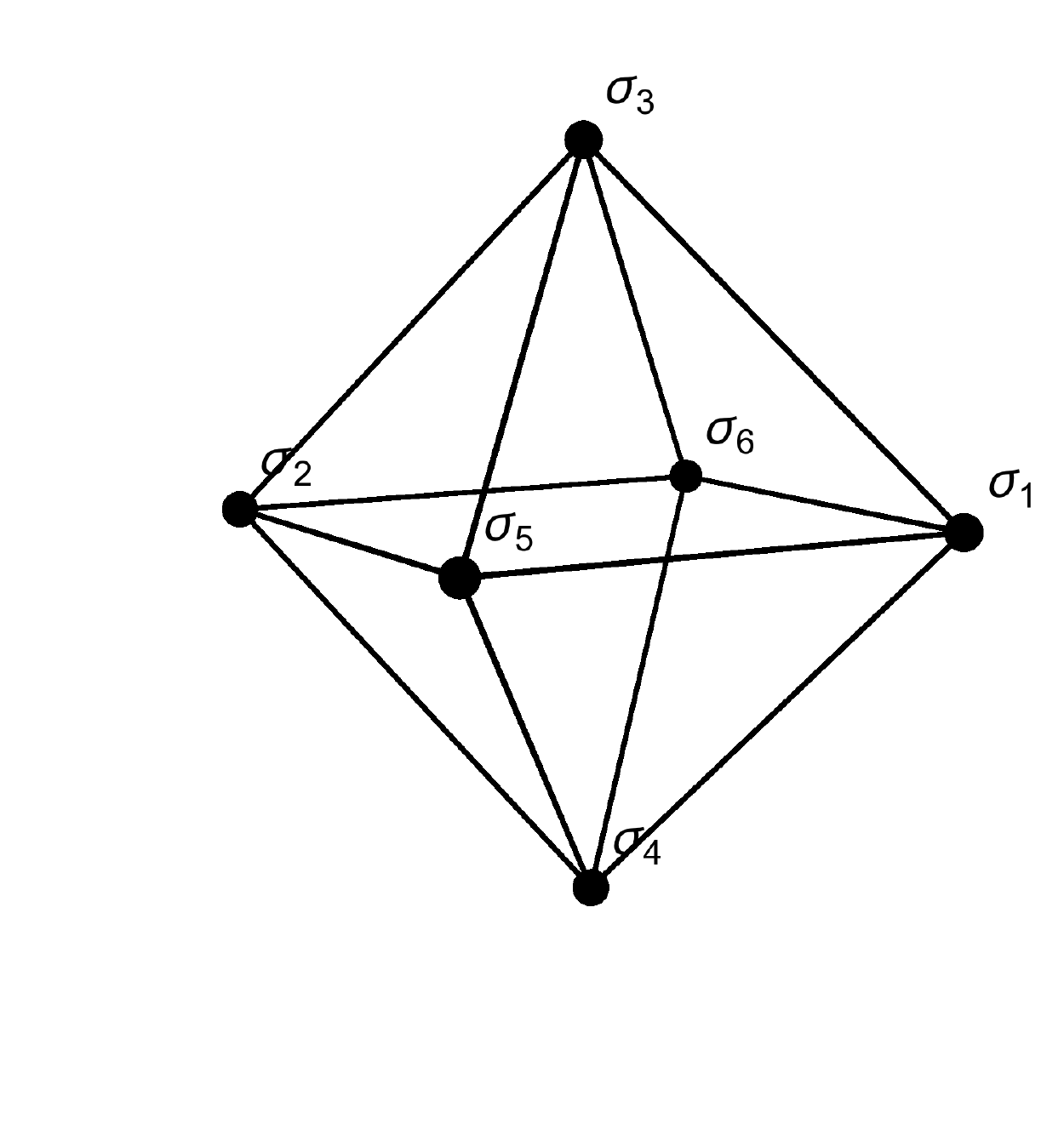}  &\includegraphics[width=5cm]{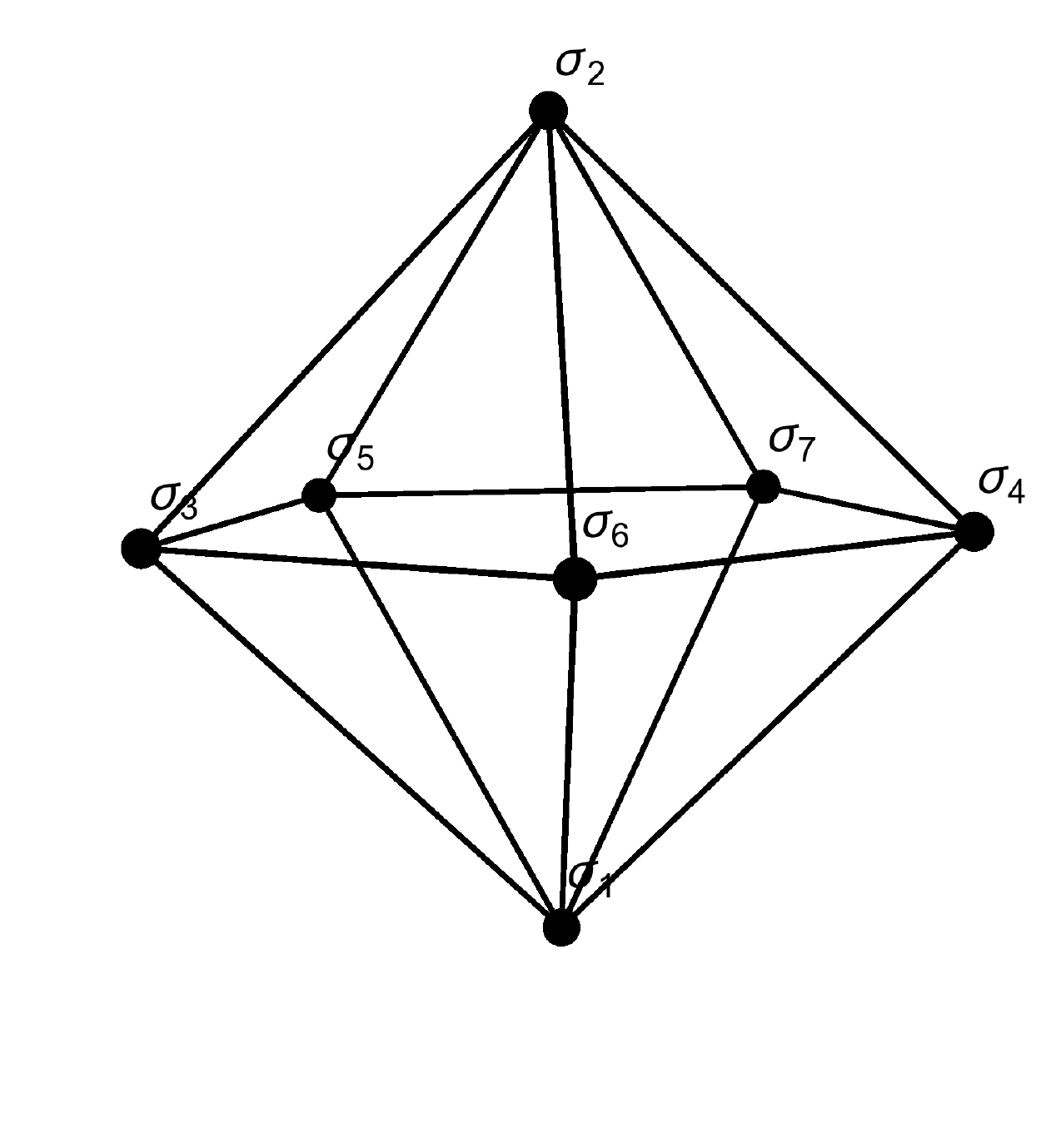}\\
      $f^2_{\{b_{ij},\sigma\}}$ &    $f^3_{\{b_{ij},\sigma\}}$\ .
    \end{tabular}
   \end{align}
An important feature of both these graphs, which we return to in the next subsection, is that they are planar, and they have the property that adding any further edge to either graph makes them non-planar.\footnote{{This property is not valid starting from four loops.}}

We should also note that although at two loops the above structure is the only possibility consistent
with conformal weights, at three loops there are three other
inequivalent topologies consistent with conformal weights. However
these do not contribute to the planar correlation function (since they
do not yield planar component correlation functions -- a requirement we
insist on, as discussed in the following subsection) and so we do not
write them out here.

\subsection{The role of planarity}\label{s32}

A further strong constraint on the polynomial $P^\ell$ in \p{3.10}  comes from the planar limit. 
We have  classified the possible $P^\ell$ having the properties discussed in Sect.~\ref{s31} above. The number can be greatly reduced by requiring that the correlator that we want to construct should correspond to the leading colour approximation in the limit $N_c \to \infty$. If we wished to compute the Born level $(4+\ell)-$point component correlator~\eqref{supercor} from standard Feynman diagrams, we would only draw {\it planar  graphs}, i.e. graphs with leading order colour factors. Here we are not using  the highly inefficient Feynman diagram technique. Instead, we wish to  {\it predict} the answer based on its elementary properties like symmetries, singularities and now planarity. Our  result should arise from the simplification of the sum of many planar Feynman graphs.
Following \cite{Eden:2012tu}, we make the natural  assumption that the final expression for any component correlator (i.e. the result of these simplifications) is itself representable as a sum of  {\it planar graphs}. These graphs are formed in the usual way with a propagator $1/x_{ij}^2$ represented by a line between points $i$ and $j$ (we can also represent numerators $x_{ij}^2$ via dashed lines, but these will not take part in the planarity criterion). We thus assert that every component correlator corresponds to a sum of planar graphs. Equivalently, every term accompanying a given $y$-structure corresponds to a sum of planar graphs. This turns out to be a strong requirement.

To illustrate the power of this we will examine in detail the
restriction from planarity on the functions $f^2_{\{111110\}}$ and
$f^2_{\{111111\}}$.

Consider the formula for the
correlator~\eqref{Gell}.
The contribution of the
coefficient function $f_{\{111110\}}^2$ to the correlator is
  \begin{align}
    \cI \times
    y_{12}^2y_{13}^2y_{14}^2y_{23}^2y_{24}^2 \times
    \frac{P^2_{\{111110\}}(x_1,\ldots,x_{6})}{x_{12}^4x_{13}^4x_{14}^4x_{23}^4x_{24}^4x_{34}^2x_{15}^2x_{25}^2x_{35}^2x_{45}^2
x_{16}^2x_{26}^2x_{36}^2x_{46}^2x_{56}^2    }\ .
  \end{align}
The powers of 4 in the denominator come from the additional propagator factors $(d_{ij})^{b_{ij}}$. They can never be removed by a numerator term (which in the two-loop case only contain $x_{ij}^2$ but not $x_{ij}^4$, see~\eqref{eq:2}). Let us concentrate on an individual term from the sum in~\eqref{eq:2}.  In order to obtain a planar contribution all three numerator factors must completely cancel  the matching factors in the
denominator. This is because each term in $f^2_{\{b_{ij}\}}$ has
the topology of an octahedron which is a planar graph, see \p{eq:9}. But the
addition of any new edges to the octahedron will produce a non-planar graph. In
$f^2_{\{b_{ij}\}}$ itself all numerators cancel denominators, and so
we conclude that any
numerator not cancelling a denominator will automatically yield a
non-planar graph.

Graphically, the multiplication by $d_{ij}$ corresponds to attaching further edges $e_{12},e_{13},e_{14},e_{23},e_{24}$ to the graph in~\eqref{eq:9} for some choice of permutation $\sigma$. This is only allowed if all these 5 edges are  already existing edges  (since as mentioned below~\eqref{eq:9}, adding a new edge results in a non-planar graph). It thus becomes apparent that the only possible term in
$P^2_{\{111110\}}$ which can yield a planar contribution is
\begin{align}
  P^2_{\{111110\}}  \propto  x_{34}^2 x_{15}^2 x_{26}^2 +x_{34}^2 x_{15}^2
  x_{26}^2\ .
\end{align}
This is indeed the only non-zero term in our  final  result given in the
penultimate line of table~\ref{tab1}. It corresponds to a single
orientation of the two-loop ladder integral.

However it is also now clear that a similar analysis in the case
$f^2_{\{111111\}}$ -- which will have an additional power of $x_{34}^2$ in
the denominator compared to the previous case -- will mean there is
no numerator that can yield a planar contribution. We
conclude that this contribution vanishes, $f^2_{\{111111\}}=0$. It is then clear that having all $b_{ij} \geq 1$ does not modify this non-planar topology, therefore planarity
 alone implies that
\begin{align}\label{f0}
  f^2_{\{b_{ij}\}}&=0\quad \text{if all } \ b_{ij} \geq 1\ .
\end{align}

The analysis at three loops is very similar. However the presence of $x^4_{ij}$ in the numerator (see~\eqref{eq:2}) means that the effect
is slightly delayed and takes place for $b_{ij}\geq 2$ 
rather than $b_{ij}\geq 1$. We find
\begin{align}
  f^3_{\{b_{ij}\}}&=0\quad \text{if all } \ b_{ij} \geq 2\ .
                    \end{align}

Note that in the above analysis we have ignored the effect of
the polynomial $\cI$ on planarity. Indeed terms in $\cI$ can cancel
denominators and this softens the ``non-planarity'' of the
result. However there are a number of terms in $\cI$ with
different $y$ factors~\eqref{intril} and all terms need to be
planar. It turns out that apart from one-loop the presence of this polynomial does not affect the conclusions. Similarly one should consider the sum over all building block
functions rather  
than focussing on each single coefficient function alone. However, again,  doing
so does not seem to change the above conclusions. In other words so far
we found that it was enough to assume that 
\begin{align}
\left( \prod_{1 \leq i < j \leq 4} (d_{ij})^{b_{ij}}\right) \,  f^\ell_{\{b_{ij}\}}(x_1,\ldots,x_{4+\ell}) \label{bb}  
\end{align}
are all given by planar expressions. This is not a necessary consequence of the above planarity requirement. When~\eqref{bb}  is inserted into the correlator expression~\eqref{Gell}, one multiplies by the  polynomial $\cI$ -- thus removing propagators -- and sums over different structures which all mix together. Thus this leaves the possibility that the contributing expressions~\eqref{bb} could be non-planar whilst still giving planar component correlators. But in actual fact this never appears to happen in practice.

The next step is to find a way of fixing the coefficients in  \p{eq:2} , i.e. in the {\it planar} ansatz for the integrand $G^{\ell}_{k_1 k_2 k_3 k_4}$ in \p{Gell}. We are going to use two criteria based on the detailed understanding of the OPE. The first amounts to comparing the singular light-like limits of two correlators $\lim_{x^2_{12}\to0}G^\ell_{k_1 k_2 k_3 k_4}$ and $\lim_{x^2_{12}\to0}G^\ell_{k_1+1, k_2+1, k_3, k_4}$. The second criterion, proposed in \cite{Eden:2012tu}, derives from the requirement that the logarithm of the correlator  have simple log divergences in the short-distance limit.

\subsection{Light-cone OPE relation}\label{s33}

We claim the existence of a powerful relation between different integrands, i.e. Born-level correlators. It is based on the structure of the OPE of two half-BPS operators  $\cO^{(k_1)}(1)$ and $\cO^{(k_2)}(2)$ in the light-cone limit $x^2_{12} \to 0$. The key property is that the leading light-cone singularity in each $SU(4)$ channel of the correlator $G^\ell_{k_1+1, k_2+1, k_3 k_4}$ is simply related, in the planar approximation $N_c \to \infty$, to that of $G^{\ell}_{k_1 k_2 k_3 k_4}$:
\begin{align}\label{310}
 \lim_{\substack{x_{12}^2,y_{12}\rightarrow 0 \\
  d_{12} \text{ fixed}}}  \, \left[ \frac{G^\ell_{k_1+1, k_2+1, k_3 k_4}}{C_{k_1+1, k_2+1, k_3 k_4}}  
	- d_{12}\times \frac{G^\ell_{k_1 k_2 k_3 k_4}}{C_{k_1 k_2 k_3 k_4}} \right] = \a\, d_{12} \,.%%\text{-linear terms}\,,
\end{align}
The limit is taken as follows: $y_{12} =\ep \vec n$, $x^2_{12} = \ep^2 $ with some complex four-vector $\vec n$ and  $\ep\to0$. The propagator factor
$d_{12}$ in the second term on the left-hand side equalises the
conformal and $SU(4)$ weights of the two terms.  The claim is that in
this limit the two correlators in \p{310} can only differ by terms proportional to  $d_{12}$. The proof is given below in Sect.~\ref{s331}.

Let us insert the general form of the correlators \p{Gell}, \p{3.10} in \p{310}:
\begin{align}\label{3.11}
 \lim_{\substack{x_{12}^2,y_{12}\rightarrow 0 \\
  d_{12} \text{ fixed}}} \, \frac{\cI}{\prod_{1\leq p<q \leq 4+\ell} x_{pq}^2}\,  d_{12}\, \sum_{\{b_{ij}\}} \left( \prod_{1 \leq i < j \leq 4} (d_{ij})^{b_{ij}}\right)  \left[ P^{\ell}_{\{b_{ij}\}}\vert_{b_{12} \to b_{12} +1} - P^{\ell}_{\{ b_{ij}\}} \right]  =0\,,  
\end{align}
where the sextuples $\{ b_{ij} \}$ correspond to the labels $k_1 k_2 k_3 k_4$ before the shift. The shift of the label $b_{12}$ accounts for the shifts of $k_1$ and $k_2$ (recall that $k_i = \sum_{j\neq i} b_{ij} + 2$). The reason for the vanishing right-hand side of \p{3.11} is that in our limit  $d_{12} \cI/x^2_{12} \to (d_{12})^2 (x^2_{13} x^2_{24} - x^2_{14} x^2_{23})^2 y^2_{13} y^2_{14} y^2_{34}$ (see \p{intril}), while we expect only $(d_{12})^1$ on the right-hand side of \p{310}.\footnote{The terms with $b_{12}=0$ in $G^\ell_{k_1+1, k_2+1, k_3 k_4}$ are not displayed in \p{3.11} because they contribute only to the right-hand side in \p{310}. } 
{Let us extract the terms with the same $y-$structure  from \p{3.11}. 
The family of sextuples $\{b^{(k)}\} = \{b_{12},b_{13}-k,b_{14}+k,b_{23}+k,b_{24}-k,b_{34}\}$ parametrised by 
an integer $k$ from the interval $[\a,\beta]\,, \a = -\min(b_{14},b_{23}) , \beta = \min(b_{13},b_{24}),$ 
corresponds to the $y-$structure $(d^2_{12})^{b_{12}}(y^2_{13})^{b_{13}+b_{23}}(y^2_{14})^{b_{14}+b_{24}}(y^2_{34})^{b_{34}}$.
We deduce the following condition on the polynomials in the ansatz} 
\begin{align}\label{321''}
&\sum_{k = \a}^{\beta} 
(x^2_{13} x^2_{24})^{k-\a} (x^2_{14}x^2_{23})^{\a+\beta-k}
\left[ P^{\ell}_{\{b^{(k)}\}}\vert_{b_{12} \to b_{12} +1}  - P^{\ell}_{\{ b^{(k)}\}}\right]_{x^2_{12}=0} =0 \,.
\end{align}

In reality, eq.~\p{321''} {in combination with planarity at two and three loops implies the stronger constraint (see Appendix \ref{planarity2})}
\begin{align}\label{321'}
\left[ P^{\ell}_{\{b_{ij}\}}\vert_{b_{12} \to b_{12} +1}  - P^{\ell}_{\{ b_{ij}\}}\right]_{x^2_{12}=0} = 0 \,.
\end{align}

The constraint \p{321'} can be applied to any pair of the four outer points of the correlators. It can be repeated iteratively, shifting the weights at the chosen pair of points by any finite amount. This results in many relations between the coefficients of the numerators $P^{\ell}_{\{b_{ij}\}}$ of correlators of different BPS weights. We use the known correlators $\cG^2_{2222}\, , \cG^3_{2222}$ (which have been obtained in \cite{Eden:2011we} by a similar method) as the starting
point of the recursion.  The planarity requirement of Sect.~\ref{s32}, in combination with this light-cone OPE relation, 
implies that it is sufficient to consider only configurations with $b_{ij}=0,1$ (two loops) or $b_{ij}=0,1,2$ (three loops), all cases with higher weights are reduced to these (see Appendix~\ref{planarity2}). Then relation \p{321'} allows us to fix all the coefficients in our ansatz at two loops and all but one at three loops. To fix the latter we need yet another OPE criterion explained in Sect.~\ref{s23}.

\subsubsection{Origin of the light-cone relation}\label{s331}

Relation \p{310} follows from  the light-cone OPE of two half-BPS operators and any third operator 
$\cO^{\Delta,S}_{[a,b,a]}$ of dimension $\Delta$, spin $S$ and in the $SU(4)$ representation with Dynkin labels $[a,b,a]$.\footnote{This OPE has been studied in \cite{Eden:2001ec,Heslop:2001gp}. For a summary see Appendix~\ref{ApB}.}
This takes the form
\begin{align}
  \cO^{(k_1)}(x_1,y_1)
  \cO^{(k_2)}(x_2,y_2) \ \sim\  &  C_{k_1 k_2 \cO^{\D,S}_{[a,b,a]}}
  \frac{(y_{12}^2)^{(k_1+k_2-2a-b)/2}}{(x_{12}^2)^{(k_1+k_2-\Delta+S)/2}}
 \left( [x_{12}]^S [y_{12}]^a\cO^{\Delta,S}_{[a,b,a]}(x_2,y_2) \ +\
  \dots\ \right)
  .\label{eq:1}
\end{align}
Here the dots denote descendant terms, both space-time descendants
($x$-derivatives of operators) and 
$SU(4)$ descendants ($y$-derivatives of operators). Importantly, these
descendant terms only appear together with 
polynomials in $x_{12}$ and $y_{12}$. We are slightly schematic in our display of indices. The square brackets
simply indicate  symmetrised tensor products, and the indices will be contracted
with those of the operator. The range of $SU(4)$ representations  on the right-hand side of \p{eq:1} is determined by the tensor product (we assume that $k_1 \geq k_2$)
\begin{equation}\label{5.2'}
  [0,k_1,0]\otimes [0,k_2,0] = \bigoplus_{r=0}^{k_2}
  \bigoplus_{a=0}^{k_2-r}[a,b,a] \qquad {\rm with} \  b=k_1+k_2-2a-2r\,.
\end{equation}
These are $SO(6)$ tensor representations of rank $2a+b=k_1+k_2-2r$. 
The OPE $\cO^{(k_1)}(1)\times \cO^{(k_2)}(2)$ contributes to the correlator \p{integrand} only those  representations which are in the overlap with the tensor product $[0,k_3,0]\otimes [0,k_4,0]$.

In $\cN=4$ super-Yang-Mills operators form multiplets. The superconformal primary operators with $r=0,1$ are protected (BPS or semishort), the unprotected operators have $r\geq 2$ (see  \cite{Eden:2001ec,Heslop:2001gp}).  Further, the protected operators in the product $\cO^{(k_1)}(1)\times \cO^{(k_2)}(2)$ do not appear in \p{integrand} because their three-point functions with $\cO^{(k_3)}(3)\,\cO^{(k_4)}(4)$ are protected \cite{Heslop:2003xu} and hence have no loop corrections. Therefore, for our purposes the first sum on the right-hand side of \p{5.2'} starts at $r=2$ corresponding to long
multiplets in the OPE. Each long multiplet contains
 a number of superdescendant operators, only some of
which appear in the OPE of two half BPS scalars. 
The superdescendants which occur  are: 
\begin{align}\label{eq:10aa}
  \begin{array}{c}
   \cO^{\D,S}_{[a,b,a]}, \\ \\
  \  \cB^{\D+1,S+1}_{[a+1,b,a+1]},\   \cB^{\D+1,S+1}_{[a-1,b+2,a-1]},\   \cB^{\D+1,S-1}_{[a-1,b+2,a-1]},\  \cA'{}^{\D+1,S-1}_{[a+1,b,a+1]}, \\ \\
  \  \cC^{\D+2,S+2}_{[a,b+2,a]}, \   \cC^{\D+2,S}_{[a,b+2,a]},\  \cC^{\D+2,S}_{[a-2,b+4,a-2]},\   \cB'{}^{\D+2,S}_{[a+2,b,a+2]},\  \cB'{}^{\D+2,S}_{[a,b+2,a]},\  \cB'{}^{\D+2,S-2}_{[a,b+2,a]},\\ \\  \   \cD^{\D+3,S+1}_{[a-1,b+4,a-1]},\   \cC'{}^{\D+3,S+1}_{[a+1,b+2,a+1]},\   \cC'{}^{\D+3,S-1}_{[a+1,b+2,a+1]},\   \cC'{}^{\D+3,S-1}_{[a-1,b+4,a-1]}, \\ \\ \   \cD'{}^{\D+4,S}_{[a,b+4,a]} \ .
  \end{array}
\end{align}
The superdescendants  are obtained by acting with the supercharges on the primary operator (HWS) $\cO^{\Delta,S}_{[a,b,a]}$. The derivation of this from analytic superspace together with more details is in appendix~\ref{ApB}.
Here we will simply consider the first and the last  terms. The highest dimension component which occurs is $\cD'{}^{\Delta+4,S}_{[a,b+4,a]}  \in Q^4 \bar Q^4 \cO^{\Delta,S}_{[a,b,a]}$. The super OPE then takes the form 
\begin{align}\label{eq:4}
& \cO^{(k_1)}(x_1,y_1)
  \cO^{(k_2)}(x_2,y_2) \notag \\\ \sim\  &  \sum_{\Delta,\, S,\, [a,b,a]} C_{k_1 k_2 \cO^{\D,S}_{[a,b,a]}}\ 
 (d_{12})^{(k_1+k_2-2a-b)/2 -2} \ 
  (x_{12}^2)^{(\Delta-S-2a-b)/2-2}\,  \times \ 
  \notag\\
& 
 \left[{y_{12}^4} [x_{12}]^S  [y_{12}]^a\cO^{\Delta,S}_{[a,b,a]}
  \ +\    \ldots 
 \ +\     x_{12}^4 [x_{12}]^S [y_{12}]^a\cD'{}^{\D+4,S}_{[a,b+4,a]}  \ +\ \dots\right]
  \ ,
\end{align}
where the dots in the middle denote terms relating to the other  superdescendants (listed in~\eqref{eq:10aa}) and the dots at the end denote conformal and $SU(4)$ descendants. The main point is that each component of the supermultiplet appears with the same OPE
coefficient $ C_{k_1 k_2 \cO^{\D,S}_{[a,b,a]}}$.

We now wish to consider the OPE in the limit in which $x_{12}^2\rightarrow 0$ and $y_{12}\rightarrow 0$ but with the ratio $d_{12}=y_{12}^2/x_{12}^2$ fixed. The leading contribution to the OPE in this limit comes from operators with the minimum value of $\Delta -S - 2a-b$. 
Well-known  superconformal unitarity bounds~\cite{Dobrev:1985qv} state
that for long representations the superconformal primary satisfies $\Delta -S - 2a-b \geq 2$.  Let us thus set $\Delta -S= 2a+b+ 2$ in  \p{eq:4}:
\begin{align}\label{322}
& \cO^{(k_1)}(x_1,y_1)
  \cO^{(k_2)}(x_2,y_2)  \sim 
 \sum_{ \Delta-S=2a+b+2} C_{k_1 k_2 \cO^{\D,S}_{[a,b,a]}}\  (d_{12})^{(k_1+k_2-2a-b)/2 -1}
  [x_{12}]^S  \left( {y_{12}^2} [y_{12}]^a\cO^{\Delta,S}_{[a,b,a]} \right.  \notag \\\ \  &  \left.  +\   [y_{12}]^{a+1}[x_{12}]\ \cB^{\Delta+1,S+1}_{[a+1,b,a+1]}  \ +\  y_{12}^2 [y_{12}]^{a-1}[x_{12}]\ \cB^{\Delta+1,S+1}_{[a-1,b+2,a-1]}  \ +\  \  
  [y_{12}]^{a}\  [x_{12}]^2 \cC^{\Delta+2,S+2}_{[a,b+2,a]} \ +\ \dots\right)
  \ ,
\end{align}
where we have displayed only the terms of leading twist. In our double limit  only one superdescendant  survives, 
 \begin{align}\label{eq:4'}
& \lim_{\substack{x_{12}^2,y_{12}\rightarrow 0 \\
  d_{12} \text{ fixed}}}  \, \cO^{(k_1)}(x_1,y_1)
  \cO^{(k_2)}(x_2,y_2) \nt
  & \qquad \sim\   \sum_{\Delta-S=b+2} C_{k_1 k_2 \cO^{\D,S}_{[0,b,0]}}\ 
 (d_{12})^{(k_1+k_2-b)/2 -1}  \ 
  [x_{12}]^{S+2} \  \cC^{\Delta+2,S+2}_{[0,b+2,0]}  \ +\ \dots 
  \ .
\end{align}
An important property of this type of operators is that they are made entirely from scalars. Indeed, they have twist $(\Delta+2)-(S+2)= b+2$ and $SU(4)$ labels $[0,b+2,0]$ implying that the length of the operator (i.e. number of constituent scalars) equals the rank of the $SO(6)$ tensor representation. Trying to replace some of the scalars by fermion bilinears or by gluons either increases the twist or modifies the representation.

The question we want to investigate now is what happens to the OPE
structure constant $C_{k_1 k_2 \cO^{\D,S}_{[0,b,0]}}$ when we increase
the BPS weights $k_1 \to k_1+1\,, \ k_2 \to k_2 + 1$. We wish to show
that\footnote{This result is consistent with the values of the structure constants for the operators in the so-called $SL(2)$ sector considered in \cite{Vieira:2013wya}. Such operators correspond to the descendants $\cC^{\Delta+2,S+2}_{[0,b+2,0]}$ in \p{eq:10aa} and they are made only from scalars.}
\begin{align}\label{319}
{C_{k_1+1, k_2+1, \cO^{\D,S}_{[0,b,0]}}} = {C_{k_1 k_2
  \cO^{\D,S}_{[0,b,0]}}} \times \frac{N_c}{2}
  \frac{(k_1+1)(k_2+1)}{k_1 k_2}\ .
\end{align}
In other words, the ratio of the two OPE coefficients is {\it
  independent of the quantum numbers of the 
  operator} $\cO^{\D,S}_{[0,b,0]}$. We call this property  of the
structure constants  in the relevant sector of our OPE {\it
  universality}.

The structure constant ${C_{k_1 k_2
  \cO^{\D,S}_{[0,b,0]}}} $ is the same for all the members of the supermultiplet. We find it advantageous to determine it from the three-point function
$\vev{ \cO^{(k_1)}(1)  \cO^{(k_2)}(2)  \cC^{\Delta+2,S+2}_{[0,b+2,0]}(3)}$
divided by the two-point function $\vev{ \cC \cC}$ (the latter drops out of the ratio \p{319}). The key point in
our argument is that the operator $ \cC^{\Delta+2,S+2}_{[0,b+2,0]}$ is made from scalars only. Our integrand \p{integrand} is a Born-level correlator. For the three-point function $\vev{ \cO(1)  \cO(2)\cC(3)}$, where all the operators are made from scalars, the Born approximation coincides with the free theory result. It is obtained by Wick contractions with free scalar propagators. A certain number $k$
of propagators connect points 1 and 2.  The remaining $k_1-k$ scalars
at point 1 are connected to point 3 and similarly for point 2. This
means that the operator $\cC^{\Delta+2,S+2}_{[0,b+2,0]}$ is made from
$k_1+k_2-2k=b+2$ scalars %{[\bf PH: $k_1+k_2-2k$ scalars, yes,  but this need not equal  $\D-S$. Eg $\partial_\mu \phi\partial_\mu\phi\phi$ is twist 5 but length 3] }
and $S$ space-time derivatives.\footnote{The number $k$ cannot be zero because otherwise
the twist of the superconformal primary $\cO^{\D,S}_{[0,b,0]}$ will be  $\Delta-S= k_1+k_2$. We have already set  $\Delta -S= b+ 2=
  k_1+k_2 +2(1-r)$, so this would imply $r=1$ in \p{5.2'}. As pointed
  out earlier, the multiplets with $r=1$ are protected and do not contribute to the integrand  \p{integrand}. } The space-time dependence of the
three-point function $\vev{\cO \cO \cC}$ is fixed by conformal symmetry, and the
structure constants are determined by the colour and combinatorial
factors.  Let us examine the colour tensor of each operator. For the
half-BPS (single-trace) operator $\cO^{(k_i)}$ it is the  trace $\tr(t_{( a_1} \ldots t_{a_{k_i} )})$ of $k_i$ generators of
$SU(N_c)$ (here $( \ldots )$ denotes weighted symmetrisation). For
$ \cC^{\Delta+2,S+2}_{[0,b+2,0]}$ the colour tensor $\gamma^{S}_{a_1 \ldots
  a_{k_1+k_2-2k}}$ depends on the details of the operator in question.\footnote{In the case of degeneracy, i.e. existence of several operators $\cC$ with the same quantum numbers made from scalars (for an example see \cite{Bianchi:2002rw}), we assume that they have been diagonalised. Then the colour tensor $\gamma^S$ corresponds to a specific eigenstate.  \label{f8} }  The combination of the three colour tensors, including the combinatorial factor,  has the form
\begin{align}\label{}
\frac{k_1 ! k_2 ! }{k!} \tr[t_{( a_1} \ldots t_{a_k } t_{b_1} \ldots t_{b_{k_1-k})}] \, \tr[t_{(a_1} \ldots t_{a_k} t_{c_1} \ldots t_{c_{k_2-k})}] \,   \gamma^{S}_{b_1 \ldots b_{k_1-k} c_1 \ldots c_{k_2-k}}\,.
\end{align}
Here the $k$ Wick contractions between points 1 and 2 are realised as contractions of the first $k$ colour tensor indices. In the large $N_c$ limit this becomes\footnote{This relation does not hold for $k=0$ but as we have explained, this case is of no relevance for us.}
\begin{align}\label{320}
 \left(\frac{N^{k-1}_c}{2^k}\,  k_1  k_2\right)   (k_1-k)! (k_2-k)!\
  \tr[t_{(b_1} \ldots t_{b_{k_1-k})} \, t_{( c_1} \ldots
  t_{c_{k_2-k})}] \gamma^S_{b_1 \ldots c_{k_2-k}} \qquad (k\geq 1).
\end{align}

Now, let us see what happens when we shift $k_1 \to k_1+1\,, \ k_2 \to
k_2 + 1$. In order to maintain the twist or equivalently the length $k_1+k_2-2k$ fixed, we need to also increase $k\to k+1$.  Only the first factor in the parentheses in \p{320} changes.  Then the ratio of the structure constants is  as given by eq.~\p{319}.

Next, let us insert the OPE limit \p{eq:4'} in the correlator \p{integrand} 
\begin{align}\label{3.21}
& \lim_{\substack{x_{12}^2,y_{12}\rightarrow 0 \\
  d_{12} \text{ fixed}}}  \,  \cG^\ell_{k_1 k_2 k_3 k_4} =\sum_{\cO^{\D,S}_{[0,b,0]}: \Delta=S+b+2} 
C_{k_1 k_2 \cO^{\D,S}_{[0,b,0]}} \ 
 d_{12}^{\, r-1} 
  [x_{12}]^{S+2} \times\notag\\
& \left\langle{\left[   \cC^{\Delta+2,S+2}_{[0,b+2,0]}   +\ \dots\right] \,\cO^{(k_3)}(3)\,\cO^{(k_4)}(4)\,
\cL(5) \ldots \cL(4+\ell)}\right\rangle_{\text{Born}}\ ,
\end{align}
and in its counterpart $\cG^\ell_{k_1+1, k_2+1, k_3 k_4}$.
%\begin{align}\label{3.22}
%& \lim_{\substack{x_{12}^2,y_{12}\rightarrow 0 \\
  %d_{12} \text{ fixed}}}  \,  \cG^\ell_{k_1+1, k_2+1, k_3 k_4} =\sum_{\cO^{\D,S}_{[a,b,a]}: \Delta=S+2a+b+2} 
%C_{k_1+1, k_2+1, \cO^{\D,S}_{[a,b,a]}} \ 
 %d_{12}^{\, r'-1} 
  %[x_{12}]^S [y_{12}]^a\times\notag\\
%& \vev{\left(d_{12} [y_{12}][x_{12}]\ \cB^{\Delta+1,S+1}_{[a+1,b,a+1]}  +\ \dots\right) \,\cO^{(k_3)}(3)\,\cO^{(k_4)}(4)\,
%\cL(5) \ldots \cL(4+\ell)}_{\text{Born}}\ .
%\end{align}
The shifted version of the tensor product \p{5.2'} is
\begin{align}\label{}
 [0,k_1 + 1 ,0]\otimes [0,k_2 + 1 ,0] = \bigoplus_{r=0}^{k_2+1}
  \bigoplus_{a=0}^{k_2+1-r}[a,k_1+k_2+2-2a-2r,a]   \,.
\end{align}
As before, the long multiplets have $r\geq 2$. Comparing this decomposition with  \p{5.2'} for $a=0$ and $r\geq 2$, we see that the shifted version contains a representation $[0,k_1+k_2 - 2,0]$ with $r=2$. This is an $SO(6)$ tensor of rank $k_1+k_2-2$. The maximal rank for long operators ($r=2$) in  \p{5.2'} is $k_1+k_2-4$.  From \p{3.21} it follows that this extra channel comes with  a prefactor $d_{12}$ which  account for the right-hand side of \p{310}. Note that such a contribution will only be visible if it also appears in the  tensor product $[0,k_3,0]\otimes [0,k_4,0]$.

Finally, taking into account the universality of the structure constants \p{319} and the normalisation factors \p{219} in \p{310}, we see that the contributions of the long supermultiplets  common for both correlators coincide in our limit. The only source of  difference are the terms with $r=2$ in the shifted version of \p{3.21}, which are responsible for  the right-hand side of \p{310}. We have thus proven this important relation.

\subsubsection{A possible stronger relation}

The examination of our two- and three-loop results in Sect.~\ref{s2.1} shows that they are compatible with a stronger light-cone relation between two planar correlators:
\begin{align}\label{newrel}
 \frac{G^\ell_{k_1+1, k_2+1, k_3 k_4}}{C_{k_1+1, k_2+1, k_3 k_4}}  
	- d_{12}\times \frac{G^\ell_{k_1 k_2 k_3 k_4}}{C_{k_1 k_2 k_3
  k_4}} = O(1/x_{12}^2)\ .
\end{align}
The claim is that in
this limit the two correlators in \p{310} can only differ by terms
of order $1/x_{12}^2$. Bearing in mind that the correlator $G^\ell_{k_1 k_2 k_3 k_4}$ can have
poles in $x_{12}^2$ up to $1/(x_{12}^2)^{(k_1+k_2-2)/2}$, this involves a remarkable cancellation of much of the correlation functions. Notice that the new relation does not involve any limit of the auxiliary $y-$variables, i.e. it applies to all the $SU(4)$ channels in the correlators. From \p{newrel} we deduce the stronger condition on the numerator polynomials in \p{3.10}
\begin{align}\label{321}
 P^{\ell}_{\{b_{ij}\}}\vert_{b_{12} \to b_{12} +1}  - P^{\ell}_{\{b_{ij}\}}= O((x_{12}^{2})^{b_{12}+1}) \,.
\end{align}
Here $b_{12} \leq \ell-2$ in order to match the conformal weights of the left- and right-hand sides. If $b_{12} > \ell-2$ the right-hand side must vanish.

How could we possibly prove such a relation? The starting point would be the conformal light-cone OPE \p{eq:1} (no need to evoke its supersymmetric version \p{eq:4} anymore). The relation would hold if the universality of the structure constants \p{319} applied to all possible operators, not just the specific $SU(4)$ channels $[0,b,0]$. To prove this we would be tempted to argue that the structure constants are determined by the free three-point function $\vev{ \cO^{(k_1)}  \cO^{(k_2)} \cO^{\D,S}_{[a,b,a]}}_{\rm free}$. Then the planar colour factor \p{320} would explain the universality, with the exception of the case $k=0$, i.e. when the operator $\cO^{\D,S}_{[a,b,a]}$ has maximal length. This would explain the right-hand side in \p{newrel}.

The problem with this argument is that for generic operators $\cO^{\D,S}_{[a,b,a]}$,  whose length does not equal the rank $2a+b$ of the $SO(6)$ representation, we cannot rule out the presence of fermions and gluons in their composition. For such ingredients the three-point function $\vev{\cO \cO \cO^{\D,S}_{[a,b,a]}}_{\rm Born}$ is not necessarily free anymore. A simple example is an operator of the type $\tr(F^2_{\mu\nu})$. It can only talk to the half-BPS scalar operators via interaction vertices, so $\vev{\cO \cO \tr(F^2)}_{\rm Born} \sim g^2$ and not $g^0$ as for an operator made of scalars. The colour factors of such three-point functions become more difficult to control. This does not mean that the universality of the structure constants stops working, but at present we cannot make a definitive claim. This issue deserves further study. 

We would like to point out that the three constraints -- the weaker \p{321''}, the intermediate \p{321'} and the stronger \p{321}, are in fact equivalent up to three loops if we assume planarity (see Appendix \ref{planarity2} for the explanation). This is however not true starting from four loops.

\subsubsection{Singularities of the integrand}\label{s332}

The OPE considerations above allow us to explain the structure of the space-time singularities of our ansatz for the integrand \p{Gell}, \p{3.10}. Consider first the singularities with respect to the four external points, i.e. for $x^2_{ij}\to0$ with $1\leq i,j\leq 4$.  They are determined by the OPE of two half-BPS operators saturating the unitarity bound, eq.~\p{322}.  We see that all the poles in $x^2_{12}$ appear as propagator factors, accompanied by an extra power of $y^2_{12}$ if the contribution comes from operators of $SO(6)$ rank $2a+b$. Comparing with  \p{Gell}, \p{3.10} and recalling the definition \p{R4}, we see exactly the same structure.

Further, the singularities between an external  and a Lagrangian insertion points are determined by the OPE 
\begin{align}\label{}
\cO^{(k)} \times \cL  \sim \frac{C_{\cO \cO \cL}}{x^4}  \cO^{(k)} + O\left(\frac1{x^2} \right)\,.
\end{align}
The leading singularity  $1/x^4$ does not really appear there because the two-point function of BPS operators is protected and hence $C_{\cO \cO \cL}=0$. So, this OPE contributes at most a singularity $1/x^2$, as in our ansatz \p{Gell}, \p{3.10}. 

Finally, the OPE $\cL \times \cL$ of two {\it chiral} Lagrangians has a leading singularity in the form of a contact term, $\delta^4(x)$. Our correlators are always considered for non-coincident points, so we can only see the subleading singularity $1/x^2$ in this OPE. 

\subsection{Double short-distance OPE}\label{s23}

As explained in Sect.~\ref{s33}, the powerful recursion relation \p{310} allows us to fix  all but one coefficient in our three-loop ansatz (and all at two loops). To fix the single remaining coefficient it is sufficient to consider the simplest correlator (cf. \p{211})
\begin{align}\label{}
\cG^{\rm loop}_{3322} = C_{3322}\, R\,  d_{12}\ \sum_{\ell \geq 1} a^\ell F_\ell\,.
\end{align}
A new independent restriction on this correlator follows from the Euclidean OPE (coincident points). Let us perform a double OPE in two inequivalent ways,
\begin{align}\label{}
& \lim_{1\to2, 3\to4}\cG_{3322} = \langle (\cO^{(3)}(1) \times \cO^{(3)}(2)) 
\,(\cO^{(2)}(3) \times \cO^{(2)}(4)) \rangle \nt
& \lim_{1\to3, 2\to4} \cG_{3322} = \langle (\cO^{(3)}(1) \times \cO^{(2)}(3)) 
\,(\cO^{(3)}(2) \times \cO^{(2)}(4)) \rangle\,.
\end{align}

In the first case $x_1 \to x_2, x_3 \to x_4$ or $u\to0\,, \ v\to 1$ in terms of the conformal cross-ratios \p{cr}.
The criterion derived in \cite{Eden:2011we} is that the function 
\be \lb{sdOPElog1}
\log\Bigl( 1 + 6 x_{13}^4 \sum_{\ell \geq 1}a^{\ell} F_{\ell}\Bigr)
%\stackrel{\stackrel{u\to0}{v\to 1}}{\longrightarrow}\ 
\ \xrightarrow{\substack{ u\to0 \\ v\to 1 }}\ \frac{\gamma_K(a)}{2} \log u + O(u^0)
\ee
diverges as a simple logarithm at all orders in $a$. Here $\gamma_K$ is the anomalous dimension of the Konishi operator, the leading non-protected operator in the overlap of the OPEs  $\cO^{(3)}\times \cO^{(3)}$ and $\cO^{(2)}\times \cO^{(2)}$. The numerical coefficient $6$ (planar limit) on the left-hand side has been worked out in \cite{Eden:2011we} for the case $\cG_{2222}$ by examination of the Born-level OPE and comparison of the free two- and three-point functions. Alternatively, knowing the correlator up to two loops  allows us to fix this coefficient by making sure that the log criterion works at two loops.

In the second case $x_1 \to x_4, x_2 \to x_3$ the criterion imposes simple 
logarithmic behaviour on the function 
\be \lb{sdOPElog2}
\log\Bigl( 1 + 4 x_{12}^4 \sum_{\ell \geq 1}a^{\ell} F_{\ell}\Bigr) 
%% \stackrel{\stackrel{v\to0}{u\to 1}}{\longrightarrow}\ 
\ \xrightarrow{\substack{ v\to0 \\ u\to 1 }}\
 \frac{\gamma_{\mathbf{6}}(a)}{2} \log v + O(v^0)\,.
\ee
Here $\gamma_{\mathbf{6}}$ is the anomalous dimension of the scalar operator of dimension 3 and in the vector representation of $SO(6)$, the leading non-protected operator in the OPE $\cO^{(2)}\times \cO^{(3)}$. On the left-hand side we used our knowledge of the two-loop correlator  from the recursion relation \p{310}  to fix the coefficient $4$ (notice that it differs from the 6 in \p{sdOPElog1}). 

Conditions \p{sdOPElog1} and \p{sdOPElog2} are to be implemented as follows (see \cite{Eden:2011we} for the detailed explanation). 
We expand the logarithms up to $a^3$
and obtain restrictions on the linear combinations of two-loop (at level $a^2$)  and three-loop (at level $a^3$)  integrals. 
The integrals beyond one loop in general diverge 
stronger than simple logarithms. In order to weaken the divergences, 
the numerator of the integrand must vanish in the singular regime where an integration point approaches an outer point.\footnote{A similar criterion for the integrand of the four-gluon amplitude was first proposed in \cite{Bourjaily:2011hi}.} For example, for the first OPE we choose $x_5 \to x_1$ or $x_5 \to x_3$,
for the second OPE we choose $x_5 \to x_1$ or $x_5 \to x_2$.

We remark that the conditions following from the OPE \p{sdOPElog1} with dominant twist two are not independent from what the light-cone relation \p{310} has already given us. Only the second OPE \p{sdOPElog2} is really useful for our purposes. It fixes the only remaining coefficient and thus fully determines all the correlators up to three loops.

\subsection{Integral identities}

Once we have fully determined the integrand of a given correlator, we need to turn it into a  set of conformal integrals by substituting the polynomials $P^\ell$ in \p{Gell} and then the integrand in \p{loops}. We find the integrals  listed in \p{eq:14}, appearing in various orientations.  We can profit from a number of identities that these integrals satisfy  \cite{Drummond:2006rz,Eden:2011we},
\begin{align}
&h_{1 2;3 4} = h_{3 4;1 2} \;,\; h_{1 2;3 4} = h_{2 1;3 4} \;,\; h_{1 2;3 4} = h_{1 2;4 3} \, \notag\\
&L_{1 2;3 4} = L_{3 4;1 2} \;,\; L_{1 2;3 4} = L_{2 1;3 4} \;,\; L_{1 2;3 4} = L_{1 2;4 3} \, \notag\\ 
%&T_{1 2;3 4} = L_{1 2;3 4} \, \notag\\
&E_{1 2;3 4} = E_{3 4;1 2} \;,\; E_{1 2;3 4} = E_{2 1;3 4} \;,\; E_{1 2;3 4} = E_{1 2;4 3} \, \notag\\
&H_{1 2;3 4} = H_{3 4;1 2} \;,\; H_{1 2;3 4} = H_{2 1;4 3} \;,\; H_{2 1;3 4} = 1/v H_{1 2;3 4} \, \notag\\ 
&H_{3 1;2 4} = u/v H_{1 3;2 4} \;,\; H_{4 1;2 3} = u H_{1 4;2 3}\,,  \label{id}
\end{align}
to bring the  answer to the form \p{loop3F}.
The final results are listed in tables \ref{tab1}, \ref{tab2}. 

Note that at the level of the {\it integrand} we distinguish the topology of the three-loop ladder integral $L_{1 2;3 4}$ defined in \p{eq:14} from that of the so-called `tennis court' integral,
\begin{align}\label{}
 T_{1 2;3 4}&={x_{34}^2\over (-4\pi^2)^3}\int 
  \frac{ d^4x_5 d^4x_6 d^4x_7 \ x_{17}^2 }{(x_{15}^2 x_{35}^2) (x_{16}^2  x_{46}^2) (x_{37}^2 x_{27}^2  x_{47}^2)
    x_{56}^2 x_{57}^2 x_{67}^2} \,.
\end{align}
The latter does not appear in our result  \p{loop3F} for the {\it integrated} four-point correlation function because of the identity $T_{1 2;3 4} = L_{1 2;3 4} $ proven in \cite{Drummond:2006rz}.

We remark that the same set of integrals was used in \cite{Eden:2011we}  to construct the three-loop correction to the correlator $\cG_{2222}$. The main difference is that in the latter case one has an enhanced permutation symmetry $S_{4+\ell}$ because all the four operators $\cO^{(2)}$ belong to the stress-tensor multiplet \p{33}.  Consequently, the freedom is reduced to a single constant per loop order, up to three loops.

\section{OPE analysis of the integrated four-point correlators}

Now let us turn to an OPE analysis of the four-point correlation functions. 
%We begin by recalling the structure of the correlation functions we study here and in particular the constraints on the form of the perturbative corrections,
%\begin{align}\
%\cG_{k_1 k_2 k_3 k_4} = \cG^0_{k_1 k_2 k_3 k_4} + C_{k_1 k_2 k_3 k_4}\, R(1,2,3,4)\frac{1}{x_{13}^2 x_{24}^2}
% \sum_{\{b_{ij}\}}\left(\prod_{1 \leq i < j \leq 4} (d_{ij})^{b_{ij}} \right) F_{\{b_{ij}\}}(u,v)\,.
%\label{corrk1k2k3k4}
%\end{align} 
%{\bf remove formula!}
We would like to discuss the constraints that the light-cone OPE places on the functions $F_{\{b_{ij}\}}$ appearing in (\ref{211}). Similar analysis has been performed in \cite{Nirschl:2004pa,Dolan:2004iy} and we follow the general discussion therein.\footnote{For more general and recent approaches see~\cite{Bissi:2015qoa,Doobary:2015gia}.} We will see that simple consistency conditions in fact require many of the coefficients in the tables \ref{tab1} and \ref{tab2} presented before  to take precisely the correct values. By performing the OPE analysis we will also be able to present detailed checks of the (derived) tree-level and (conjectured) one-loop formulae for three-point functions of two half-BPS and one long operator presented in \cite{Vieira:2013wya}. Moreover we will be able to check the recently presented three-loop formulae \cite{Eden:2015ija,Basso:2015eqa} for the same three-point functions in the case where the long operator has twist two.

In order to have a uniform discussion of the light-cone OPE for the four-point correlators discussed in this paper, we choose to consider the expansion around the limit $x_{12}^2 x_{34}^2 \rightarrow 0$, or equivalently $u \rightarrow 0$ with $v$ fixed. Then, instead of discussing different expansions of a given correlator, we consider our preferred expansion of the various correlators obtained by permuting the ordering of the external operators.

Without loss of generality we can always pick $k_4$ to be the largest of the weights and order the weights so that $k_1 \leq k_2$. We define the quantity $E$ via
\be
E = \tfrac{1}{2}(k_1 + k_2 + k_3 - k_4)\,.
\ee
We then find it convenient to rewrite the correlation functions as follows (we use the notation $k_{ij}=k_i-k_j$),
\be
\cG_{k_1 k_2 k_3 k_4} = \cG^0_{k_1 k_2 k_3 k_4} + C_{k_1 k_2 k_3 k_4}\, d_{14}^a d_{24}^b d_{12}^c d_{34}^{k_3} u^{\tfrac{1}{2}k_{34}}\mathcal{S}(u,v;\sigma,\tau) \mathcal{H}(u,v;\sigma,\tau)\,. \label{Guvform}
\ee
In equation (\ref{Guvform}) we have introduced the variables
\be
\sigma = \frac{ y_{13}^2 y_{24}^2}{y_{12}^2 y_{34}^2}\,, \quad \tau = \frac{ y_{14}^2 y_{23}^2}{y_{12}^2 y_{34}^2}\,
\ee
while the powers on the propagator factors are given by
\be
a = k_1-E \,, \qquad b= k_2 -E \,, \qquad c = E\,.
\ee
Finally the function $\mathcal{S}$ in (\ref{Guvform}) is a simple polynomial obtained from $R(1,2,3,4)$ and is given by
\begin{align}
\mathcal{S}(u,v;\sigma,\tau) &= R(1,2,3,4)\frac{x_{12}^2 x_{34}^2 x_{14}^2 x_{23}^2}{x_{13}^2 x_{24}^2 y_{12}^4 y_{34}^4} \notag \\
&= v + \sigma^2 u v + \tau^2 u + \sigma v (v-1-u) + \tau (1-u-v) + \sigma \tau (u - 1 -v)\,.
\end{align}
We recall \cite{Eden:2000bk,Nirschl:2004pa} that the presence of the factor $\mathcal{S}$ in (\ref{Guvform}) or $R(1,2,3,4)$ in (\ref{211}) is a reflection of the fact that the quantum loop corrections to the full correlator can only come from intermediate operators in the OPE which belong to long supermultiplets. The free correlator $G^0_{k_1 k_2 k_3 k_4}$ on the other hand receives contributions both from protected operators and long operators,
\be
\cG^0_{k_1 k_2 k_3 k_4} = \cG^{\rm protected}_{k_1 k_2 k_3 k_4} + C_{k_1 k_2 k_3 k_4}\, d_{14}^a d_{24}^b d_{12}^c d_{34}^{k_3} u^{\tfrac{1}{2}k_{34}}\mathcal{S}(u,v;\sigma,\tau) \mathcal{H}^{(0)}(u,v;\sigma,\tau)\,. 
\label{freeG}
\ee

In order to understand the OPE expansion of the correlator $G_{k_1 k_2 k_3 k_4}$, we first expand the function $\mathcal{H}$ into eigenmodes of the $su(4)$ Casimir acting at points 1 and 2 as follows 
\be
\mathcal{H}(u,v;\sigma,\tau) = \sum_{L \leq m \leq n \leq U} A_{nm}(u,v) Y^{(a,b)}_{nm}(\sigma,\tau)\,.
\label{su4exp}
\ee
The channel with labels $n,m$ corresponds to an exchanged supermultiplet with superconformal primary in the representation with $su(4)$ Dynkin labels $[n-m,a+b+2m,n-m]$.

The bounds on the summation region are given by
\be
L= \max(0,E-k_1) \, , \qquad U=\min(k_3,E)-2\,.
\ee
The functions $Y^{(a,b)}_{nm}$ are given in terms of Jacobi polynomials via
\be
Y^{(a,b)}_{nm}(\sigma,\tau) = \frac{P^{(a,b)}_{n+1}(y) P^{(a,b)}_{m \phantom{1}}(\bar{y}) - P^{(a,b)}_{m\phantom{1}}(y) P^{(a,b)}_{n+1}(\bar{y})}{y-\bar{y}}\,,  
\ee
where the variables $y$ and $\bar{y}$ are defined via
\be
\sigma = \tfrac{1}{4}(1+y)(1+\bar{y})\,, \quad \tau = \tfrac{1}{4}(1-y)(1-\bar{y})\,.
\ee

We recall that the Jacobi polynomials are given by a finite hypergeometric series which can be usefully expressed via Rodrigues's formula as follows,
\be
P_n^{(\alpha,\beta)}(z) = \frac{(-1)^n}{2^n n!} (1-z)^{-\alpha} (1+z)^{-\beta} \frac{d^n}{dz^n}\Bigl[ (1-z)^\alpha (1+z)^\beta (1-z^2)^n\Bigr]\,.
\ee
The function $\mathcal{H}^{(0)}$ appearing in (\ref{freeG}) has an expansion directly analogous to (\ref{su4exp}) with coefficients $A_{nm}^{(0)}$ whose precise form will not be important in what follows.

Now the functions $A_{nm}$ appearing in the expansion (\ref{su4exp}) can themselves be expanded in terms of conformal blocks describing the quantum loop corrections to the contributions of conformal primary operators of dimension $\Delta$ and spin $l$.\footnote{In this section we denote the spin by $l$ and the `t Hooft coupling by $\lambda$.} Specifically we have
\be
A_{nm}^{(0)}(u,v) + A_{nm}(u,v) = \sum_{\Delta,l} a_{nm}^{\Delta l} G_\Delta^{(l)}(u,v;k_{21},k_{43})\,.
\label{confblockexp}
\ee
We are interested in the form of the perturbative quantum corrections, so we will need to expand the above sum over conformal blocks order by order in the Yang-Mills coupling. The order $g^0$ term will correspond to the contribution of $A_{nm}^{(0)}$, while all higher orders come from the $A_{nm}$ which are themselves directly obtained from the explicit results of the previous sections.

The conformal blocks are given by \cite{Dolan:2000ut}
\be
G_\Delta^{(l)}(u,v;\delta,\tilde{\delta}) = \frac{u^{\tfrac{1}{2}(\Delta-l)}}{x-\bar x}\Bigl( x \bigl(-\tfrac{1}{2}x\bigr)^l f^{\delta,\tilde{\delta}}_{\Delta+l}(x) f^{\delta,\tilde{\delta}}_{\Delta-l-2}(\bar{x}) - \bar{x} \bigl(-\tfrac{1}{2}\bar{x}\bigr)^l f^{\delta,\tilde{\delta}}_{\Delta+l}(\bar{x}) f^{\delta,\tilde{\delta}}_{\Delta-l-2}(x)\Bigr)\,,
\label{blocks}
\ee
with
\be
f^{\delta,\tilde{\delta}}_\rho(z) = {}_2F_{1}\Bigl(\tfrac{1}{2}(\rho+\delta),\tfrac{1}{2}(\rho-\tilde{\delta});\rho;z\Bigr)\,.
\ee
In (\ref{blocks}) we employ the variables
\be
u = x \bar{x}\,, \qquad v= (1-x)(1-\bar{x})\,.
\ee

For our purposes here it will be sufficient to consider only the leading power in the expansion for small $u$ for each function $A_{nm}(u,v)$, keeping any powers of $\log u$. This corresponds to keeping the leading twist\footnote{The twist $T$ of an operator is the dimension minus the spin $T=\Delta-l$.} contribution to each distinct $su(4)$ channel in the expansion of the correlation functions. The limit may be achieved by taking $\bar{x}\rightarrow 0$ with $x$ fixed. In this case we may drop any power suppressed terms from the conformal blocks,
\be
G_\Delta^{(l)}(u,v;\delta,\tilde{\delta}) = u^{\tfrac{1}{2}(\Delta-l)}(-\tfrac{1}{2}x)^l f^{\delta,\tilde{\delta}}_{\Delta+l}(x) + O(\bar{x})\,.
\ee

Our task is now to match the explicit expressions for the leading powers in the $\bar{x}$ expansions of $A_{nm}$, obtained from the limits of the correlation functions $G_{k_1,k_2,k_3,k_4}$, with the perturbative expansion of the sum over conformal blocks given in (\ref{confblockexp}). We write the scaling dimension $\Delta$ of a superconformal primary as $\Delta = \Delta_0 + \gamma(\lambda)$, where $\gamma$ is the anomalous dimension. The free scaling dimension is not a good label for different operators since in the free theory many operators of a given spin may have the same $\Delta_0$. We therefore label operators of a given spin $l$ which have degenerate free scaling dimensions with an extra index $I$. We find
\be
A_{nm}^{(0)}(u,v) + A_{nm}(u,v) = u^p \sum_{I,l} a_{nm,I,l} u^{\eta_{I,l}} \bigl(-\tfrac{1}{2}x\bigr)^l f_{2p+2l+2\eta_{I,l}}^{k_{21},k_{43}}(x) + O(u^{p+1})
\label{leadingTexp}
\ee
where $\eta_{I,l} = \tfrac{1}{2}\gamma_{I,l}$ and
\be
p= \tfrac{1}{2}\bigl(\max(k_{21},k_{43}) + 2 + 2(n - \max(0,E-k_1))\bigr)
\ee
is half the free twist of the leading twist operators in a given $su(4)$ channel with labels $m,n$. 
The coupling dependence in (\ref{leadingTexp}) is in the quantities $A_{mn}(u,v)$ on the LHS and $a_{nm,I,l}$ and $\gamma_{I,l}$ on the RHS.  They admit perturbative expansions of the form
\be
A_{nm}(u,v) = \sum_{r=1}^{\infty} \lambda^r A_{nm}^{(r)}(u,v)\,, \qquad \eta_{I,l}(\lambda) = \sum_{r=1}^\infty \lambda^r \eta^{(r)}_{I,l}\,, \qquad a_{nm,I,l}(\lambda) = \sum_{r=0}^\infty \lambda^r a_{nm,I,l}^{(r)}\,.
\label{pertexps}
\ee
The functions $A_{nm}^{(r)}(u,v)$ exhibit logarithmic corrections in their expansions for small $u$ of the form
\be
A_{nm}^{(r)}(u,v) = u^p \sum_{s=0}^r (\log u)^s g^{(r)}_{nm,s}(x) + O(u^{p+1})\,.
\label{Anmexp}
\ee

Now we may expand both sides of eq.~(\ref{leadingTexp}) in the coupling. This leads us to expressions of the following general form for the functions $g_{nm,s}^{(r)}$,
\be
g_{nm,s}^{(r)}(x) = \sum_{I,l} \bigl(-\tfrac{1}{2} x \bigr)^l \bigl[\mathcal{O}_{nm,I,l,s}^{(r)} f_{2t+2l}^{k_{21},k_{43}}(x)\bigr]_{t=p}\,,
\label{gfromO}
\ee
where $\mathcal{O}_{nmI,l,s}^{(r)}$ is in general a differential operator (in $t$) acting on the function $f$.
To simplify the notation a little we suppress the indices $n$ and $m$  (in other words we write $\mathcal{O}_{nm,I,l,s}^{(r)}(x) \equiv \mathcal{O}^{(r)}_{I,l,s}(x)$ and $a_{nm,I,l}^{(r)} \equiv a_{I,l}^{(r)}$). At leading order we simply have a multiplicative operator,
\begin{align}
\mathcal{O}_{I,l,0}^{(0)} &= a_{I,l}^{(0)}\,.
\label{O00}
\end{align}
At order $\lambda$ we have
\begin{align}
\mathcal{O}_{I,l,1}^{(1)}&= a_{I,l}^{(0)}  \eta^{(1)}_{I,l}\,, \label{O22}\nt
\mathcal{O}_{I,l,0}^{(1)}&=  a_{I,l}^{(1)}  + a_{I,l}^{(0)}  \eta^{(1)}_{I,l} \partial_t \,.
\end{align}
At order $\lambda^2$ we find
\begin{align}
\mathcal{O}_{I,l,2}^{(2)} &= a_{I,l}^{(0)}  \tfrac{1}{2} (\eta^{(1)}_{I,l})^2 \,,\nt
\mathcal{O}_{I,l,1}^{(2)} &= a_{I,l}^{(1)} \eta^{(1)}_{I,l} + a_{I,l}^{(0)} \eta^{(2)}_{I,l} + a_{I,l}^{(0)} (\eta^{(1)}_{I,l})^2 \partial_t \,, \nt
\mathcal{O}_{I,l,0}^{(2)}&=  a_{I,l}^{(2)} + a_{I,l}^{(1)} \eta^{(1)}_{I,l} \partial_t + a_{I,l}^{(0)}\bigl[\eta^{(2)}_{I,l} \partial_t  + \tfrac{1}{2} (\eta^{(1)}_{I,l})^2 \partial_t^2\bigr] \,.
\end{align}
Finally, at order $\lambda^3$ we have
\begin{align}
\mathcal{O}_{I,l,3}^{(3)} &= a_{I,l}^{(0)}  \tfrac{1}{6} (\eta^{(1)}_{I,l})^3 \,,\nt
\mathcal{O}_{I,l,2}^{(3)} &= a_{I,l}^{(1)} \tfrac{1}{2} (\eta^{(1)}_{I,l})^2 + a_{I,l}^{(0)} \bigl[\eta^{(2)}_{I,l} \eta^{(1)}_{I,l} + \tfrac{1}{2} (\eta^{(1)}_{I,l})^3 \partial_t \bigr]\,, \nt
\mathcal{O}_{I,l,1}^{(3)} &= a_{I,l}^{(2)} \eta^{(1)}_{I,l}+ a_{I,l}^{(1)} \bigl[ \eta^{(2)}_{I,l}\ + (\eta^{(1)}_{I,l})^2 \partial_t \bigr] + a_{I,l}^{(0)} \bigl[\eta^{(3)}_{I,l} + 2\eta^{(2)}_{I,l}\eta^{(1)}_{I,l} \partial_t + \tfrac{1}{2} \bigl(\eta^{(1)}_{I,l}\bigr)^2 \partial_t^2 \bigr]\,, \nt
\mathcal{O}_{I,l,0}^{(3)} &= a_{I,l}^{(3)} + a_{I,l}^{(2)} \eta_{I,l}^{(1)} \partial_t  + a_{I,l}^{(1)}\bigl[ \eta_{I,l}^{(2)} \partial_t  + \tfrac{1}{2} (\eta_{I,l}^{(1)})^2 \partial_t^2 \bigr] + a_{I,l}^{(0)} \bigl[ \eta_{I,l}^{(3)} \partial_t + \eta_{I,l}^{(2)} \eta_{I,l}^{(1)} \partial_t^2 + \tfrac{1}{6} ( \eta_{I,l}^{(1)} )^3 \partial_t^3\bigr]\!.
\label{O30}
\end{align}

These leading-twist OPE expansions are to be compared to the explicit leading-twist results for the perturbative expansion of the four-point correlation functions. As we have seen, these correlation functions are expressed purely in terms of one-, two- and three-loop ladder integrals, as well as, at three-loops, the Easy and Hard integrals.

Let us now compare the expressions (\ref{freeandloop}) and (\ref{211}) with the form (\ref{Guvform}) for the four-point correlator. We find
\be
\mathcal{H}(u,v;\sigma,\tau) = u^{\tfrac{1}{2}k_{43}+1} v^{-1}\sum_{\{ b_{ij} \}}\sigma^{b_{13}} \tau^{b_{23}} u^{b_{13} + b_{23}} v^{-b_{23}} F_{\{b_{ij}\}}(u,v)\,.
\label{HtoFpoly}
\ee
For fixed $k_1,k_2,k_3,k_4$ we may regard $b_{13}$ and $b_{23}$ as free variables while the other $b_{ij}$ are related to them via
\be
b_{14} = b_{23} + a\,, \quad b_{24} = b_{13} + b\,, \quad b_{12} = E -2-b_{13}-b_{23}\,, \quad b_{34} = k_3 -2 - b_{13} - b_{23}\,.
\ee
The bounds on $b_{13}$ and $b_{23}$ follow from the fact that all the $b_{ij}$ are non-negative. Defining $n=b_{13}+b_{23}$ we may rewrite (\ref{HtoFpoly}) as
\be
\mathcal{H}(u,v;\sigma,\tau) = u^{\tfrac{1}{2}k_{43}+1} v^{-1}\sum_{L\leq b_{23} \leq n \leq U}\sigma^{n-b_{23}} \tau^{b_{23}} u^{n} v^{-b_{23}} 
%F_{\{E-2-n,n-b_{23},b_{23}+a,b_{23},n-b_{23}+b,k_3-2-n\}}
F_{n,b_{23}}(u,v)\,.
\label{Hexpfreebij}
\ee
with 
\be
L = \max(0,E-k_1)\,, \qquad U= \min(k_3,E)-2\,.
\ee

We recall that each $F_{\{b_{ij}\}}(u,v)$ is given by a perturbative expansion given in (\ref{loop3F}).
%{\bf suppress and refer}
%\be
%F_{\{b_{ij}\}}(u,v) = \sum_{r=1}^{\infty} \lambda^r F_{\{b_{ij}\}}^{(r)} (u,v)
%\ee
%and up to three loops we have for each $F$ (dropping the labels $\{b_{ij}\}$),
%\begin{eqnarray}
%\frac{F^{(1)}(u,v)}{x_{13}^2 x_{24}^2} & =&  g_{1234}\, , \notag
%\\[2mm] \notag
%\frac{F^{(2)}(u,v)}{x_{13}^2 x_{24}^2} & =& c_h^1 h_{1 2;3 4} + c_h^2 h_{2 3;1 4}    +c_h^3 h_{1 3;2 4}  
%  + 
% \frac12 \lr{c_{gg}^1 x_{12}^2x_{34}^2+ c_{gg}^2 x_{13}^2 x_{24}^2+c_{gg}^2  x_{14}^2x_{23}^2}  [g_{1 2 3 4}]^2\, , \notag
%\\[2mm] \notag
%\frac{F^{(3)}(u,v)}{x_{13}^2 x_{24}^2} & = & c_{gh}^1 x^2_{12} x^2_{34} \, (g\times h)_{1 2;3 4} \, + \,
%c_{gh}^2 x^2_{13} x^2_{24} \, (g\times h)_{1 3;2 4} \, + \, c_{gh}^3  x^2_{14} x^2_{23} \, (g\times h)_{1 4;2 3} 
%\nonumber \\[2mm]
%& + & c_L^1 L_{1 2;3 4} \, + \, c_L^2 L_{1 3;2 4} \, + \, c_L^3 L_{1 4;2 3}   \, + \,  c_E^1 E_{1 2;3 4} \, + \, c_E^2 E_{1 3;2 4} \, + \, c_E^3 E_{1 4;2 3}  
% \nonumber \\[2mm] \label{loop3Fagain} 
%& + &  \frac12 (c_H^1 + c_H^2 1/v) H_{1 2;3 4} \, + \,  \frac12(c_H^3 +c_H^4 u/v) H_{1 3;2 4} \, + \,\frac12(c_H^5 +c_H^6 u) H_{1 4;2 3}  \, .
%\end{eqnarray}
We will now consider a few examples of the above expansions on the correlators under consideration.

\subsection{Equal weights $(kkkk)$}

In the first instance we specialise to the case where all weights are equal, $k_1=k_2=k_3=k_4=k$, which was studied extensively in \cite{Dolan:2004iy}. Note that having all $k_i$ equal implies 
\be
b_{14}=b_{23}\,, \quad b_{24}=b_{13}\,, \quad b_{12} = b_{34}\,, \quad b_{34} = k-2 - b_{13} - b_{23}\,.
\ee
In this case the correlator simplifies to
\be
\cG_{k k k k} = \cG^0_{k k k k} + C_{k k k k}\, (d_{12})^k (d_{34})^{k} \mathcal{S}(u,v;\sigma,\tau) \mathcal{H}(u,v;\sigma,\tau)\,. \label{Guvformequalk}\,
\ee
The expansion in (\ref{su4exp}) above reduces to an expansion in terms of Legendre polynomials,
\be
\mathcal{H}(u,v;\sigma,\tau) = \sum_{0 \leq m \leq n \leq k-2} A_{nm}(u,v) Y^{(0,0)}_{nm}(\sigma,\tau)\,.
\label{su4equalk}
\ee
The leading twist of the exchanged operators in the OPE in a given $su(4)$ channel is $2+2n$ (hence $p=1+n$). The expansion (\ref{Hexpfreebij}) takes the form
\be
\mathcal{H}(u,v;\sigma,\tau) = (u/v)\sum_{0\leq b_{23} \leq n \leq k-2}\sigma^{n-b_{23}} \tau^{b_{23}} u^{n} v^{-b_{23}} F_{n,b_{23}}(u,v)\,.
\label{HtoFpolyequalk}
\ee

Now we consider the first few values of $k$.

\subsubsection*{2222}

The case $(2222)$ corresponds to the well-studied case of four stress-tensor multiplets. The one-loop and two-loop results were derived in \cite{Chalmers:1998xr,Eden:1998hh,Eden:1999kh,Eden:2000mv,Bianchi:2000hn} and the OPE analysis was performed in \cite{Dolan:2004iy}. The form of the correlator in terms of three-loop integrals was obtained in \cite{Eden:2011we} and the asymptotics necessary for the leading twist OPE analysis were derived in \cite{Eden:2012rr}. From these results the full two-variable kinematical dependence of the integrals was reconstructed in \cite{Drummond:2013nda}. From the expansion (\ref{su4equalk}) we have only a single $su(4)$ channel whose leading twist is 2,
\be
\mathcal{H}(u,v;\sigma,\tau) = A_{00}(u,v) = \frac{u}{v} F_{\{0,0,0,0,0,0\}}(u,v)\,,
\ee
where the second equality comes from inserting the only allowed values of the $\{b_{ij}\} = \{0,0,0,0,0,0\}$ into (\ref{HtoFpoly}).

Expanding the above function for small $u$ and keeping only the leading power in $u$ we have
\be
A_{00}^{(r)}(u,v) = u \sum_{s=0}^r (\log u)^s g^{(r)}_{00,s}(x) + O(u^{2})\,.
\label{A00exp2222}
\ee
The explicit forms of the $g^{(r)}_{00,s}(x)$ may then be read off up to three loops from the explicit expression for the function $F_{\{0,0,0,0,0,0\}}$ in terms of the known integrals. The case of leading twist equal to two is special in that there is only a single operator for each spin $l$. The sum over operators labelled by $I$ in the relations (\ref{gfromO}) - (\ref{O30}) may thus be dropped. From the knowledge of the free theory three-point functions for two weight-two protected operators and one twist-two long operator the anomalous dimensions and normalisations may be constructed up to three loops. These have already been explicitly worked out in \cite{Dolan:2004iy,Eden:2012rr}. Following \cite{Eden:2012rr} we reorganise the final expansion in (\ref{pertexps}) to express it as
\begin{align}
a_{00,l}(\lambda) = 2^{l-1} & \biggl[\frac{\Gamma(l+1+\eta_l(\lambda))^2}{\Gamma(2l+1+2\eta_l(\lambda))}\biggr]_{\rm rational} \notag\\ 
& \times \biggl[1 + \lambda c^{(1)}_{2,l} + \lambda^2\bigl(c^{(2)}_{4,l}+ \zeta_3 c^{(2)}_{1,l}\bigr) + \lambda^3\bigl(c^{(3)}_{6,l} + \zeta_3 c^{(3)}_{3,l} + \zeta_5 c^{(3)}_{1,l}\bigr)+ \ldots\biggr]\,.
\label{Burkform}
\end{align}
The subscript `rational' on the first factor in brackets denotes the fact that one should discard all zeta-value contributions arising from expanding the Gamma functions in the $\eta_l(\lambda)$.
Here let us note only the explicit expression of the coefficients $c$ up to two loops,
\begin{align}
c^{(1)}_{2,l} &= -h_2\,, \label{Bc1loop} \\
c^{(2)}_{1,l} &= 3h_1\,, \nt
c^{(2)}_{4,l} &= \tfrac{5}{2} h_{-4} + h_{-2}^2 + 2 h_{-3} h_{1} + 
 h_{-2} h_{2} + h_{2}^2 + 2 h_{1} h_{3} + \tfrac{5}{2} h_{4} - 2 h_{-3, 1} - h_{-2, 2} - 2 h_{1, 3}\,. \notag
\end{align}
In the above equations we use the notation $h$ to denote a harmonic sum with argument $l$. We refer the reader to \cite{Eden:2012rr} for the explicit three-loop formulae.

\subsubsection*{3333}

In the case (3333) we have three $su(4)$ channels,
\begin{align}
\mathcal{H}(u,v;\sigma,\tau) &= \sum_{0\leq m\leq n\leq 1} A_{nm}(u,v) Y_{nm}(\sigma,\tau) \nt
& = \frac{u}{v}\Bigl(F_{\{1,0,0,0,0,1\}} + \sigma u F_{\{0,1,0,0,1,0\}} + \tau \frac{u}{v} F_{\{0,0,1,1,0,0\}}\Bigr)\,.
\end{align}
The relations between the different expansions are given by
\begin{align}
A_{00} &= \frac{u}{v} \Bigl(F_{\{1,0,0,0,0,1\}} + \frac{u}{6}\bigl(F_{\{0,1,0,0,1,0\}} + \frac{1}{v} F_{\{0,0,1,1,0,0\}}\bigr)\Bigr)\,, \label{3333A00}\nt
A_{10} &= \frac{1}{6}\frac{u^2}{v}\Bigl(F_{\{0,1,0,0,1,0\}} -\frac{1}{v} F_{\{0,0,1,1,0,0\}}\Bigr)\,, \nt
A_{11} &= \frac{1}{6}\frac{u^2}{v}\Bigl(F_{\{0,1,0,0,1,0\}} +\frac{1}{v} F_{\{0,0,1,1,0,0\}}\Bigr)\,.
\end{align}

The first channel $A_{00}$ has leading twist two. To keep only the leading twist contribution we keep only the first term in (\ref{3333A00}). From the explicit expression for $F_{\{1,0,0,0,0,1\}}$ up to three loops we may then read off the normalisations $a_{00,l}(\lambda)$, corresponding to the squares of the three-point functions for two weight-three protected operators and one twist-two long operator up to three loops. We again write the perturbative expansion in the form (\ref{Burkform}) and we note that the anomalous dimensions $\eta_l$ are identical to the (2222) case because the exchanged operators are the same.

At one loop the expression for $c^{(1)}_{2,l}$ is identical to the weight 2 case given in (\ref{Bc1loop}). This follows from the fact that there is only a single one-loop integral, namely $g_{1234}$. At two loops we find 
\begin{align}
c^{(2)}_{1,l} =& 0\,, \nt
c^{(2)}_{4,l} =& \tfrac{5}{2} h_{-4} + \tfrac{1}{2} h_{-2}^2 + 3 h_{-3} h_{1} + h_{-2} h_{1}^2 + h_{-2}h_{2} + h_{2}^2 +  2 h_{1} h_{3} + 2 h_{4} - h_{-3, 1} - h_{1, 3} \notag \\
& - 2 h_{-2, 1, 1} - 2 h_{1, -2, 1}\,.
\end{align}
At three loops we have not found an expression valid for arbitrary $l$ analogous to the ones found in \cite{Eden:2012rr} for the weight 2 case. We expect that a relatively simple formula in terms of harmonic sums, similar to the one for the (2222) case, will reproduce the results we have for each spin. However we may still check our results against the expressions found in \cite{Eden:2015ija,Basso:2015eqa} for the leading spins and we find perfect agreement with the predictions coming from the integrability approach of \cite{Basso:2015zoa}. In particular we reproduce\footnote{We need to rescale their coupling $g^2$ by a factor of 4, i.e. $g^2|_{\rm there} = 4 \lambda|_{\rm here}$ and rescale their coefficients by a global factor of $1/4$ to match our conventions.} the table at the top of page 9 of \cite{Eden:2015ija}. The parameter $\eta$ in that table should be set to $\tfrac{1}{2}$.

Let us now turn to the channel $A_{11}$ which has leading twist four. In this case the supermultiplets of twist four which are exchanged in the OPE contain super descendants which are pure scalar operators in the $sl_2$ sector studied in \cite{Vieira:2013wya}. Thus, from this channel we can compare to the predictions made in \cite{Vieira:2013wya} for the small $x$ expansions of the functions $g^{(r)}_{11,s}(x)$ defined in (\ref{Anmexp}). We pull out some simple prefactors to aid comparison:
\begin{align}
g_{11,1}^{(1)}(x) &= -\frac{1}{3 . 2 . 4\, x^4}\Bigl(8x^4 + 16 x^5 + \frac{74}{3} x^6 + 34 x^7 + \frac{659}{15} x^8 + \ldots\Bigr)\,,\nt
g_{11,0}^{(1)}(x) &= -\frac{1}{3 . 4\, x^4}\Bigl(-8x^4 - 14 x^5 - \frac{179}{9} x^6 - \frac{155}{6} x^7 - \frac{28663}{900} x^8 + \ldots\Bigr)\,,\nt
g_{11,2}^{(2)}(x) &= \frac{1}{3 . 2^2 . 4\, x^4}\Bigl(24x^4 + 48 x^5 + \frac{712}{9} x^6 + \frac{352}{3} x^7 + \frac{72953}{450} x^8 + \ldots\Bigr)\,,\nt
g_{11,1}^{(2)}(x) &= \frac{1}{3 . 2 . 4\, x^4}\Bigl(-64x^4 - 116 x^5 - \frac{1619}{9} x^6 - \frac{2287}{9} x^7 - \frac{9094423}{27000} x^8 + \ldots\Bigr)\,,\nt
%g_{11,0}^{(2)}(x) &= \frac{1}{3 . 4\, x^4}\Bigl(8(7+6\zeta_3)x^4 +96(1+\zeta_3) x^5 + \Bigl(\frac{7805}{54} +148\zeta_3\Bigr) x^6 \notag\nt
%&\qquad \qquad  +\Bigl(\frac{7193}{36} + 204 \zeta_3\Bigr) x^7 +\Bigl(\frac{421996591}{1620000} + \frac{1318}{5} \zeta_3\Bigr) x^8 + \ldots\Bigr)\,,\nt
g_{11,3}^{(3)}(x) &= -\frac{1}{3 . 2^3 . 4\, x^4}\Bigl( \frac{160}{3}x^4 + \frac{320}{3} x^5 + \frac{15688}{81} x^6 + \frac{8488}{27} x^7 + \frac{4732363}{10125} x^8  + \ldots\Bigr)\,,\nt
g_{11,2}^{(3)}(x) &= -\frac{1}{3 . 2^2 . 4\, x^4}\Bigl(-256x^4 - 472 x^5 - \frac{65954}{81} x^6 - \frac{34132}{27} x^7 - \frac{122032589}{67500} x^8   + \ldots\Bigr)\,.
\label{g32for3333}
\end{align}
The terms in the parentheses reproduce precisely\footnote{Apart from the third term in the final line in (\ref{g32for3333}) which appears to be a simple typo in \cite{Vieira:2013wya}.} the predicted expansion coming from the conjectured form of the one-loop twist-four structure constants in \cite{Vieira:2013wya}.

Finally we note that we also have explicit data for the channel $A_{10}$ which is also of leading twist four. 
%Unlike for the channel $A_{11}$ there does not appear to be any explicit calculation beyond the two-loop analysis of \cite{} to which we can compare.

\subsubsection*{4444}
In the case (4444) we have six $su(4)$ channels,
\begin{align}
\mathcal{H}(u,v;\sigma,\tau) &= \sum_{0\leq m\leq n\leq 1} A_{nm}(u,v) Y_{nm}(\sigma,\tau) \nt
& = \frac{u}{v}\Bigl(F_{\{2,0,0,0,0,2\}} + \sigma u F_{\{1,1,0,0,1,1\}} + \tau \frac{u}{v} F_{\{1,0,1,1,0,1\}} \notag \\
&\quad \qquad +  \sigma^2 u^2 F_{\{0,2,0,0,2,0\}} + \sigma \tau \frac{u^2}{v} F_{\{0,1,1,1,1,0\}} + \tau^2 \frac{u^2}{v^2} F_{\{0,0,2,2,0,0\}}\Bigr)\,.
\end{align}
The distinct $su(4)$ channels are given by
\begin{align}
A_{00} &= \frac{u}{60 v^3}(3 u^2 F_{\{0, 0, 2, 2, 0, 0\}} +
   u^2 v F_{\{0, 1, 1, 1, 1, 0\}} + 3 u^2 v^2 F_{\{0, 2, 0, 0, 2, 0\}} +
   10 u v F_{\{1, 0, 1, 1, 0, 1\}} \label{A00for4444} \notag \\
   &\qquad \quad + 10 u v^2 F_{\{1, 1, 0, 0, 1, 1\}} +
   60 v^2 F_{\{2, 0, 0, 0, 0, 2\}}) \,, \notag \\
A_{10}  &=-\frac{u^2}{12 v^3}(u F_{\{0, 0, 2, 2, 0, 0\}} - u v^2 F_{\{0, 2, 0, 0, 2, 0\}} + 
 2 v F_{\{1, 0, 1, 1, 0, 1\}} - 2 v^2 F_{\{1, 1, 0, 0, 1, 1\}})\,, \notag\\
A_{11} &= \frac{u^2}{60v^3}(6 u F_{\{0, 0, 2, 2, 0, 0\}} + u v F_{\{0, 1, 1, 1, 1, 0\}} + 
 6 u v^2 F_{\{0, 2, 0, 0, 2, 0\}} + 10  v F_{\{1, 0, 1, 1, 0, 1\}} +
 10  v^2 F_{\{1, 1, 0, 0, 1, 1\}})\,, \notag \\
 A_{20} &= \frac{u^3}{60v^3}(2 F_{\{0, 0, 2, 2, 0, 0\}} - v F_{\{0, 1, 1, 1, 1, 0\}} + 
 2 v^2 F_{\{0, 2, 0, 0, 2, 0\}})\,,\notag \\
 A_{21} &= \frac{u^3}{20v^3}(-F_{\{0, 0, 2, 2, 0, 0\}} + v^2 F_{\{0, 2, 0, 0, 2, 0\}})\,, \notag \\
 A_{22} &= \frac{u^3}{60 v^3}(F_{\{0, 0, 2, 2, 0, 0\}} + v F_{\{0, 1, 1, 1, 1, 0\}} + 
 v^2 F_{\{0, 2, 0, 0, 2, 0\}})\,.
\end{align}

The channel $A_{00}$ again has leading twist two. If we keep only the leading twist contributions, the only term which contributes is the last one on the RHS of the first equation in (\ref{A00for4444}). Up to two loops the function $F_{\{2,0,0,0,0,2\}}$ is identical to $F_{\{1,0,0,0,0,1\}}$. Thus the results for $a_{0,0,l}(\lambda)$ are identical up to two loops with the (3333) case. At three loops we may again compare with the results of \cite{Eden:2015ija,Basso:2015eqa}. Again the relevant data is given in the table on page 9 of \cite{Eden:2015ija}. This time we must set the variable $\eta$ in that table to zero. We can see that the variable $\eta$ in table 9 of \cite{Eden:2015ija} is varying exactly in accordance with the coefficient of the single integral $L_{12;34}$, which is finite in the OPE limit $x_{12}^2 \rightarrow 0$.

We may also examine the channel $A_{22}$ where the leading twist is six. Again we can compare to the data presented in \cite{Vieira:2013wya} in table 4 in Appendix C. Expanding $A_{22}$ in the leading twist sector we find perfect agreement with the coefficients for leading spins detailed in that table.

\subsubsection*{Higher k and wrapping corrections}

We have previously noted that there is a uniformity of the functions $F_{\{b_{ij}\}}$ in that there is only one contributing integral, $g_{1234}$ at one loop while functions with any given $b_{ij}\geq 1$ are identified at two loops and those with any given $b_{ij} \geq 2$ are identified at three loops. In particular the leading twist-two channel of the correlator $(kkkk)$ is of the form
\be
A_{00}(u,v) = \frac{u}{v} F_{\{k-2,0,0,0,0,k-2\}} + O(u^2)\,.
\ee
We see that the uniform behaviour of the functions $F_{\{k-2,0,0,0,0,k-2\}}$ for all $k$ at one loop, $k\geq 3$ at two loops\footnote{This uniformity (called `degeneracy') was first remarked in \cite{Arutyunov:2003ae,Arutyunov:2003ad}.} and $k \geq 4$ at three loops implies that the normalisations $a_{00,l}(\lambda)$ will also exhibit such a uniform behaviour. Thus the normalisations $a_{00,l}(\lambda)$ in the cases (5555), (6666) etc. will all be the same as those of (4444) up to three loops. This uniformity in $k$ is precisely what is expected from the nature of possible wrapping contributions to the three-point functions of two protected operators and one twist-two long operator in the approach of \cite{Basso:2015zoa}.

\subsection{Weights $(kkk'k')$ and an averaging rule}
\label{avgrule}

Let us now consider correlation functions with weights $(kkk'k')$ for $k$ and $k'$ not equal. Note that by our assumptions on the ordering of the weights we have $k'>k$. The $b_{ij}$ are related as follows,
\be
b_{14}=b_{23}\,, \quad b_{24}=b_{13}\,, \quad b_{12} = k-2 - b_{13} - b_{23}\,, \quad b_{34} = k'-2 - b_{13} - b_{23}\,.
\ee
The correlator simplifies to
\be
\cG_{k k k' k'} = \cG^0_{k k k' k'} + C_{k k k' k'}\, (d_{12})^k (d_{34})^{k'} \mathcal{S}(u,v;\sigma,\tau) \mathcal{H}(u,v;\sigma,\tau)\,. \label{Guvformequalk}\,
\ee
The expansion in (\ref{su4exp}) above reduces to an expansion in terms of Legendre polynomials just as in the case of equal weights (\ref{su4equalk}).
The leading twist of the exchanged operators in the OPE in a given $su(4)$ channel is again $2+2n$ (hence $p=1+n$). 

Since the normalisations $a_{nm,I,l}$ are products of three-point functions we expect them to obey
\be
a_{nm,I,l}^{(kkk'k')} = \sqrt{ a_{nm,I,l}^{(kkkk)} a_{nm,I,l}^{(k'k'k'k')} }\,.
\label{sqrts}
\ee
Let us now focus on the case of the twist-two operators (where we can drop the additional index $I$ as there is no operator mixing). We know in this case that, in the free theory and at one loop, the normalisations $a_{00,l}^{(kkkk)}$ are in fact independent of $k$. We write this explicitly as follows,
\be
a_{00,l}^{(kkkk)} = a_{00,l}^{(0)}(1 + \lambda b_{l}^{(1)} + \lambda^2 b_{l,k}^{(2)} + \lambda^3 b_{l,k}^{(3)}+\ldots)\,.
\ee

This means that if we perturbatively expand (\ref{sqrts}) we find
\be
a_{00,l}^{(kkk'k')} = a_{00,l}^{(0)}\Bigl(1+ \lambda b_l^{(1)} + \lambda^2 \tfrac{1}{2} \bigl(b_{l,k}^{(2)} + b_{l,k'}^{(2)} \bigr) + \lambda^3 \tfrac{1}{2} \bigl(b_{l,k}^{(3)} + b_{l,k}^{(3)} \bigr)\Bigr) + \ldots
\ee
In other words, to three loops, we find that the normalisations $a_{00,l}(\lambda)$ for the case $(kkk'k')$ are the average of those for the cases $(kkkk)$ and $(k'k'k'k')$. In particular this means that the leading twist contributions to $A_{00}^{(r)}$ obey
\be
A_{00}^{(r),(kkk'k')} = \frac{1}{2}\Bigl(A_{00}^{(r),(kkkk)}+A_{00}^{(r),(k'k'k'k')}\Bigr) \,, \qquad r=1,2,3\,.
\label{A00avgrule}
\ee
Since the leading twist-two contributions to the $A_{00}$ channels for the correlators $(kkk'k')$ are all given by the functions $(u/v)F_{\{k-2,0,0,0,0,k'-2\}}$ we conclude that in the limit of small $u$ we have
\be
F_{\{k-2,0,0,0,0,k'-2\}} = \frac{1}{2}\bigl(F_{\{k-2,0,0,0,0,k-2\}} +F_{\{k'-2,0,0,0,0,k'-2\}}\bigr) + O(u)
\label{twist2Favgrule}
\ee
up to three loops.

Of the integral functions appearing in the expansion (\ref{loop3F}) of correlators up to three loops, the functions with coefficients $c^1_{gg}$ at two loops and $c^1_{gh}$, $c^4_H$, and $c^6_H$ at three loops are all power-suppressed in the OPE limit $x_{12}^2 \rightarrow 0$. The remaining functions can all contribute in the limit to the leading twist expansion of any given $su(4)$ channel of a correlation function. Moreover, the remaining functions are all linearly independent in the limit, as can be verified from their explicit expressions \cite{Drummond:2013nda}. This means that, for those functions which are not power suppressed, one may apply the averaging rule (\ref{twist2Favgrule}) directly at the level of the individual integral coefficients.

In fact the only reason we have restricted ourselves to three loops in equation (\ref{A00avgrule}) is the dependence on $k$ in $a_{00,l}^{(kkkk)}$ at two loops. However, as we have seen from the analysis of the preceding section, this dependence is very mild indeed. In fact $a_{00,l}^{(kkkk)}$ is again independent of $k$ at two loops as long as $k\geq 3$ and even at three loops as long as $k \geq 4$. This means that we can actually extend our average rule beyond three loops in these cases. This may prove a useful tool in higher loop explorations of the correlation functions of $\tfrac{1}{2}$-BPS operators.

\subsection{General consistency checks}

In fact we may use the OPE of the four-point correlation functions to cross-check many of our results from tables \ref{tab1} and \ref{tab2}. If we allow ourselves to make an ansatz in terms of the known ladder integrals and, at three loops, the Easy and Hard integrals, we find that many of their coefficients are fixed by consistency of the OPE expansion. The reason is that at $\ell$ loops the operators $\mathcal{O}_{nm,I,l,s}^{(\ell)}$ appearing in (\ref{gfromO}) are explicitly known for $s>1$ if all the lower loop data $a_{nm,I,l}^{(\ell')}$ and $\eta_{I,l}^{(\ell')}$ are known for $\ell'<\ell$. Moreover if $\eta_{I,l}^{(\ell)}$ is known then one also knows $\mathcal{O}_{nm,I,l,1}^{(\ell)}$. In the case of the exchange of twist-two operators these data may be read off from the lower loop correlators themselves, while the anomalous dimensions to the relevant order are well known. We find that matching such constraints from the form of the OPE fixes many of the constants in the ansatz. As an example, for the correlator (3333), such OPE consistency checks fix all but one coefficient, namely the coefficient $c_{L}^1$ in an ansatz for $F_{\{1,0,0,0,0,1\}}$. This final coefficient can then be determined from the averaging rule we described above in Sect. \ref{avgrule}, assuming the correlator (2233) is known. The fact that the coefficients obtained in tables \ref{tab1} and  \ref{tab2} are all consistent with what is essentially an independent check based on the forms of the actual integrated functions further increases our confidence that the values are correct.

\section{Conclusions} 

In this paper we have shown how to construct the four-point correlation functions of half-BPS operators of arbitrary weights, in the planar limit and up to three loops. Our construction uses only elementary properties of the {\it integrand} of the loop corrections, viewed as a rational correlator at Born level. Knowing its symmetries and singularity structure, we are able to write down a relatively concise ansatz in the planar limit. The unknown coefficients are then determined from a chain of relations between correlators with different weights, following from comparing their light-cone OPEs.  Interestingly, we need to consider the set of all such correlators and all the relations between them, in a kind of bootstrap procedure.

We have used the known correlator $\cG_{2222}$ as the starting point of the recursion. The three-loop correlator $\cG_{2222}$ was found in \cite{Eden:2011we} with the help of another, exceptional property -- the hidden permutation symmetry between external and Lagrangian insertion points for the operators $\cO^{(2)}$. We have also used the same symmetry in the present work, if one or more of the four operators are of the type $\cO^{(2)}$. So, it may seem that this symmetry is an essential ingredient of the whole construction of integrands. However, we have experimented with a more general ansatz where the symmetry is not taken into account. Using the light-cone OPE consistency conditions from this paper we were able to determine all but two coefficients at two loops and seven at three loops. Then we applied the Minkowski and Euclidean logarithmic divergence criteria from \cite{Eden:2012tu} and succeeded in fixing {\it all the coefficients}, including those in the correlator $\cG_{2222}$. So, the hidden symmetry of  \cite{Eden:2011we} may be very helpful but is not indispensable, at least up to three loops. 

We would like to emphasise the role of the planar limit in our construction. Not only it greatly reduces the size of our ansatz for the integrand, but most importantly, it is responsible for the universality property of the OPE structure constants discussed in Sect.~\ref{s33}. Without the OPE relation \p{310} that follows from this universality we would not be able to go very far in the non-planar case. We interpret this as a sign of some new type of integrability for the correlation functions of half-BPS operators.

To further elucidate the predictability (or integrability) of these correlators we have to see what happens at higher loops. We have some preliminary encouraging results at four loops. We hope that they can be useful for checking the recent integrability predictions for the OPE structure constants \cite{Eden:2015ija,Basso:2015eqa}. However, we can only provide the answer in terms of four-loop conformal integrals, which will have to be evaluated by some modern techniques \cite{Drummond:2013nda,Caron-Huot:2014lda}.

Ultimately, the goal would be to try to construct all correlators of half-BPS operators with an arbitrary number of points and at arbitrary loop levels, using only basic properties of the integrands. In case of success this can shed new and very nontrivial light on the origin of the remarkable properties of scattering amplitudes/Wilson loops. The latter are known to be light-like limits of correlation functions \cite{Alday:2010zy,Adamo:2011dq}. This duality is most easily seen at the level of their integrands \cite{Eden:2010zz,Eden:2011yp,Eden:2011ku}.

\section*{Acknowledgements}

E.S. is grateful to Sergio Ferrara, Vasco Gon\c{c}alves, Yassen Stanev and Pedro Vieira for helpful discussions. We acknowledge discussions with Burkhard Eden at the early stages of this project. D.C. and E.S. acknowledge
partial support by the French National Agency for Research (ANR) under contract StrongInt (BLANC-SIMI-4-2011). The work of D.C. has been supported by the ``Investissements d'avenir, Labex ENIGMASS''. P.H. acknowledges support from the STFC Consolidated Grant ST/L000407/1 and the Marie Curie network GATIS (\href{http://gatis.desy.eu}{gatis.desy.eu}) of the European Union's Seventh Framework Programme FP7/2007-2013/ under REA Grant Agreement No 317089. J.M.D acknowledges support from ERC grant 648630.

\appendix

\section{Uniformity of the correlators}
\label{planarity2}

As discussed in Sect.~\ref{s32}, the planarity of component terms  is a
powerful restriction on the correlation functions. The light-cone OPE
condition relating different correlators is another powerful
condition. In combination they are enough to determine all three-loop
correlation functions up to a single unfixed coefficient. However
even before performing the detailed analysis leading to this
conclusion we can use the two restrictions to deduce  a uniform structure
for correlation functions of sufficiently high weights.

More concretely, they combine to imply that at two loops
\begin{align}
  \label{eq:6}
  f^2_{\{b_{12},b_{13},b_{14},b_{23},b_{24},b_{34}\}} =
  f^2_{\{1,b_{13},b_{14},b_{23},b_{24},b_{34}\}} \quad \text{for all }
  b_{12}\geq 1, b_{13},\dots b_{34} \geq 0\ ,
\end{align}
and similarly at three loops
\begin{align}
  f^3_{\{b_{12},b_{13},b_{14},b_{23},b_{24},b_{34}\}} =
  f^3_{\{2,b_{13},b_{14},b_{23},b_{24},b_{34}\}} \quad \text{for all }
  b_{12}\geq 2, b_{13},\dots b_{34} \geq 0\ .
\end{align}
The choice of $b_{12}$ here is simply for convenience and similar
equations apply for  any other $b_{ij}$. These equations imply that we
can restrict our attention  to the set $b_{ij} \in \{0,1\}$ at two loops
and $b_{ij} \in  \{0,1,2\}$ at three loops: all other cases will
reduce to these cases. For example the above equation implies that 
$f^3_{\{7,1,5,0,1,8\}} = f^3_{\{2,1,2,0,1,2\}}$.

To see where this comes from we first consider two loops. The
light-cone OPE in the form of~\eqref{321'}  implies that $P^{\ell}_{\{b_{ij}\}}\vert_{b_{12} \to b_{12} +1}  - P^{\ell}_{\{b_{ij}\}} = O(x_{12}^2)$.  Since at two loops the numerator has the form~\eqref{eq:2}, this means the difference between the two numerators has only two unfixed terms:
\begin{align}
  \label{eq:7}
   P^{2}_{\{b_{ij}\}}\vert_{b_{12} \to b_{12} +1}-P^{2}_{\{
  b_{ij}\}} =  x_{12}^2 \Big(a_1 x_{34}^2 x_{56}^2 +a_2 (x_{35}^2
  x_{46}^2+x_{36}^2 x_{45}^2)\Big)\ .
\end{align}

We can then ask what values of $a_1$ and $a_2$ are consistent with  planarity.
Inserting $P^{2}_{\{b_{ij}\}}$ into the expression for the corresponding correlation function~\eqref{Gell} and then replacing it with the right-hand side of~\eqref{eq:7} we see that the constants $a_1,a_2$  
appear as
 \begin{align}
   \label{eq:8}
  & \cI \times \prod_{ij \neq 12} d_{ij}^{b_{ij}} \times
   \frac{x_{12}^2 \Big(a_1 x_{34}^2 x_{56}^2 +a_2 (x_{35}^2
  x_{46}^2+x_{36}^2 x_{45}^2)\Big) } {(x_{12}^2)^{b_{12}} \prod_{1\leq
   p<q\leq 6} x_{pq}^2} \ .
 \end{align}
If $b_{12}\geq 1$ these terms are non-planar (since the $x_{12}^2$ in
the numerator does not cancel that in the denominator). Thus planarity requires $a_1=a_2=0$ and we deduce~\eqref{eq:6}.

The three-loop proof is very similar, the difference being that the presence of the additional power in the  numerator at three loops (see~\eqref{eq:2}) delays the universal structure by one level.

{ Now we briefly explain why \p{321'} follows from \p{321''} and planarity at two and three loops.
We will prove a more general statement from which this one follows. 
Consider the equation}
\begin{align} \lb{A5}
\sum_{k = 0}^{n}(x^2_{13} x^2_{24})^{k} (x^2_{14}x^2_{23})^{n-k} Q_k(x) = 0
\end{align}
{ for a set of polynomials $Q_k(x)$ of the form \p{eq:2} at two and three loops, correspondingly, and $n > 0$.
We assume that each contribution in \p{A5} is planar, i.e. for each $k$ the rational function}
\begin{align} \lb{A6}
Q_k / \bigl( (x_{13}^2 x_{24}^2)^{n-k} (x^2_{14}x^2_{23})^{k} \prod_{i<j} x_{ij}^2 \bigr) 
\end{align}
{ corresponds to a set of planar graphs. Then $Q_k = 0, k = 0 , 1 , \ldots n,$ is the only solution of \p{A5}.}

{ Indeed, the different terms on the left-hand side of \p{A5} can possibly cancel each other only if 
$Q_k = x_{13}^2 x_{24}^2 f_k(x) + x_{14}^2 x_{23}^2 g_k(x)$ with some polynomials $f_k(x),g_k(x)$.
Substituting this in \p{A6}, we see that all contributions correspond to nonplanar graphs
according to the argument around \p{eq:8}. Let us mention that 
this statement relies upon a special property of the two-loop and three-loop
planar graphs from \p{eq:9}: adding any further edge to either
of them makes them non-planar. This property is not valid at four loops and higher.}

{ Finally we show that at two and three loops the strongest criterion \p{321} is a consequence
of \p{321'} and planarity. At two loops \p{321} is equivalent to \p{321'} if $b_{12} = 0$,
and if $b_{12} \geq 1$ we can evoke the uniformity property formulated above. At three loops the relationship between the two criteria is more involved. We need to show that
$P^3_{\{b_{ij}\}}|_{b_{12} = 2} - P^3_{\{b_{ij}\}}|_{b_{12} = 1} = O(x_{12}^2)$ and planarity imply that
$P^3_{\{b_{ij}\}}|_{b_{12} = 2} - P^3_{\{b_{ij}\}}|_{b_{12} = 1} = O(x_{12}^4)$.
Indeed, let $P^3_{\{b_{ij}\}}|_{b_{12} = 2} - P^3_{\{b_{ij}\}}|_{b_{12} = 1} = x_{12}^2 f(x)$ with a
polynomial $f(x)$ such that $f(x)|_{x_{12}^2 = 0} \neq 0$. Then following the argument around \p{eq:8}
we conclude that
\begin{align}
(x_{12}^2 f(x)) / \bigl((x^2_{12})^{b_{12}}\prod_{i<j} x_{ij}^2 \bigr)
\end{align}
with $b_{12} = 1$ necessarily produces nonplanar graphs. The reason is that the $x^2_{12}$ in the numerator cannot cancel the $x_{12}^4$ in the denominator.

In conclusion, the three constraints -- the weaker \p{321''}, the intermediate \p{321'} and the strongest \p{321}, are in fact equivalent up to three loops if we assume planarity. This is however not true starting from four loops. 
}

\section{The superconformal OPE}\label{ApB}

In~\cite{Heslop:2001gp} a manifestly superconformal form for the OPE
in $\cN=4$ SYM was written down in analytic superspace which has
coordinates
\begin{align}\label{eq:3}
  X^{AA'}= \left(
  \begin{array}{cc}
x^{\alpha\dot\alpha}&0\\
0&y^{aa'}\ .
  \end{array}
\right).
\end{align}
The indices $A,A'$ are superindices carrying representations of
$SL(2|2)$ and they split as $A=(\alpha|a)$ and $A'=(\dot\alpha|a')$
with $\alpha,\dot\alpha$ carrying the left and right spinor
representations of the Lorentz group and $a,a'$ two different $SU(2)$
subgroups of $SU(4)$. We have dropped  all superspace coordinates
whcih is why we have 0's in the off-diagonal blocks
in~(\ref{eq:3}) but these can  be easily put back in.

Then the  OPE of two
half BPS operators takes the form
\begin{align}\label{eq:5}
   \cO_p(X_1) \cO_q(X_2) = \sum_{\cO} C_{pq\cO} 
 (d_{12})^{{1\over2}(p+q-L)} (X_{12})^{\underline A  \underline A'}
 \cO^L_{\underline A \underline A'}(2) +\ldots\ .
\end{align}
Here the sum is over all operators in the theory, $L=\Delta-S$ is the twist
of the operator and the underlined index is a multi-index indicating a
tensor in a representation determined by the operator. The operators
of most interest  here are the semi-short operators whose highest
weight states have spin $S$, lie in the $SU(4)$ representation space with
Dynkin labels $[MNM]$ and which have  twist 
\begin{align}
L=2M+N+2\ .
\end{align}
For such
operators the $SL(2|2)$ representation which the indices $\underline
A,\underline A'$ lie in is given by a hook-shaped Young tableau with
top row of length $S+2$ and left column of height $M+1$. There are
thus $M+S+2$ boxes in total.

\usetikzlibrary{arrows}
\usetikzlibrary{positioning}

\newcount\tableauRow
\newcount\tableauCol
\def\tableauDim{0.4}
\newenvironment{Tableau}[1]{%
  \tikzpicture[scale=0.4,draw/.append style={thick,black},
                      baseline=(current bounding box.center)]
    % now draw the tableau
    \tableauRow=-1.5
    \foreach \Row in {#1} {
       \tableauCol=0.5
       \foreach\k in \Row {
         \draw(\the\tableauCol,\the\tableauRow)rectangle++(1,1);
         \draw(\the\tableauCol,\the\tableauRow)+(0.5,0.5)node{$\k$};
         \global\advance\tableauCol by 1
       }
       \global\advance\tableauRow by -1
    }
}{\endtikzpicture}
\newcommand\tableau[1]{\begin{Tableau}{#1}\end{Tableau}}

\begin{center}
  \begin{Tableau}{{,,,,,,,,,,,,,,,,,,,},{ },{ },{ },{ },{ },{ },{ },{ }}
\draw[black,<->]
    (-0.35,-1)--node[anchor=east]{$ M $}(-0.35,-8.9)  ;
\draw[black,<->]
    (2,.35)--node[above=.2em ,anchor=south]{$S$}(20,.35)  ;
 \end{Tableau}
\end{center}

Since these are Young tableaux involving superindices, horizontal boxes
correspond to  symmetrisation of 
$\alpha,\beta$ indices, but anti-symmetrisation of  $a,b$ indices, and
vice versa for
vertical boxes.
The HWS is obtained by filling as many of the boxes as possible with
$a$ indices. Since the index only takes on 2 values, at most two  $a$
indices can be found in the same row (more than two in a row would
correspond to antisymetrising 3 indices and thus vanish). So the HWS has
the first two columns filled with $a$'s and all other columns filled
with $\alpha$'s. Other index choices with fewer $a$'s and more
$\alpha$s correspond to acting with $Q_{\alpha a}$ on the HWS. 

The primed indices follow a similar story (and are in the same
representation as the unprimed indices for a nonvanishing OPE of half BPS
operators).

Let us consider a simple example,
the Konishi operator has $L=2, S=M=N=0$ and the
corresponding Young tableau is simply the symmetric representation, so
$\cK_{(AB)(A'B')}$. Writing out all bosonic terms we see that it
decomposes into the bosonic terms
  \begin{align}
    \cA  &\qquad   \tableau{{a,b}} \otimes \tableau{{a',b'}} \notag\\
    \cB_{\beta \dot \beta aa'} = Q_{\beta a} \bar Q_{\dot \beta a'} \cA 
     &\qquad   \tableau{{a,\beta}} \otimes \tableau{{a',\dot \beta}}
       \notag \\
    \cC  = Q^2_{\alpha \beta } \bar Q^2_{\dot \alpha \dot \beta} \cA   &\qquad   \tableau{{\alpha,\beta}} \otimes \tableau{{\dot\alpha,\dot\beta}}.
  \end{align}

We would expect more terms where we make different choices
for the two $SL(2|2)$ reps. However these are the only terms we will
obtain when switching off the superspace variables since in the OPE we
contract with the block diagonal $X$ given in~(\ref{eq:3}), thus tying together the
primed and unprimed indices.

The general case is similar. Consider any operator with twist
$L=2M+N+2$, $\cO_{\underline A  \underline A'}$. There are $S+M+2$ unprimed superindices and $S+M+2$ primed superindices, symmetrised according to the above hook-shaped Young Tableau. Splitting the superindices into the $SL(2)$ subgroups we obtain the following components
\begin{align}
  &L{-}2M{-}N\notag \quad \text{Component operator}\\
&  \ \  2\qquad \qquad   \cA^{\D,S}_{[MNM]}  = \cO_{bb'}^{bb'}{}_{\alpha(S)\dot \alpha(S) a(M)a'(M)}   \notag\\
 &   \left.
   \begin{array}{l}
 0\qquad \qquad    \cB^{\D+1,S+1}_{[(M+1)N(M+1)]} = \cO_{\alpha(S+1)\dot \alpha(S+1) a(M+1)a'(M+1)}\\
 2\qquad \qquad    \cB^{\D+1,S+1}_{[(M-1)(N+2)(M-1)]} = \cO_{bb'}^{bb'}{}_{\alpha(S+1)\dot \alpha(S+1) a(M-1)a'(M-1)}\\
 4\qquad \qquad    \cB^{\D+1,S-1}_{[(M-1)(N+2)(M-1)]} = \cO_{\beta\dot\beta bb'}^{\beta\dot \beta bb'}{}_{\alpha(S-1)\dot \alpha(S-1) a(M-1)a'(M-1)}
   \end{array}\right\}
\in  Q \bar Q  \cA^{\D,S}_{[MNM]}   \notag \\
 &   \left.
   \begin{array}{l}
  0\qquad \qquad   \cC^{\D+2,S+2}_{[M(N+2)M]} =\cO_{\alpha(S+2)\dot \alpha(S+2) a(M)a'(M)} \\
  2\qquad \qquad   \cC^{\D+2,S}_{[M(N+2)M]} =\cO_{\beta\dot\beta}^{\beta\dot\beta}{}_{\alpha(S)\dot \alpha(S) a(M)a'(M)} \\
  4\qquad \qquad   \cC^{\D+2,S}_{[(M-2)(N+4)(M-2)]} = \cO_{bb'\beta\dot\beta}^{bb'\beta\dot\beta}{}_{\alpha(S)\dot \alpha(S) a(M-2)a'(M-2)}
   \end{array}\right\}
                                         \in   Q^2 \bar Q^2  \cA^{\D,S}_{[MNM]}  \notag \\
&  \  \  2\qquad \qquad  \cD^{\D+3,S+1}_{[(M-1)(N+4)(M-1)]} = \cO_{\beta\dot \beta}^{\beta\dot \beta}{}_{\alpha(S+1)\dot \alpha(S+1) a(M-1)a'(M-1)}   \in  Q^3 \bar Q^3 \cA^{\D,S}_{[MNM]}\ .
  \end{align}

  These are all the components which occur in the free OPE.
Writing out the super OPE~(\ref{eq:5}) in components we then get
\begin{align}\label{eq:5a}
   \cO_p(X_1) \cO_q(X_2) &= \sum_{\cO} C_{pq\cO} 
 (d_{12})^{{1\over2}(p+q-L)} (x_{12})^{\alpha(S)\dot
  \alpha(S)}(y_{12})^{a(M-1)a'(M-1)}\times\notag \\
&\quad \left(
 y_{12}^2 y_{12}^{a_Ma'_M} \cA_{\alpha(S)\dot \alpha(S) a(M)a'(M)}(2)
  +\dots + x_{12}^2 x_{12}^{\alpha_{S+1} \dot \alpha_{S+1}}
  \cD_{\alpha(S+1)\dot \alpha(S+1) a(M-1)a'(M-1)}(2) \right) \ .
\end{align}

  However in the interacting theory the semi-short multiplet combines with three others to form a long multiplet.  
In fact only one of these three multiplets contributes to the
OPE. This is another semi-short multiplet with highest weight state
$M\rightarrow M+1, S\rightarrow S-1, N\rightarrow N$,
$(\cA')^{\D+1,S-1}_{[(M+1),N,(M+1)]}$. Whilst in the free theory $\cA,
\cA'$ are both superconformal primaries of independent multiplets, in
the interacting theory $\cA'$ becomes a descendant of $\cA$,
$(\cA')^{\D+1,S-1}_{[(M+1),N,(M+1)]} \in Q \bar Q \cA^{\D,S}_{[M,N,M]}
$. In total therefore in the interacting theory we have the following
components contributing to a long supermultiplet 
\begin{align}\label{eq:10}
  &L{-}2M{-}N\notag \quad \text{Component operator}\\
&  \ \  2\qquad \qquad   \cA^{\D,S}_{[MNM]}   \notag\\
 &   \left.
   \begin{array}{l}
 0\qquad \qquad    \cB^{\D+1,S+1}_{[(M+1)N(M+1)]}\\
 2\qquad \qquad    \cB^{\D+1,S+1}_{[(M-1)(N+2)(M-1)]}\\
     4\qquad \qquad    \cB^{\D+1,S-1}_{[(M-1)(N+2)(M-1)]}\\
     2\qquad \qquad    (\cA')^{\D+1,S-1}_{[(M+1)N(M+1)]}
   \end{array}\right\}
\in  Q \bar Q  \cA^{\D,S}_{[MNM]}   \notag \\
 &   \left.
   \begin{array}{l}
  0\qquad \qquad   \cC^{\D+2,S+2}_{[M(N+2)M]}\\
  2\qquad \qquad   \cC^{\D+2,S}_{[M(N+2)M]} \\
     4\qquad \qquad   \cC^{\D+2,S}_{[(M-2)(N+4)(M-2)]} \\
      0\qquad \qquad    (\cB')^{\D+2,S}_{[(M+2)N(M+2)]}\\
 2\qquad \qquad    (\cB')^{\D+2,S}_{[M(N+2)M]}\\
     4\qquad \qquad    (\cB')^{\D+2,S-2}_{[M(N+2)M]}
   \end{array}\right\}
                                         \in   Q^2 \bar Q^2  \cA^{\D,S}_{[MNM]}  \notag \\
&   \left.
  \begin{array}{l}
  2\qquad \qquad  \cD^{\D+3,S+1}_{[(M-1)(N+4)(M-1)]}\\
  0\qquad \qquad   (\cC')^{\D+3,S+1}_{[(M+1)(N+2)(M+1)]} \\
  2\qquad \qquad   (\cC')^{\D+3,S-1}_{[(M+1)(N+2)(M+1)]}\\
     4\qquad \qquad   (\cC')^{\D+3,S-1}_{[(M-1)(N+4)(M-1)]}
   \end{array}\right\}
  \in  Q^3 \bar Q^3 \cA^{\D,S}_{[MNM]}  \notag \\
  &  \ \   2\qquad \qquad  (\cD')^{\D+4,S}_{[M(N+4)M]} \in   Q^4 \bar Q^4  \cA^{\D,S}_{[MNM]}
  \ .  
\end{align}

Writing out the full super OPE is now equivalent to summing two copies of the semi-short OPE ~(\ref{eq:5a}) with appropriate quantum numbers. In components we then get
\begin{align}\label{eq:5b}
   \cO_p(X_1) \cO_q(X_2) &= \sum_{\cO} C_{pq\cO} 
 (d_{12})^{{1\over2}(p+q-L)} (x_{12})^{\alpha(S)\dot
  \alpha(S)}(y_{12})^{a(M-1)a'(M-1)}\times\notag \\
&\quad {x_{12}^2\over y_{12}^2} \left(
 y_{12}^4 y_{12}^{a_Ma'_M} \cA_{\alpha(S)\dot \alpha(S) a(M)a'(M)}(2)
  +\dots + x_{12}^4 x_{12}^{\alpha_{S} \dot \alpha_{S}}
  (\cD')_{\alpha(S)\dot \alpha(S) a(M)a'(M)}(2) \right) \ .
\end{align}

Note that this long OPE  can also be derived directly using superindices rather than by summing two semi-short OPEs as we have done here.   

Note also that in the free theory one can write down explicit forms for the single trace
operators in question. They have the schematic form
\begin{align}
  \cO_{\underline A \underline A'} = Tr ( \partial^{a+s+2}_{\underline
  A \underline A'} W^L)
\end{align}
where the derivatives can act on any of the different $W$s and in
general will be linear combinations of such terms.

Finally in the main text we will be interested in considering this OPE in the limit in which both $x_{12}^2$ and $y_{12}\rightarrow 0$ (but with the ratio fixed). In this limit only the operators from the above list with $L-2M-N=0$ survive.

%%%%%%%%%

%%%%%%%%%%%%%%%%%%

%%%%%%%%%%%%%%%%%%

%%%%%%%%%%%%%%%%%%

%%%%%%%%%

\end{document}